\documentclass[structabstract]{aa}
\usepackage{txfonts}
\usepackage{natbib}
\usepackage{graphicx}
\usepackage{lscape}
\usepackage{longtable}

\newcommand{\name}{1E~0102.2-7219}
\newcommand{\NH}{\mbox {$N_{\mathrm H}$}}

\newcommand{\chan}{{\em Chandra}}
\newcommand{\rosat}{{\em ROSAT}}
\newcommand{\xmmn}{XMM-{\em{Newton}}}
\newcommand{\suzaku}{{\em Suzaku}}
\newcommand{\swift}{{\em Swift}}

\def\keV{\thinspace \rm{keV}}
\def\eV{\thinspace \rm{eV}}

\begin{document}
\titlerunning{X-ray Cross-calibration with SNR 1E~0102.2-7219}
\authorrunning{P.P. Plucinsky et al.}

\title{SNR \name~as an X-ray Calibration Standard in the 0.5--1.0 keV
Bandpass and Its Application to the CCD Instruments aboard \chan,
\suzaku, \swift\/ and \xmmn }
\subtitle{}
\author{
Paul~P. Plucinsky\inst{1}
\and
Andrew P. Beardmore\inst{2}
\and
Adam Foster\inst{1}
\and
Frank Haberl\inst{3}
\and
Eric D. Miller\inst{4}
\and
A.M.T. Pollock\inst{5}
\and
Steve Sembay\inst{2}
}
\institute{
Harvard-Smithsonian Center for Astrophysics, MS-3, 60 Garden Street, Cambridge, MA 02138, USA
\and
Department of Physics and Astronomy, University of Leicester, Leicester LE1 7RH, United Kingdom
\and
Max-Planck-Institut f\"ur Extraterrestrische Physik, Giessenbachstra{\ss}e , 85748 Garching, Germany
\and
MIT Kavli Institute for Astrophysics and Space Research, Cambridge, MA
 02139,
\and 
University of Sheffield, Department of Physics and Astronomy, Hounsfield Road, Sheffield S3 7RH, United Kingdom
}
\date{Received 3 May 2016 / Accepted 30 June 2016}
\abstract
{The flight calibration of the spectral response of CCD instruments 
below 1.5 keV is difficult in general because of the  lack of strong 
lines in the on-board calibration sources typically available. This
calibration is also a function of time due to the effects of radiation
damage on the CCDs and/or the accumulation of a contamination layer on
the filters or CCDs.
}
{We desire a simple comparison of the absolute effective areas of the
current generation of CCD instruments onboard the following observatories:
\chan\/ ACIS-S3, \xmmn\/ (EPIC-MOS and EPIC-pn), \suzaku\/
XIS, and \swift\/ XRT and a straightforward comparison of the
time-dependent response of these instruments across their respective 
mission lifetimes. 
}
{We have been using \name, the brightest 
supernova remnant in the Small Magellanic Cloud, to evaluate  and
modify the 
response models of these instruments. \name\/ has strong lines of 
O, Ne, and Mg below 1.5 keV and little or no Fe emission to complicate the spectrum. The spectrum of \name\/ has been
well-characterized using the RGS gratings
instrument on \xmmn\/ and the HETG
 gratings instrument on \chan\/.
As part of the activities of the {\em International Astronomical
Consortium for High Energy Calibration} (IACHEC),
we have developed a standard spectral model for \name\/ and fit this model to
the spectra extracted from the CCD instruments.  The model is
empirical in that it includes Gaussians for the identified lines, an
absorption component in the Galaxy, another absorption component
in the SMC, and two thermal continuum components with different temperatures.
In our fits, the model is highly constrained in that only the 
normalizations of the four brightest lines/line complexes (the 
\ion{O}{vii}~He$\alpha$~triplet, \ion{O}{viii}~Ly$\alpha$ line, the 
\ion{Ne}{ix}~He$\alpha$~triplet, and the \ion{Ne}{x}~Ly$\alpha$ line) 
and an overall
normalization are allowed to vary, while all other components are
fixed. We adopted this approach to provide a straightforward comparison
of the measured line fluxes at these four energies. We have examined
these measured line fluxes as a function of time for each instrument
after applying the most recent calibrations that account for the
time-dependent response of each instrument.
 }
{We perform our effective area comparison with representative, early
mission data when the radiation damage and contamination layers were
at a minimum, except for the \xmmn\/ EPIC-pn instrument which is
stable in time.
We find that the measured fluxes of the \ion{O}{vii}~He$\alpha$~{\em r} line,
the \ion{O}{viii}~Ly$\alpha$ line, the \ion{Ne}{ix}~He$\alpha$~{\em r} line, and
the \ion{Ne}{X}~Ly$\alpha$ line generally agree to within $\pm10\%$
for all instruments, with 38 of our 48 fitted normalizations within
$\pm10\%$ of the IACHEC model value.  
We then fit all available observations of \name\/ for the CCD
instruments close to the on-axis position to characterize the time
dependence in the 0.5--1.0~keV band.
We present the measured line normalizations as a function of time for
each CCD instrument so that the users 
may estimate the uncertainty in their measured line fluxes for the 
epoch of their observations.  


}
%
{}
\maketitle

\section{Introduction}\label{intro}

  This paper reports the progress of a working group within the {\em 
International Astronomical Consortium for High Energy Calibration}
(IACHEC) to develop a calibration standard for X-ray astronomy in the 
bandpass from 0.3 to 1.5 keV.  An introduction to the IACHEC
organization, its objectives and meetings, may be found at the web
page {\tt http://web.mit.edu/iachec/}.  Our working group was tasked with
selecting celestial sources with line-rich spectra in the 0.3--1.5~keV
bandpass which would be suitable cross-calibration targets for the
current generation of X-ray observatories.  The desire for strong
lines in this bandpass stems from the fact that the
quantum efficiency and spectral resolution of the current CCD-based
instruments is changing rapidly from 0.3 to 1.5~keV but the on-board
calibration sources currently in use typically have strong lines at
only two energies, 1.5~keV (Al~K$\alpha$) and 5.9~keV (Mn~K$\alpha$). 
The only option
available to the current generation of flight instruments to calibrate
possible time variable responses in this bandpass
is to use celestial sources.  The missions which
have been represented in this work are the {\em Chandra X-ray
Observatory} (\chan)~\citep{weiss2000,weiss2002}, the {\em X-ray Multimirror
Mission} (\xmmn)~\citep{jansen2001}, the {\em ASTRO-E2 Observatory} ({\em
Suzaku}), and  the {\em Swift} Gamma-ray Burst Mission 
(\swift)~\citep{gehrels2004}.  Data from the following instruments have
been included in this analysis: the {\em High-Energy Transmission Grating}
(HETG)~\citep{canizares2005} and the {\em Advanced CCD Imaging Spectrometer} 
(ACIS)~\citep{bautz98,garmire03,garmire92}  on \chan, the  {\em Reflection Gratings Spectrometers} (RGS)~\citep{denherder2001}, the 
{\em European Photon Imaging Camera} (EPIC) 
{\em Metal-Oxide Semiconductor} (EPIC-MOS)~\citep{turner2001}  CCDs and the 
EPIC p-n junction (EPIC-pn)~\citep{strueder2001}  CCDs on \xmmn, the 
{\em X-ray Imaging Spectrometer} (XIS) on {\em Suzaku}, and the 
{\em X-ray Telescope} (XRT)~\citep{burrows2005,godet2007}  on {\em Swift}.

  Ideal calibration targets would need to possess the following 
qualities.  The source would need to be constant in time, 
to have a simple spectrum defined by a few bright lines
with a minimum of line-blending, and to be extended
so that ``pileup'' effects in the CCDs are minimized but not so
extended that the off-axis response of the telescope dominates
the uncertainties in the response.  Our working group focused on supernova
remnants (SNRs) with thermal spectra and without a central source such as a
pulsar, as the class of source which had the greatest likelihood of
satisfying these criteria.  We narrowed our list to the Galactic SNR
Cas~A, the Large Magellanic Cloud remnant N132D and the Small
Magellanic Cloud remnant \name~(hereafter E0102).  We discarded 
Cas~A since it is relatively young ($\sim350$~yr), with significant brightness
fluctuations in the X-ray, radio, and optical over the past three
decades \citep{patnaude2007,patnaude2009,patnaude2011}, it contains
a faint (but apparently variable) central source,
and it is relatively large (radius $\sim3.5$~arcminutes).  We
discarded N132D because it has a complicated, irregular morphology in
X-rays \citep{borkowski2007}
and its spectrum shows strong, complex Fe emission \citep{behar2001}. 
The spectrum of N132D is significantly more complicated in the
0.5--1.0~keV bandpass than the spectrum of E0102.
We therefore settled on E0102 as the
most suitable source given its relatively uniform morphology, small
size (radius $\sim0.4$~arcminutes), and comparatively simple X-ray
spectrum.

We presented preliminary results from this effort in
\cite{plucinsky2008} and \cite{plucinsky2012} using a few 
observations with the calibrations available at that time.  
In this paper, we present an updated analysis of the representative
data acquired early in the various missions and
expand our investigations to include a characterization of the time
dependence of the response of the various CCD instruments. The low
energy responses of some of the instruments (ACIS-S3, EPIC-MOS, \& XIS)
included in this analysis have a complicated time dependence due to the
time-variable accumulation of a contamination layer. A primary
objective of this paper is to inform the Guest Observer communities of
the respective missions on the current accuracy of the calibration at
these low energies.

\medskip

\section{The SNR 1E~0102.2-7219}\label{e0102}

\begin{figure}[hbtp]
 \begin{center}
  \resizebox{\hsize}{!}{\includegraphics{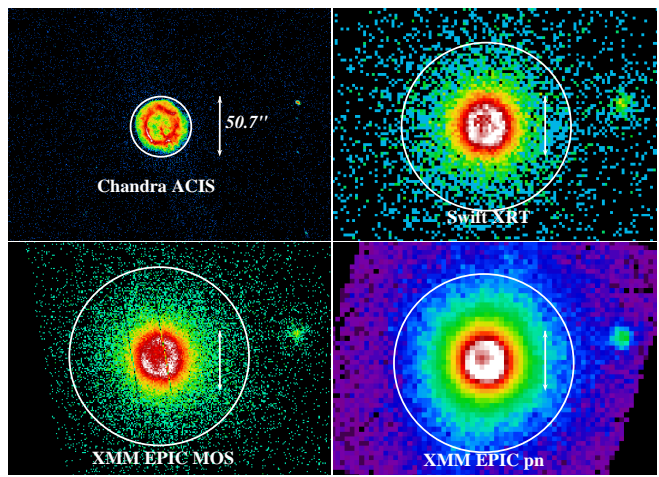}}
 \end{center}
 \caption{ \label{fig:image4}   Images of E0102 from ACIS-S3 (top left),
EPIC-MOS (bottom left), XRT (top right), EPIC-pn (bottom right). 
The white circles indicate the extraction regions used for the
spectral analysis. The fine structure in E0102 is evident in the
\chan\/ image. Note that the \chan\/ extraction region is the smallest.}
 %
\end{figure}

\begin{figure}[hbtp]
 \begin{center}
  \resizebox{\hsize}{!}{\includegraphics{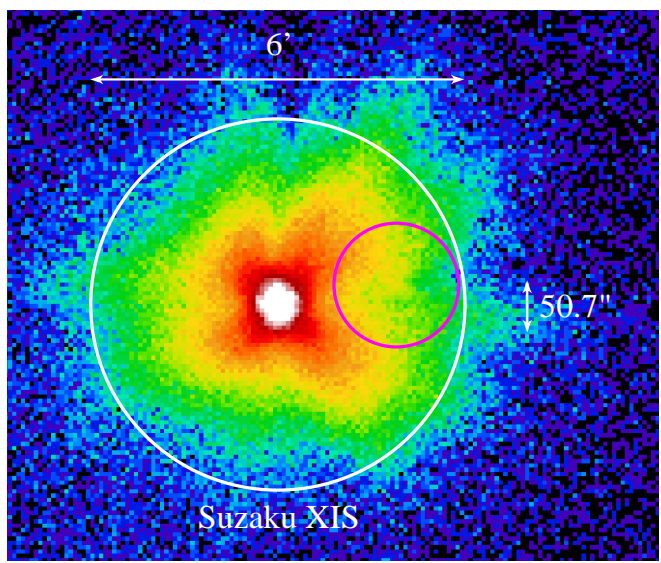}}
 \end{center}
 \caption{ \label{fig:image_xis}  XIS image of E0102. 
The white circle
indicates the extraction region used for the spectral analysis, a 6 arcmin
diameter circle.  The magenta circle indicates the region excluded due to
the contaminating point source RXJ0103.6-7201, which can also be seen in
the images from the other instruments shown in Fig. \ref{fig:image4}.
The vertical white line indicates the size of the ACIS-S3 extraction region
for comparison. }
 %
\end{figure}

 The SNR E0102 was discovered 
by the {\em Einstein Observatory}~\citep{seward1981}.  It is the
brightest SNR in X-rays in the {\em Small Magellanic Cloud} (SMC).
E0102 has been extensively imaged by \chan~\citep{gaetz2000,hughes2000} and 
\xmmn~\citep{sasaki2001}.  
Figures~\ref{fig:image4} and~\ref{fig:image_xis}
 show images of E0102 with the relevant
spectral extraction regions for each of the instruments included in this
analysis.  E0102 is classified as an ``O-rich'' SNR based on the
optical spectra acquired soon after the X-ray
discovery~\citep{dopita1981} and confirmed by followup observations~\citep{tuohy1983}. 
The age was estimated as $\sim1,000$~yr by \cite{hughes2000} based on
the expansion deduced from comparing \chan\/ images to \rosat\/
images, but \cite{finkelstein2006} estimate an age of $\sim2,050$~yr
based on twelve filaments observed during two epochs by the 
{\em  Hubble Space Telescope} (HST). \cite{blair1989} presented the first UV
spectra of E0102 and argued for a progenitor mass between 15 and
25~$M_\odot$ based on the derived O, Ne, and Mg abundances.  
\cite{blair2000} refined this argument with {\em Wide Field
and Planetary Camera 2} and {\em Faint Object Spectrograh} data from HST to
suggest that the precursor was a Wolf-Rayet star of between  25 and
35~$M_\odot$ with a large O mantle that produced a Type Ib supernova.
\cite{sasaki2006} compared the UV spectra from the {\em Far Ultraviolet Spectroscopic Explorer}
to the CCD spectra from \xmmn\/ to conclude that a single
ionization timescale cannot fit the O, Ne, and Mg emission lines,
possibly indicating a highly structured ejecta distribution in which
the O, Ne, and Mg have been shocked at different
times. \cite{vogt2010} argued for an asymmetric, bipolar structure in
the ejecta based on spectroscopy of the [\ion{O}{III}] filaments.
The {\em Spitzer} Infrared Spectrograph detected strong lines of O and
Ne in the infrared (IR) \citep{rho2009}.
In summary, all available spectral data in the optical, UV, IR, and 
X-ray bands indicate
significant emission from O, Ne, \& Mg with very little or no emission
from Fe or other high Z elements.

\begin{figure}[hbtp]
 \begin{center}
  \resizebox{\hsize}{!}{\includegraphics[angle=0]{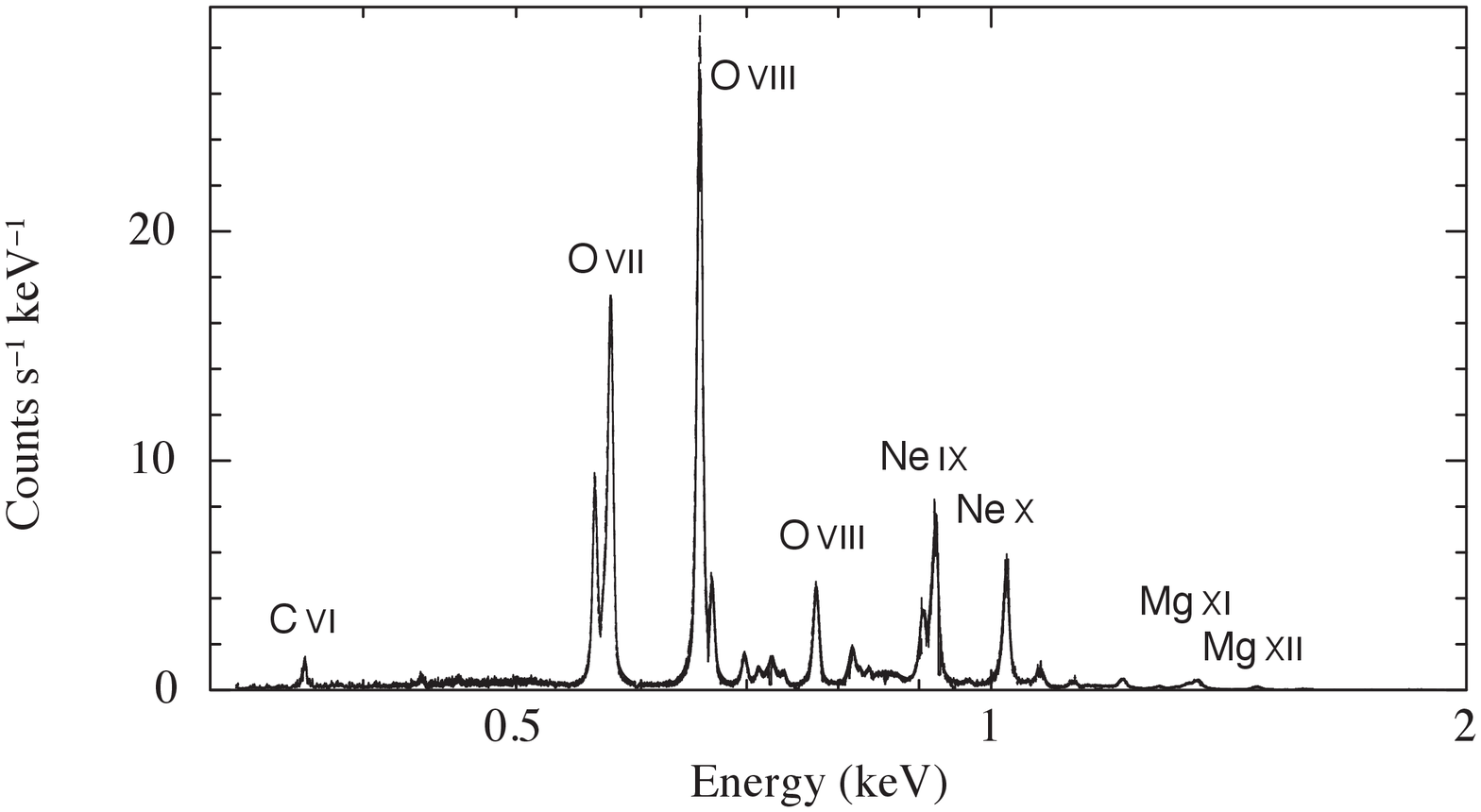}}
  \resizebox{\hsize}{!}{\includegraphics[angle=0]{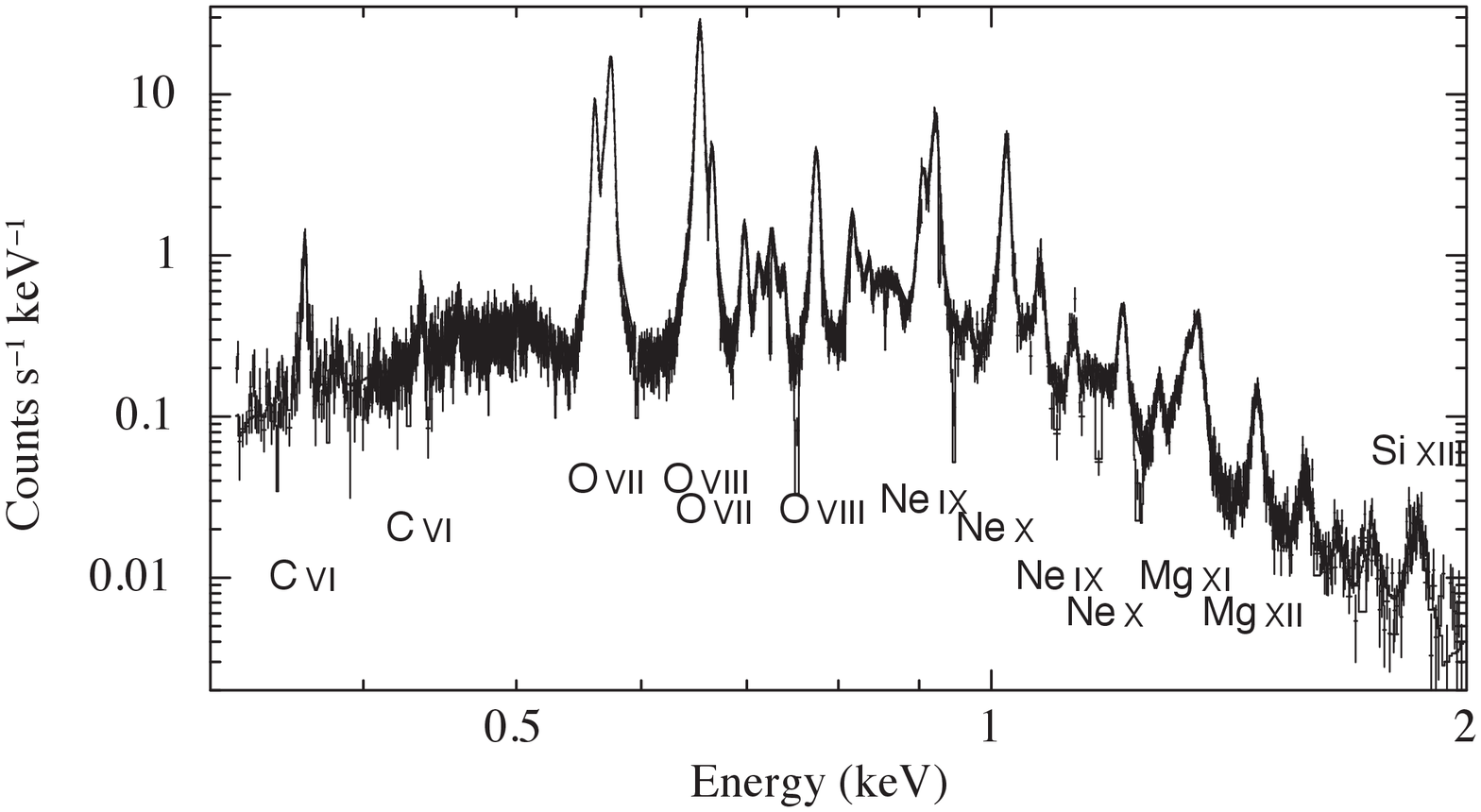}}
 \end{center}
 \caption{\label{fig:rgs_spec_lin} {(top) RGS1/RGS2 spectrum of E0102 from a combination
of 23 observations. Note that the bright lines of O and Ne
dominate the flux in this band.
(bottom) Same as top figure except with a logarithmic Y axis to emphasize the
continuum and the weakest lines. Note the absence of Fe~L emission in
this spectrum. 
} }
 %
\end{figure}

The diameter of E0102 is small enough
such that a high resolution spectrum may be
acquired with the HETG on \chan\/ and the RGS on \xmmn.
The HETG spectrum~\citep{flanagan2004} and the RGS spectrum~\citep{rasmussen2001}
both show strong lines of O, Ne, and Mg with little or no Fe,
consistent with the spectra at other wavelengths.
E0102's spectrum is relatively simple compared to a typical SNR
spectrum.  
Figure~\ref{fig:rgs_spec_lin} displays the RGS spectrum from E0102.
The strong, well-separated lines in the energy range 0.5 to
1.5~keV make this source a useful calibration target for CCD
instruments with moderate spectral resolution in this bandpass.
The source is extended enough to reduce the effects of photon pileup,
which distorts a spectrum. 
Although some pileup is expected in all the non-grating instruments 
when observed in modes with relatively long frame times.
The source is also bright enough to
provide a large number of counts in a relatively short observation.
Given these characteristics, E0102 has become a standard calibration source
that is observed repeatedly by all of the current generation of X-ray
observatories.

\section{Spectral Modeling and Fitting}\label{spectra}

\subsection{Construction of The Spectral Model}\label{model}

  Our objective was to develop a model which would be useful in
calibrating and comparing the response of the CCD instruments;
therefore, the model presented here is of limited value for
understanding E0102 as a SNR.  Our approach was to rely upon the
high-resolution spectral data from the RGS and HETG to identify and
characterize the bright lines and the continuum  
in the energy range from 0.3--2.0 keV and the moderate-spectral
resolution data from the EPIC-MOS and EPIC-pn to characterize the lines and
continuum above 2.0~keV.  Since our objective is
calibration, we decided against using any of the available plasma
emission models for several reasons. First, the \chan\/
results on E0102~\citep{flanagan2004,gaetz2000,hughes2000} have shown
there are significant spectral variations with position in the SNR,
implying that the plasma conditions are varying throughout the
remnant.  Since the other missions considered here have poorer angular
resolution than \chan, the emission from these regions 
is mixed such that the unambiguous interpretation of
the fitted parameters of a plasma emission model is difficult if not
impossible.  Second, the available parameter space in the more complex
codes is large, making it difficult to converge on a single best fit
which represents the spectrum.  We therefore decided to construct a
simple, empirical model based on interstellar absorption components, 
Gaussians for
the line emission, and continuum components which would be appropriate 
for our limited calibration objectives.

   We assumed a two component absorption model using the 
{\tt tbabs}~\citep{wilms2000} model in {\tt XSPEC}. The first component
was held fixed at $5.36\times10^{20}~{\mathrm cm^{-2}}$ to account for
absorption in the Galaxy.  The second component was allowed to vary
in total column, but with the abundances fixed to the lower abundances
of the SMC~\citep{russell1989,russell1990,russell1992}.  
We modeled the continuum using a modified version of the {\tt APEC}\  
plasma emission model~\citep{smith2001} called the {\tt ``No-Line''}
model.   
This model excludes all line emission, while retaining all continuum  
processes including bremsstrahlung, radiative recombination continua  
(RRC), and the two-photon continuum from hydrogenic and helium-like  
ions (from the strictly forbidden ${}^2S_{1/2} 2s \rightarrow$\ gnd  
and ${}^1S_0 1s2s \rightarrow$\ gnd transitions, respectively).   
Although the bremsstrahlung continuum dominates the X-ray spectrum in  
most bands and at most temperatures, the RRCs can produce observable  
edges while the two-photon emission creates 'bumps' in specific energy  
ranges.  The {\tt ``No-Line''} model assumes collisional equilibrium 
and so may  
overestimate the RRC edges in an ionizing plasma or have the wrong  
total flux in some of the two-photon continua.  However, the available  
data did not justify the use of a more complex model, while the  
simpler bremsstrahlung-only model showed residuals in the RGS spectra  
that were strongly suggestive of RRC edges.
The RGS data were adequately fit by a single continuum component, but
the HETG, EPIC-MOS, and EPIC-pn data showed an excess at energies above 2.0~keV.
We therefore added a second continuum component to account for this
emission.  

The lines were modeled as simple Gaussians in {\tt XSPEC}.
The lines were identified in the RGS and HETG data in a hierarchical
manner, starting with the brightest lines and working down to
the fainter lines. 
We have used the ATOMDB~v2.0.2~\citep{foster2012} database 
to identify the transitions which produce the observed lines. 
The RGS spectrum from 23 observations totaling
708/680~ks for RGS1/RGS2 is shown in Fig.~\ref{fig:rgs_spec_lin}(top) 
with a linear Y axis
to emphasize the brightest lines.  The spectrum is dominated by the  
\ion{O}{vii}~He$\alpha$ triplet at 560-574~eV, the \ion{O}{viii}~Ly$\alpha$
line at 654~eV, the \ion{Ne}{ix}~He$\alpha$ triplet at 905-922~eV, and the
\ion{Ne}{X}~Ly~$\alpha$ line at 1022~eV. This figure demonstrates  
the lack of strong Fe emission in the spectrum of E0102.  
  The
identification of the lines obviously becomes more difficult as the
lines become weaker.   Figure~\ref{fig:rgs_spec_lin}(bottom) shows the same
spectrum but with a logarithmic Y
axis.  In this figure, one is able to see the weaker lines more
clearly and also the shape of the continuum.  Lines were added to the
spectrum at the known energies for the dominant elements, C, N, O, Ne,
Mg, Si, S, and Fe and the resulting decrease in the reduced $\chi^2$
value was evaluated to determine if the addition of the line was
significant.  The list of lines identified in the RGS and HETG data
were checked for consistency. 
The identified lines were 
compared against representative spectra from the {\tt vpshock} model
(with lines) to ensure that no strong lines were missed.

 In this manner a list of lines in the
0.3--2.0~keV bandpass was developed based upon the RGS and HETG data.
In addition, the temperature and normalization were determined for the
low-temperature {\tt APEC  ``No-Line''} continuum component.  These
model components were then frozen and the model compared to the EPIC-pn,
EPIC-MOS, and XIS data.  Weak lines above 2.0~keV were evident in the
EPIC-pn, EPIC-MOS, and XIS data and also what appeared to be an
additional continuum component above 2.0~keV.  Several lines were
added above 2.0~kev and a high-temperature continuum component with 
kT~$\sim1.7$~keV was added.
Once the model components above 2.0~keV had been determined, the RGS
data were re-fit with components above 2.0~keV frozen to these values
and the final values for the SMC \NH\/ and the low-temperature
continuum were determined. In practice, this was an iterative process
which required several iterations in fitting the RGS and EPIC-MOS/EPIC-pn/XIS
data.  Once the absorption and continuum components were determined,
the parameters for those components were frozen and the final
parameters for the line emission were determined from the RGS data.
We included 52 lines in the final model and these lines are described in
Table~\ref{tab:model}. When fitting the RGS data, the line energies
were allowed to vary by up to 1.0~eV from the expected energy to
account for the shifts when an extended source is observed by
the RGS.  Shifts of less than 1.0~eV are too small to be significant
when fitting the CCD instrument data.  The line widths were also
allowed to vary.  
In most cases the line widths are small but non-zero,
consistent with the Doppler widths seen in the RGS~\citep{rasmussen2001}
and HETG~\citep{flanagan2004} data, $\sigma_E \approx 0.003 \times E$;  but
in a few cases noted in the table the widths are larger than this value.
This is most likely due to weak,
nearby lines which our model has ignored. We do not have an
identification for the line-like feature at 1.4317~keV, but we note
that it is weak.
  
  As noted above, the identification of the lines becomes less certain
as the line fluxes get weaker.  Our primary purpose is to characterize
the flux in the bright lines of O and Ne.  Any identification of
a line with flux less than $1.0\times10^{-4}~{\mathrm{
photons~cm^{-2}~s^{-1}}}$ in Table~\ref{tab:model} should be considered
tentative.  The Fe lines in  Table~\ref{tab:model} warrant special
discussion. There are nine Fe lines included in our model from
different ions.  
We have not verified the self-consistency of the Fe lines included 
in this model. As our objective is calibration and not the
characterization of the plasma that might produce these Fe lines, the
possible lack of consistency does not affect our analysis.
Of particular note is the \ion{Fe}{XIX}
line at 917~eV with zero flux.  We went through several iterations of
the model with this line included and excluded.  Unfortunately this
line is only 2~eV away from the \ion{Ne}{ix}~He$\alpha$~i line
at 915~eV and neither the RGS nor the HETG has the resolution to
separate lines this close together.  We have decided to
attribute all the flux in this region to the
\ion{Ne}{IX}~He$\alpha$~i line but have retained the \ion{Fe}{XIX}
line for
future investigations.  It is possible that some of the emission
which we have identified as Fe emission is due to other elements.  For
our calibration objective this is not important because all of the Fe
lines are weak and they do not have a significant effect on the fitted
parameters of the bright lines of O and Ne.  We hope that future
instruments will have the resolution and sensitivity to uniquely
identify the weak lines in the E0102 spectrum.

\begin{table*}
\centering
\caption[]{{\bf {Spectral Lines Included in the E0102 Reference Model  
(v1.9)} }}

\label{tab:model}

\begin{tabular}{lllr|clllr}\hline\hline
Line ID & E (keV)\,$^{a}$ & $\lambda$~(\AA)\,$^{a}$ & Flux\,$^{b}$ &  &
Line ID & E (keV)\,$^{a}$ & $\lambda$~(\AA)\,$^{a}$ & Flux\,$^{b}$ \\
\hline
\ion{C}{VI}~~Ly$\alpha$ & 0.3675 & {\small 33.737} & 175.2 & \quad &  
\ion{Ne}{IX}~~He$\alpha$~~i & 0.9148 & {\small 13.553} & 249.6 \\
\ion{Fe}{XXIV} & 0.3826 & {\small 32.405} & 18.4 & \quad &  
\ion{Fe}{XIX} & 0.9172 & {\small 13.517} & 0.0 \\
\ion{S}{XIV} & 0.4075 & {\small 30.425} & 11.8 & \quad & 
\ion{Ne}{IX}~~He$\alpha$~~r & 0.922 & {\small 13.447} & 1380.5 \\
\ion{N}{VI}~~He$\alpha$~~f & 0.4198 & {\small 29.534} & 6.8 & \quad & 
\ion{Fe}{XX} & 0.9668\,$^{c}$ & {\small 12.824} & 120.5 \\
\ion{N}{VI}~~He$\alpha$~~i & 0.4264 & {\small 29.076} & 2.0 & \quad & 
\ion{Ne}{X}~~Ly$\alpha$ & 1.0217 & {\small 12.135} & 1378.3 \\
\ion{N}{VI}~~He$\alpha$~~r & 0.4307 & {\small 28.786} & 10.5 & \quad &  
\ion{Fe}{XXIII} & 1.0564 & {\small 11.736} & 24.2 \\
\ion{C}{VI}~~Ly$\beta$ & 0.4356 & {\small 28.462} & 49.5 & \quad &  
\ion{Ne}{IX}~~He$\beta$ & 1.074 & {\small 11.544} & 320.7 \\
\ion{C}{VI}~~Ly$\gamma$ & 0.4594 & {\small 26.988} & 27.3 & \quad &  
\ion{Ne}{IX}~~He$\gamma$ & 1.127 & {\small 11.001} & 123.1 \\
\ion{O}{VII}~~He$\alpha$~~f & 0.561 & {\small 22.1} & 1313.2 & \quad &  
\ion{Fe}{XXIV} & 1.168\,$^{c}$ & {\small 10.615} & 173.5 \\
\ion{O}{VII}~~He$\alpha$~~i & 0.5686 & {\small 21.805} & 494.4 & \quad &  
\ion{Ne}{X}~~Ly$\beta$ & 1.211 & {\small 10.238} & 202.2 \\
\ion{O}{VII}~~He$\alpha$~~r & 0.5739 & {\small 21.603} & 2744.7 & \quad &  
\ion{Ne}{X}~~Ly$\gamma$ & 1.277 & {\small 9.709} & 78.5 \\
\ion{O}{VIII}~~Ly$\alpha$ & 0.6536 & {\small 18.969} & 4393.3 &  \quad & 
\ion{Ne}{X}~~Ly$\delta$ & 1.308 & {\small 9.478} & 37.1 \\
\ion{O}{VII}~~He$\beta$ & 0.6656 & {\small 18.627} & 500.9 & \quad &  
\ion{Mg}{XI}~~He$\alpha$~~f & 1.3311 & {\small 9.314} & 108.7 \\
\ion{O}{VII}~~He$\gamma$ & 0.6978 & {\small 17.767} & 236.1 & \quad  &
\ion{Mg}{XI}~~He$\alpha$~~i & 1.3431 & {\small 9.231} & 27.5 \\
\ion{O}{VII}~~He$\delta$ & 0.7127 & {\small 17.396} & 124.9 & \quad  &
\ion{Mg}{XI}~~He$\alpha$~~r & 1.3522 & {\small 9.169} & 231.0 \\
\ion{Fe}{XVII} & 0.7252 & {\small 17.096} & 130.9 & \quad & ~? 
                      & 1.4317 & {\small 8.659} & 8.1 \\
\ion{Fe}{XVII} & 0.7271\,$^{c}$ & {\small 17.051} & 165.9 & \quad &  
\ion{Mg}{XII}~~Ly$\alpha$ & 1.4721 & {\small 8.422} & 110.2 \\
\ion{Fe}{XVII} & 0.7389 & {\small 16.779} & 82.3 & \quad &  
\ion{Mg}{XI}~~He$\beta$ & 1.579\,$^{c}$ & {\small 7.852} & 50.6 \\
\ion{O}{VIII}~~Ly$\beta$ & 0.7746 & {\small 16.006} & 788.6 & \quad  &
\ion{Mg}{XI}~~He$\gamma$ & 1.659 & {\small 7.473} & 16.0 \\
\ion{Fe}{XVII} & 0.8124\,$^{c}$ & {\small 15.261} & 90.5 & \quad &  
\ion{Mg}{XII}~~Ly$\beta$ & 1.745\,$^{c}$ & {\small 7.105} & 29.7 \\
\ion{O}{VIII}~~Ly$\gamma$ & 0.817 & {\small 15.175} & 243.1 & \quad  &
\ion{Si}{XIII}~~He$\alpha$~~f & 1.8395 & {\small 6.74} & 13.8 \\
\ion{Fe}{XVII} & 0.8258 & {\small 15.013} & 65.1 & \quad &  
\ion{Si}{XIII}~~He$\alpha$~~i & 1.8538 & {\small 6.688} & 3.4 \\
\ion{O}{VIII}~~Ly$\delta$ & 0.8365 & {\small 14.821} & 62.7 & \quad  &
\ion{Si}{XIII}~~He$\alpha$~~r & 1.865 & {\small 6.647} & 34.6 \\
\ion{Fe}{XVIII} & 0.8503\,$^{c}$ & {\small 14.581} & 407.3 & \quad &  
\ion{Si}{XIV}~~Ly$\alpha$ & 2.0052 & {\small 6.183} & 11.2 \\
\ion{Fe}{XVIII} & 0.8726\,$^{c}$ & {\small 14.208} & 89.6 & \quad &  
\ion{Si}{XIII}~~He$\beta$ & 2.1818 & {\small 5.682} & 4.3 \\
\ion{Ne}{IX}~~He$\alpha$~~f & 0.9051 & {\small 13.698} & 690.2 & \quad &  
\ion{S}{XV}~~He$\alpha$~~f,i,r & 2.45 & {\small 5.06} & 12.7 \\
\hline

\end{tabular}

\begin{flushleft}
$^{a}$~Theoretical rest energies; wavelengths are $hc/E$. \\
$^{b}$~Flux in $10^{-6}$~photons~cm$^{-2}$\,s$^{-1}$ \\
$^{c}$~This line is broader than the nominal width, see text\\
\end{flushleft}

\end{table*}


\subsection{Fitting Methodology}\label{fitmethod}

The spectral data were fit using the {\tt XSPEC} software package
\citep{arnaud1999} with the modified Levenberg-Marquardt minimization
algorithm and the C statistic \citep{cash1979} as the fitting
statistic.  We fit the data in the energy range from 0.3--2.0~keV since
that is the energy range in which E0102 dominates over
the background.  We adopted the C statistic as the fitting
statistic to avoid the well-known bias with the $\chi^2$ statistic with a low
number of counts per bin \citep[see][]{cash1979,nousek1989}
and the bias that persists even with a relatively large number of
counts per bin \citep[see][]{humphrey2009}. 
Given how bright E0102 is compared to the typical instrumental
background, the low number of counts per bin bias should only affect
the lowest and highest energies in the 0.3--2.0~keV bandpass.
The EPIC-pn spectra were fit with both the C statistic and the $\chi^2$
statistic and the derived parameters were nearly identical.  The EPIC-pn
spectra have the largest number of counts and the count rate is stable
in time over the mission.  We performed the final fits for the EPIC-pn with
$\chi^2$ as the fit statistic.
The source extraction regions for each of the CCD instruments are shown
in Figures~\ref{fig:image4} \& \ref{fig:image_xis}.  The source and
background spectra were not binned in order to preserve the maximal
spectral information.  Suitable backgrounds were selected for each 
instrument nearby E0102 where there was no obvious enhancement in the 
local diffuse emission. 
If the C statistic is used and the user does not supply an explicit 
background model, {\tt XSPEC} computes a background model based on
the background spectrum provided in 
place of a user-provided background model.
{\tt XSPEC} does not subtract the background spectrum
from the source spectrum in this case, rather the source and background
spectra are both modeled.   This is referred to in the {\tt XSPEC}
documentation as the so-called ``W statistic'' \footnote{see
https://heasarc.gsfc.nasa.gov/xanadu/xspec/manual/\\qXSappendixStatistics.html}.
Although this approach is suitable for our analysis objectives, it may
not be suitable if the source is comparable to or only slightly brighter than the
background. In such a case, it might be beneficial to specify an
explicit background model with its own free parameters and fit
simultaneously with the source spectral model. Given how bright the
\ion{O}{vii}~He$\alpha$~triplet, \ion{O}{viii}~Ly$\alpha$ line, the 
\ion{Ne}{ix}~He$\alpha$~triplet, and the \ion{Ne}{x}~Ly~$\alpha$ line are
compared to the background, our determination of these line fluxes is
rather insensitive to the background modeling method.

 Some of the spectral data sets for the various instruments showed
evidence of gain variations from one observation to another.  Our
analysis method is sensitive to shifts in the gain since
our model spectrum has strong, well-separated lines and the line
energies are frozen in our fitting process. These gain shifts
could be due to a number of factors
such as uncertainties in the bias or offset calculation at the beginning of the
observation, drifts in the gain of the electronics, or variable
particle background.  Since our
objective is to determine the most accurate normalization for a line
at a known energy, it is important that the line be well-fitted. We
experimented with the {\tt gainfit} command in {\tt XSPEC} for the
data sets that showed evidence of a possible gain shift and determined
that the fits to the lines improved significantly in some cases. 
One disadvantage of the {\tt gainfit} approach is that the value of the
effective area is then evaluated at a different energy and this
introduces a systematic error in the determination of the line
normalization.  We determined that for gain shifts of 5~eV or less,
the error introduced in the derived line normalization is less than
2\% which is typically smaller than our statistical uncertainty on a
line normalization from a single observation. The gain shifts for the EPIC-MOS
and EPIC-pn spectra were small enough that {\tt gainfit} could be used.
The gain shifts for ACIS-S3, XIS, and XRT could be larger for some
observations, on the order of $\pm10$~eV.
Therefore, we adopted the approach of applying the indicated
gain shift to the event data outside of {\tt XSPEC}, re-extracting
the spectra from the modified events lists, and then
fitting the modified spectrum to determine the normalization of the
line. The ACIS-S3, XIS, \& XRT data had gain shifts applied to their data
in this manner.

   The number of free parameters needed to be significantly reduced
before fitting the CCD data in order to reduce the possible parameter
space.  In our fits, we have frozen the line energies and widths,
the SMC \NH\/, and the low-temperature {\tt APEC ``No-Line''}
continuum to the RGS-determined values.  The high-temperature
{\tt APEC ``No-Line''} component was frozen at the values determined from the
EPIC-pn and EPIC-MOS. The fixed absorption and continuum components are listed
in Table~\ref{tab:abs_cont}.
Since the CCD instruments lack the spectral resolution to 
resolve lines which are as close to each other as
the ones in the \ion{O}{vii}~He$\alpha$ triplet and the 
\ion{Ne}{ix}~He$\alpha$ triplet,
we treated nearby lines from the same ion as a ``line complex'' by
constraining the ratios of the line normalizations to be those
determined by the RGS and by constraining the line energies to the
known separations.  In practice, we would typically link the
normalization and energy of the {\em f} and {\em i} lines of
the triplet to the {\em r} line (except for \ion{O}{vii} for which
we linked the other lines to the {\em f} line).  
Since we also usually freeze the
energies of the lines, this means that the three lines in the triplet
would have only one free parameter, the normalization of the Resonance
line.  We have constructed the model in {\tt XSPEC} such that it would
be easy to vary the energy of the {\em r}  line in the triplet (and
hence also the {\em f} and {\em i}) to examine the gain
calibration of a detector at these energies.  Our philosophy is to
treat 
nearby lines
as a  complex which can adjust together in normalization and energy.
In this paper, we focus on adjusting the normalization of the line
complexes only.  Since most of the power in the spectrum is
 in the bright line
complexes, we froze all the normalizations of the weaker lines.
The only normalizations which we allowed to vary were the 
\ion{O}{vii}~He$\alpha$~{\em f}, \ion{O}{viii}~Ly$\alpha$ line, the
\ion{Ne}{ix}~He$\alpha$~{\em r}, and the \ion{Ne}{X}~Ly$\alpha$ line
normalizations.  In addition, we found it useful to introduce a
constant scaling factor of the entire model to account for the fact
that the extraction regions for the various instruments were not
identical. 
In this manner, we restricted a model with more than 200 parameters 
to have only 5 free parameters in our
fits.  The final version of this model in the {\tt XSPEC .xcm} file
format is available on the IAHCEC web site, on the Thermal SNRs
Working Group page:
{\em https://wikis.mit.edu/confluence/display/iachec/Thermal+SNR}.
We will refer to this at the IACHEC standard model for E0102 or the
``IAHCEC model.''


\begin{table}[h]
\centering

\caption[ ]{\bf {Fixed Absorption and Continuum Components}}

\label{tab:abs_cont}
\begin{tabular}{l|l}
\hline
\hline
Model Component      & Value \\
\hline
Galactic absorption  & \NH\/$=5.36\times10^{20}~{\rm cm^{-2}}$ \\
SMC absorption       & \NH\/$=5.76\times10^{20}~{\rm cm^{-2}}$ \\
{\tt APEC ``No-Line''} temperature \#1 & kT=0.164~keV \\
{\tt APEC ``No-Line''} normalization \#1 & $3.48\times10^{-2}~{\rm cm^{-5}}$ \\
{\tt APEC ``No-Line''} temperature \#2 & kT=1.736~keV \\
{\tt APEC ``No-Line''} normalization \#2 & $1.85\times10^{-3}~{\rm cm^{-5}}$ \\\hline

\end{tabular}

\end{table}


\section{Observations}\label{obs}

 E0102 has been routinely observed by  \chan, \suzaku\/, \swift\/, and \xmmn,
as a calibration target to monitor the response at energies below
1.5~keV.  The IACHEC standard model was developed using 
primarily RGS and HETG data as described in Sect.~\ref{model}. 
We include a description of the RGS and HETG data in this section for
completeness, but our primary objective is to improve the calibration
of the CCD instruments. Therefore, the RGS and HETG data were analyzed
with the calibration available at the time the model was finalized.
We continue to update the processing and analysis of the CCD
instrument data as new software and calibration files become available.
For this paper, we have selected a subset of these CCD instrument
observations for the comparison of the absolute effective areas.
We have selected data from the timeframe and the instrument mode for
which we are the most confident in the calibration and used those data
in this comparison. We have also
analyzed all available E0102 data from a given instrument in the same
mode, close to on-axis in order to
characterize the time dependence of the response of the individual instruments.
We now describe the data processing and calibration issues for each instrument
individually.

\subsection{\xmmn\/ RGS}
\subsubsection{Instruments}

\xmmn\/ has two essentially identical
high-resolution dispersive grating spectrometers, RGS1 and RGS2,
that share telescope mirrors with the EPIC instruments MOS1 and MOS2 
and operate
 between 6 and 38 \AA\, or 0.3 and 2.0 keV. The size of its 
9 CCD detectors along the Rowland circle define apertures of about 5 arcminutes
within which E0102 fits comfortably. Each CCD
has an image area of $1024\times384$ pixels, integrated on the chip
into bins of $3\times3$ pixels. The data consist of individual 
events whose wavelengths are determined by
the grating dispersion angles calculated from the
spatial positions at which they were detected. Overlapping orders are 
separated 
through the event energies assigned by the CCDs. The RGS instruments
have suffered
the build-up of a contamination layer of carbon included automatically
in the calibration. The status of the RGS calibration is summarized in
\citet{devries2015}.
Built-in redundancies have ensured complete
spectral coverage despite the loss
early in the mission of one CCD detector each in RGS1 and RGS2.

\subsubsection{Data}

E0102 has been a regular \xmmn\/ calibration source with over 30
observations (see Table~\ref{table:rgsobs}) made at 
initially irregular
intervals and variety of position angles between 2000 April 16 and
2011 November 04. All of these data have been used in the analysis
reported here
using spectra calculated on a fixed wavelength grid by SAS v11.0.0
separately as
normal for RGS1 and RGS2 and for 1st and 2nd orders.
An initial set of 23 observations before the end of 2007 was combined using the 
SAS task \texttt{rgscombine} to
give spectra of high statistical weight with exposure times of
708080~s for RGS1 and 680290~s for RGS2 and used at an early stage to
define the IACHEC model discussed above.

\begin{table*}
\centering

\caption{XMM RGS Observations of E0102} 
\label{table:rgsobs}

\begin{tabular}{cccc}
\hline
\hline
rev & ObsID & DATE & Exposure(ks) \\
\hline
0065  & 0123110201 & 2000-04-16 & 22.7 \\
0065  & 0123110301 & 2000-04-17 & 21.7 \\
0247  & 0135720601 & 2001-04-14 & 33.5 \\ 
0375  & 0135720801 & 2001-12-25 & 35.0 \\ 
0433  & 0135720901 & 2002-04-20 & 35.7 \\ 
0447  & 0135721001 & 2002-05-18 & 34.1 \\ 
0521  & 0135721101 & 2002-10-13 & 27.2 \\ 
0552  & 0135721301 & 2002-12-14 & 29.0 \\ 
0616  & 0135721401 & 2003-04-20 & 45.5 \\ 
0711  & 0135721501 & 2003-10-27 & 30.5 \\ 
0721  & 0135721701 & 2003-11-16 & 27.4 \\ 
0803  & 0135721901 & 2004-04-28 & 35.2 \\ 
0888  & 0135722401 & 2004-10-14 & 31.1 \\
0894  & 0135722001 & 2004-10-26 & 31.9 \\ 
0900  & 0135722101 & 2004-11-06 & 49.8 \\ 
0900  & 0135722201 & 2004-11-07 & 31.9 \\ 
0900  & 0135722301 & 2004-11-07 & 31.9 \\ 
0981  & 0135722501 & 2005-04-17 & 37.1 \\ 
1082  & 0135722601 & 2005-11-05 & 30.4 \\ 
1165  & 0135722701 & 2006-04-20 & 30.5 \\ 
1265  & 0412980101 & 2006-11-05 & 32.4 \\ 
1351  & 0412980201 & 2007-04-25 & 36.4 \\ 
1443  & 0412980301 & 2007-10-26 & 37.1 \\ 
1531  & 0412980501 & 2008-04-19 & 29.9 \\ 
1636  & 0412980701 & 2008-11-14 & 28.9 \\ 
1711  & 0412980801 & 2009-04-13 & 28.9 \\ 
1807  & 0412980901 & 2009-10-21 & 28.9 \\ 
1898  & 0412981001 & 2010-04-21 & 30.5 \\ 
1989  & 0412981301 & 2010-10-18 & 32.0 \\ 
2081  & 0412981401 & 2011-04-20 & 35.1 \\ 
2180  & 0412981501 & 2011-11-04 & 30.2 \\ 
\hline
\end{tabular}

\end{table*}


\begin{table*}
\centering

\caption[ ]{ACIS and ACIS/HETG Observations of E0102 }

\label{tab:acisobs}
\begin{tabular}{cllccl}
\hline
\hline
OBSID & Instrument & DATE & Exposure & Counts\tablefootmark{a} & Mode
\\
 &  &  & (ks) & (0.5-2.0 keV) &  \\
\hline
\phantom{112}120\tablefootmark{b}   & ACIS-HETG & 1999-09-28 &   \phantom{1}87.9  &  \phantom{1}38917  & TE, Faint, 3.2
s frametime \\
\phantom{112}968\tablefootmark{b}    & ACIS-HETG & 1999-10-08 &   \phantom{1}48.4  &  \phantom{1}22566  & TE, Faint, 3.2
s frametime \\
\phantom{12}3828\tablefootmark{b}   & ACIS-HETG & 2002-12-20 &  135.9  &  \phantom{1}49599  & TE, Faint, 3.2 s frametime \\
12147  & ACIS-HETG & 2011-02-11 &  148.9  &  \phantom{1}44341  & TE, Faint, 3.2
s frametime \\
\hline
\phantom{12}3545\tablefootmark{b}    & ACIS-S3 & 2003-08-08 & \phantom{12}7.9  &  \phantom{1}57111 & TE, 1/4 subarray, 1.1~s frametime, node 1 \\
\phantom{12}6765\tablefootmark{b}    & ACIS-S3 & 2006-03-19 & \phantom{12}7.6  & \phantom{1}51745  & TE, 1/4 subarray, 0.8~s frametime, node 0 \\
\phantom{1}8365    & ACIS-S3 & 2007-02-11 & \phantom{1}21.0 & 138685 & TE, 1/4 subarray, 0.8~s frametime, node 0 \\    
\phantom{1}9694    & ACIS-S3 & 2008-02-07 & \phantom{1}19.2 & 124795 & TE, 1/4 subarray, 0.8~s frametime, node 0 \\  
10654   & ACIS-S3 & 2009-03-01 & \phantom{12}7.3 & \phantom{1}45534  & TE, 1/4 subarray, 0.8~s frametime, node 0 \\
10655   & ACIS-S3 & 2009-03-01 & \phantom{12}6.8 & \phantom{1}43227  & TE, 1/8 subarray, 0.4~s frametime, node 0 \\
10656   & ACIS-S3 & 2009-03-06 & \phantom{12}7.8 & \phantom{1}48601  & TE, 1/4 subarray, 0.8~s frametime, node 1 \\
11957   & ACIS-S3 & 2009-12-30 & \phantom{1}18.5 & 112423 & TE, 1/4 subarray, 0.8~s frametime, node 0 \\
13093   & ACIS-S3 & 2011-02-01 & \phantom{1}19.1 & 108286 & TE, 1/4 subarray, 0.8~s frametime, node 0 \\
14258   & ACIS-S3 & 2012-01-12 & \phantom{1}19.1 & 102048 & TE, 1/4 subarray, 0.8~s frametime, node 0 \\
15467   & ACIS-S3 & 2013-01-28 & \phantom{1}19.1 &  \phantom{1}92610 & TE, 1/4 subarray, 0.8~s frametime, node 0 \\
16589   & ACIS-S3 & 2014-03-27 & \phantom{12}9.6 &  \phantom{1}40194 & TE, 1/4 subarray, 0.8~s frametime, node 0 \\
17380	& ACIS-S3 & 2015-02-28 & \phantom{1}17.7 &  \phantom{1}65809 & TE, 1/4 subarray, 0.8~s frametime, node 0 \\
17688	& ACIS-S3 & 2015-07-17 & \phantom{12}9.6 &  \phantom{1}33972 & TE, 1/4 subarray, 0.8~s frametime, node 0 \\

\hline

\end{tabular}
\tablefoot{\tablefoottext{a}{Counts for "ACIS-HETG" are the sum of MEG
    $\pm1$ order events, $0.5-2$ keV. } \\
\tablefoottext{b}{Observation included in the comparison of
    effective areas discussed in Section~\ref{fitresults}. } }

\end{table*}


\subsubsection{Processing}

As E0102 is an extended source, it required special treatment with SAS
 v11.0.0 
 whose usual procedures are designed for the analysis
of point sources. This simply involved the definition of custom
rectangular source and background regions taking into account both 
the size of the SNR and the cross-dispersion
instrumental response caused by scattering from the gratings. 
In cross-dispersion angle from the SNR centre, the source regions
were $\pm0.75'$ and the background
regions were between $\pm1.42'$ and $\pm2.58'$. An individual 
measurement was thus encapsulated in a pair of simultaneous spectra, 
one combining source and background,
the other the background only.

\subsection{Chandra HETG}
\subsubsection{Instruments}

\begin{figure*}
 \begin{center}
\resizebox{\hsize}{!}{\includegraphics[width=6.5in,angle=0]{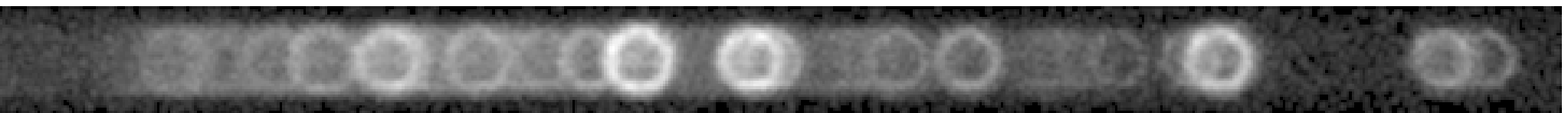}}  \\
   \vspace{0.05in}
   \resizebox{\hsize}{!}{\includegraphics[width=6.5in,angle=0]{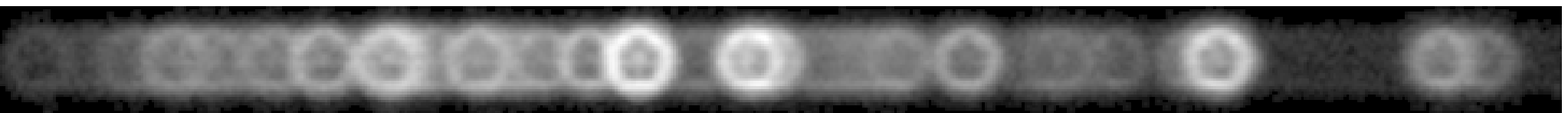}}
 \end{center}
 \caption{ \label{fig:MEG}
Images of the
MEG-dispersed E0102 data (top) and a synthesized model (bottom).
The $\pm$1 MEG orders from all three epochs have been combined
and displayed in the range from 4.6\AA\ to 23\AA\ (left-to-right, 2.7--0.54 keV).
The data clearly show bright rings of line emission for many lines; the
very brightest lines just left of center are from
\ion{Ne}{x}~Ly$\alpha$ 
and \ion{Ne}{ix}~He$\alpha$~triplet.
The simulated spectral image (bottom) was created using the IACHEC standard
model and does not include any background events.
The cross-dispersion range for each image is $\pm$28~arcsec.
}
%
\end{figure*}

The HETG is one of two transmission gratings on \chan\/ which can be
inserted into the converging X-ray beam just behind the {\em High
  Resolution Mirror Assembly} (HRMA).
When this is done the resulting HRMA--HETG--ACIS-S configuration is the
high-energy transmission grating spectrometer (HETGS, often used
interchangeably with just HETG).  The HETG and its operation are 
described as a part of  \chan\/ \citep{weiss2000,weiss2002}
and in HETG-specific publications \citep{canizares2000,canizares2005}.

The HETG consists of two distinct sets of gratings, the medium-energy
gratings (MEGs) and the high-energy-gratings (HEGs) each of which
produces plus-- and minus-- order dispersed spectral images
with the dispersion angle nearly proportional to
the photon wavelength.  The result is that a point source produces a
non-dispersed ``zeroth-order'' image (the same as if the HETG were not
inserted, though with reduced throughput) as well as four
distinct linear spectra forming the four arms of a shallow ``X''
pattern on the ACIS-S readout;
see Fig. 1 of both \citet{canizares2000} \& \citet{canizares2005}.

Hence, an HETG observation yields four first-order spectra,
the MEG $\pm$1 orders and the HEG $\pm$1 orders \footnote{There are
also higher-valued orders, $m=2,3,$\ldots, but their throughput is
much below the first-orders'; the most useful of these are the 
MEG $\pm$3 and the HEG $\pm2$ orders each with $\sim$\,$\times$0.1 the
throughput of the first orders.}.
Because the dispersed photons are spread out and detected along the
ACIS-S, the calibration of the HETG involves more than a single ACIS CCD:
the minus side orders fall on ACIS CCDs S2, S1, and S0, while the
plus side orders are on S3, S4, and S5.  Hence the calibration of all
ACIS-S CCDs is important for the HETG calibration.

When the object observed with the HETG is not a point source, 
the dispersed images become something like a convolution of the spatial
and spectral distributions of the
source; see \citet{dewey2002} for a brief elaboration of these issues.
The upshot is that for the extended-source case the 
{\em response  matrix function} (rmf)  of the spectrometer
is determined by the spatial characteristics of the source
and the position angle of the dispersion direction on the sky, set by the
observation roll angle.  These considerations guide the HETG analyses that
follow.

\subsubsection{Data}

E0102 was observed as part of the HETG GTO program
at three epochs (see Table~\ref{tab:acisobs}).: 
Sept.--Oct.\ 1999 (obsids 120 \& 968, 
   $t=$ 1999.75, exp $=$ 88.2$+$49.0\,ks, roll $=$ 11.7$^{\circ}$), in 
December of 2002 (obsid 3828, 
   $t=$ 2002.97, exp $=$ 137.7\,ks, roll $=$ 114.0+180$^{\circ}$),
and most recently in 
February of 2011 (obsid 12147, 
   $t=$ 2001.11, exp $=$ 150.8\,ks, roll $=$ 56.5+180$^{\circ}$).
The roll angles of these epochs were deliberately chosen to differ with a view
toward future spectral-tomographic analyses.
The HETG view of E0102 is presented in \citet{flanagan2004} 
using the first epoch observations: the bright ring of E0102 is
dispersed and shows multiple ring-like images due to the prominent 
emission lines
in the spectrum.  The combination of all three epoch's MEG data is
shown in Fig.~\ref{fig:MEG}.

In principle, one can analyze the 2D spectral images directly
\citep{dewey2002} to get the most information from the data.
This involves doing forward-folding of spatial-spectral models
to create simulated 2D images which are compared with the data \citep{dewey2009}.
The lower image of Fig.~\ref{fig:MEG} shows
such a simulated model for the combined MEG data sets based on the
observed E0102 zeroth-order events, the IACHEC standard model,
and {\em CIAO}-generated  ARFs.
However, for the limited purpose of fitting the 5-parameter IACHEC 
model to the HETG data we can collapse the data to 1D and use the standard HETG 
extraction
procedures (next section).

\subsubsection{Processing}

The first steps in HETG data analysis are the extraction of 1D spectra
and the creation of their corresponding ARFs (as mentioned above
the point-source RMFs are not applicable to E0102.)
Because of differences in the pointing of the two first-epoch
observations, they are separately analyzed and so we extract the 4 HETG
spectra from each of the 4 obsids available.
The archive-retrieved data were processed using {\em TGCat ISIS}
scripts \citep{huenemoerder2011};
these provide a useful wrapper to execute the {\em CIAO} 
extraction tools.
Several customizations were specified before executing {\em TGCat}'s
do-it-all {\tt run\_pipe()} command:
\begin{itemize}
\item The extraction center was manually input and 
chosen to be at the center of a 43 sky-pixel radius 
circle that approximates the outer blastwave location.  For the
recent-epoch obsid 12147, this is at RA 01:04:02.11 and Dec
-72:01:52.2 (J2000 coordinates.s).
This location is 0.5 sky-pixels East and 7 sky-pixels North of the
centroid of the bright blob at the inner end of the ``Q-stroke'' feature.
This offset from a feature in the data was used to
determine the equivalent center location in the other obsids.
\item The cross-dispersion widths of the MEG and HEG spectral extractions
were set to cover a range of $\pm$55 sky-pixels about the dispersion axes.
\item The order-sorting limits were explicitly set to a large constant
value of $\pm$0.20 \footnote{For the first epoch, early in the
\chan\/ mission, the ACIS focal plane temperature was at
$-110^{\circ}$\,C.  For these obsids we
used a somewhat larger order-sorting range of $\pm$0.25.}.
\end{itemize}
Because E0102 covers a large range in the cross-dispersion direction
compared with the $\pm$16 pixel dither range, we generated for each
extraction a set of 7 ARFs spaced to cover the 110 pixels of the
cross-dispersion range.  In making the ARFs we set {\tt osipfile=none}
because of our large order-sorting limits.
Finally, background extractions were made as for the data but with
the extraction centers shifted by 120 pixels in the cross-dispersion
direction.

\begin{figure}
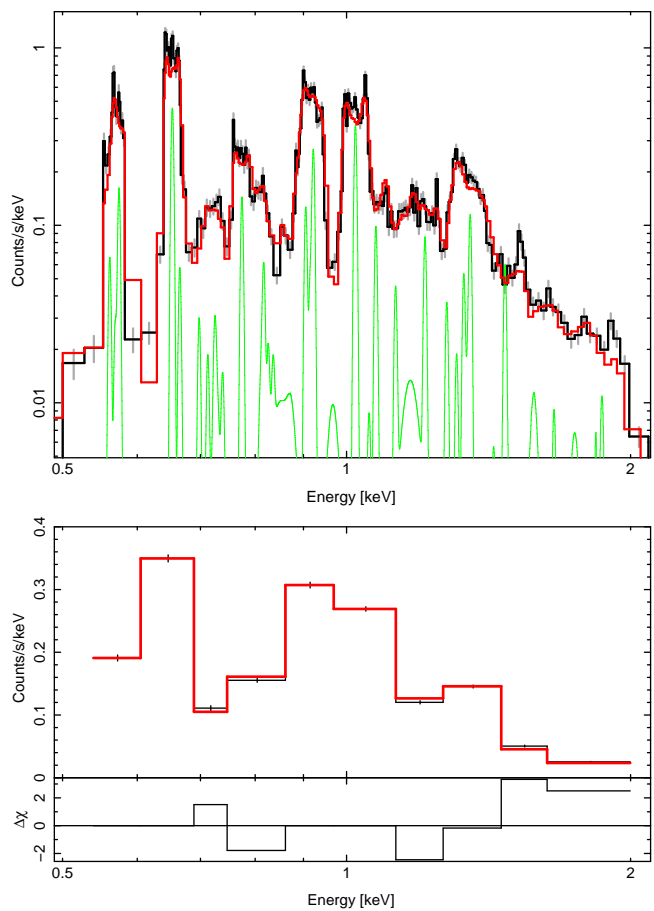

 \begin{center}
 \resizebox{\hsize}{!}{\includegraphics[angle=270]{fig5a.ps}}  
 \\
 \resizebox{\hsize}{!}{\includegraphics[angle=270]{fig5b.ps}}
 \end{center} \caption{ \label{fig:hetgfit}
 Fit to the HETG MEG$-1$ data.
 {\em Top}:  The observed MEG$-1$ data-minus-background counts are
 shown in black.  For reference,
 the IACHEC model is multiplied by the ARF and is shown
 in green; it has been scaled by 0.1 for clarity.
 The red curve shows the IACHEC  model when it is further folded
 through an RMF that approximates the spatial extension of E0102 (see
 text).  {\em Bottom}:  The data (black) and model
 (red) counts are rebinned to a set of 10 coarse bins which are used
 for the 5-parameter model fits.
}
\end{figure}

The fitting of the HETG extractions
generally follows the methodology outlined in Sect.~\ref{fitmethod}
with some adjustments because of the extended nature of E0102, and, secondarily,
because of the use of the {\em ISIS} platform \citep{houck2002}.
For each obsid and grating--order
we read in the extracted source and background spectra (PHA files)
and, after binning (below),
the background counts are subtracted bin-by-bin from the source counts.
The corresponding set of ARFs that span E0102's cross-dispersion
extent are read in, averaged, and assigned to the data. An RMF that
approximates the spatial effects of E0102
is created and assigned as well, see Fig.~\ref{fig:hetgfit}.
Finally the model is defined in ISIS and its 5 free parameters are
fit and their confidence ranges determined.


The ARFs for the HETG contain two general types of features:
those that depend on the {\em photon energy}, such as
mirror reflectivity, grating efficiency and detector QE,
and other effects that depend on the {\em specific location} on the readout
array where the photon is detected, such as bad pixels and chip gaps.
For a point source there is very nearly a one-to-one mapping of
photon energy and location of detection, hence the two terms are
combined in the overall ARF, the {\tt SPECRESP} values in the FITS file.
The latter term is, however, available separately via the
{\tt FRACEXPO} values and this is used to remove the location-specific
contribution from the ARF.
The RMF is made in-software
using ISIS's {\tt load\_slang\_rmf()} routine.
The resulting RMF approximates the 1D projected shape of E0102
and appropriately includes the {\tt FRACEXPO} features.

As shown in Fig.~\ref{fig:hetgfit}, the HETG 1D extracted spectra are
reasonably approximated by the model folded through the approximate-RMF.
However, because the RMF is not completely accurate, we reduce its influence
on the fitting
by defining a coarse binning of 10 bins from 0.54 to 2.0~keV
(23 -- 6.2~\AA).  The boundaries of the bins are chosen to be
between the brightest lines, and the three lowest-energy bins are not used when
fitting an HEG spectrum.

\subsection{\xmmn\/  EPIC-pn}\label{epicpn}
\subsubsection{Instruments}

%
%
%
%

\begin{figure*}
   \resizebox{0.48\hsize}{!}{\includegraphics[clip=]{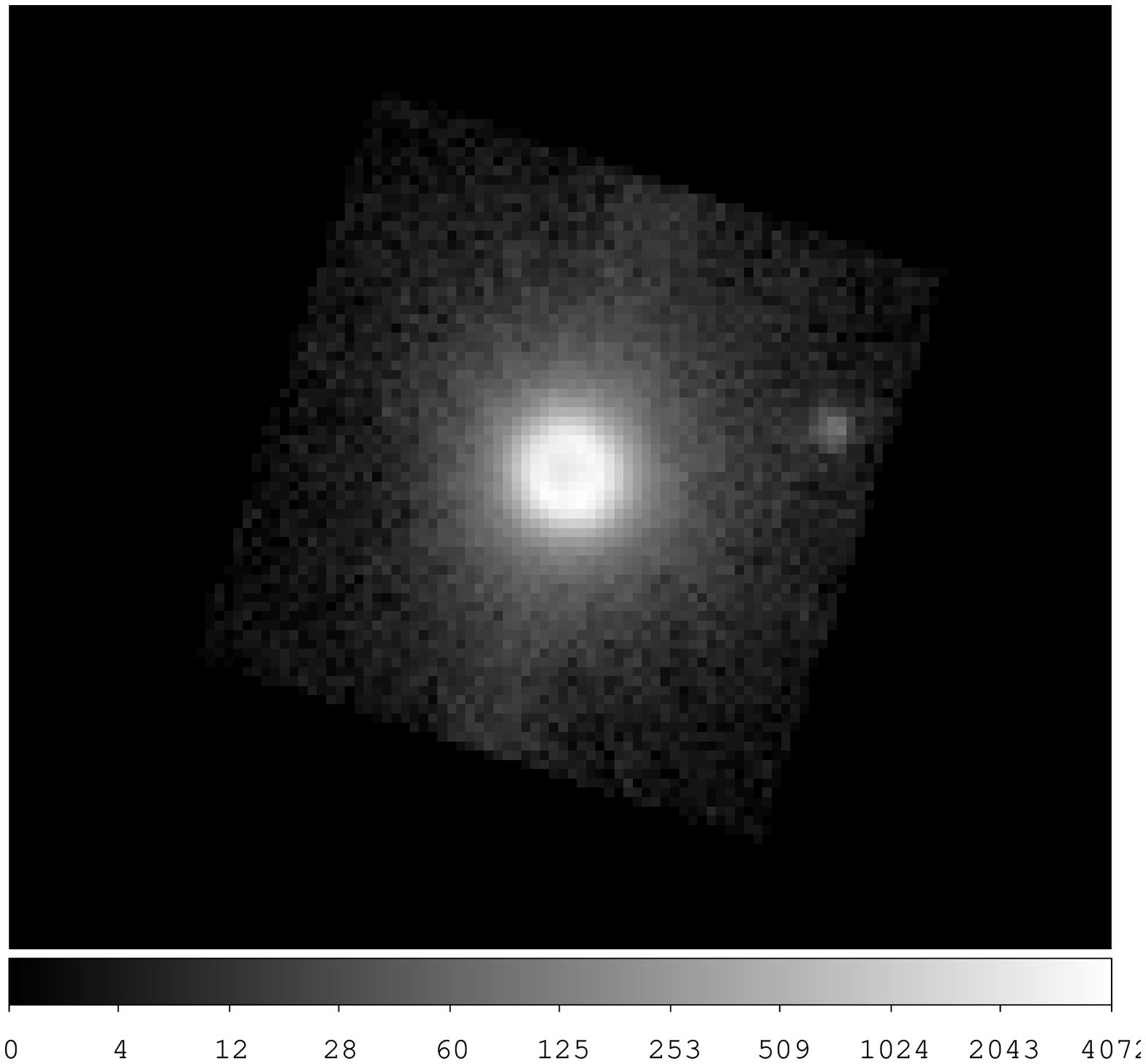}}
   \resizebox{0.48\hsize}{!}{\includegraphics[clip=]{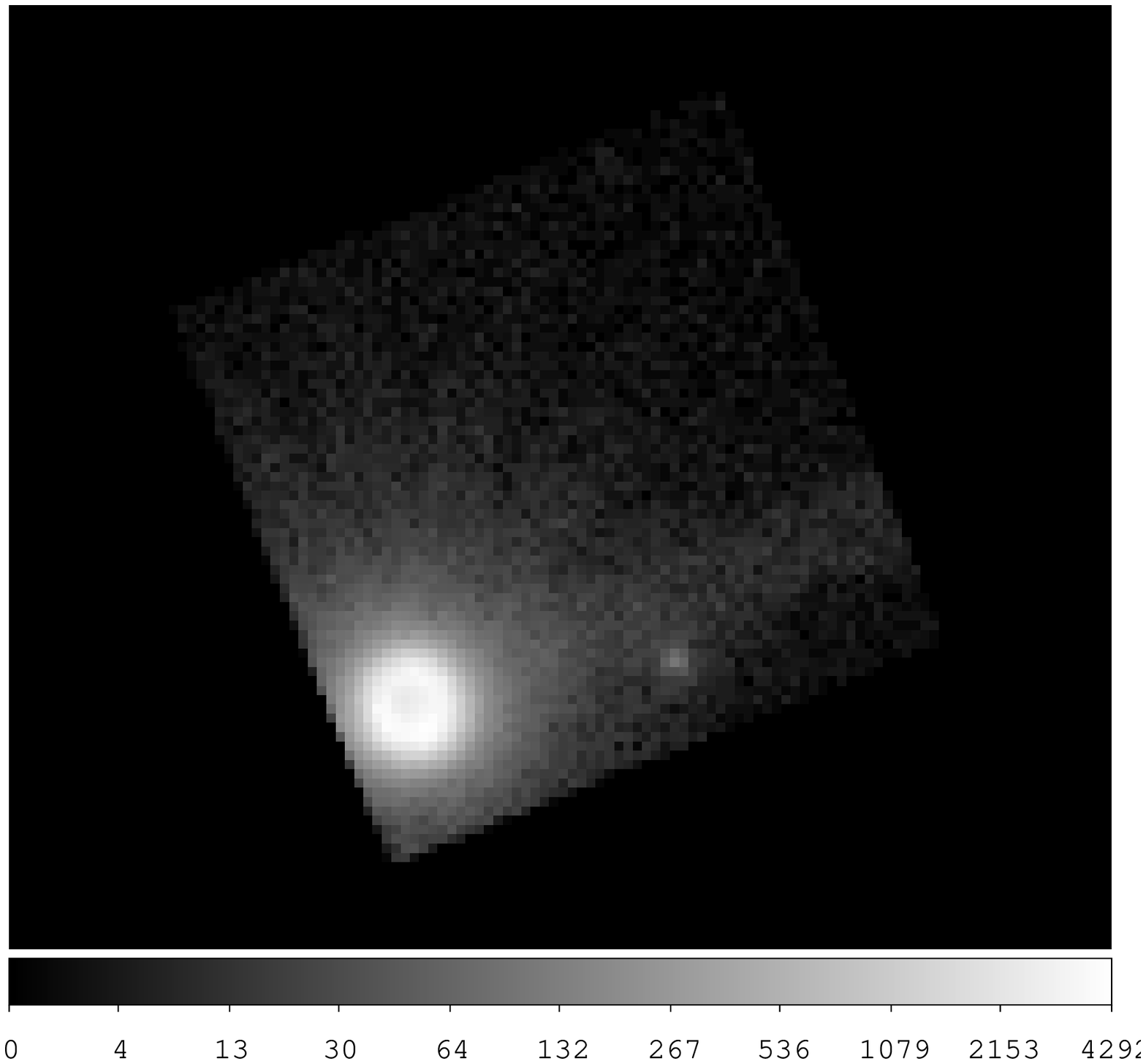}}
   \caption{EPIC-pn images of E0102 observed in SW mode. 
   {\it Left:} Observation 0135720801 with the SNR centered in the SW readout area;
   {\it Right:} Observation  0412981401 with the SNR at the nominal boresight position.
   }
   \label{fig-pnimages}
\end{figure*}

The EPIC-pn instrument is based on a back-illuminated $6 \times 6$
cm$^2$ monolithic X-ray CCD array covering the 0.15--12 keV energy band. 
Four individual quadrants each having three EPIC-pn-CCD subunits with a
format of $200 \times 64$ pixels are operated in parallel covering a 
$\sim$13\farcm6$\times$4\farcm4 rectangular region.  Different
CCD-readout modes are available which allow faster readout of
restricted CCD areas, with frame times from 73 ms for the full-frame 
(FF), 48 ms for large-window (LW), and 6 ms for the small window (SW) mode, the 
fastest imaging mode \citep{strueder2001}.

\subsubsection{Data}

\xmmn\/ observed E0102 with EPIC-pn in all imaging readout modes
(FF, LW and SW) and all available optical blocking filters.  To rule
out photon pile-up effects we only used spectra from SW mode data for
our analysis.  Between 2001-12-25 and 2015-10-30 (satellite revolution
375 to 2910) 
E0102 was observed by \xmmn\/ with EPIC-pn in small
window (SW) mode 24 times (see Table~\ref{tab-obs-epicpn}).  Two 
observations were performed with the thick filter while for 11 (11) 
observations the thin (medium) filter was used. One set of
observations placed the source at the nominal boresight position which
is close to a border of the $4.4\arcmin \times 4.4\arcmin$ read-out 
window of EPIC-pn-CCD 4 and therefore only a relatively small extraction
radius of 30\arcsec\ is possible. During 14 observations the source
was centered in the SW area which allows an extraction with 75\arcsec\ 
radius. For comparison we show in Fig.~\ref{fig-pnimages} the images
binned to $4\arcsec \times 4\arcsec$ pixels from observation 
0135720801 (with the SNR centered) and 0412981401 (nominal target
position).


\begin{table*}
\caption[]{Summary of EPIC pn SW mode observations.}
\begin{center}
\begin{tabular}{lllcrl}
\hline
\hline
Observation & Instrument & Date & Exposure & Count Rate\tablefootmark{a} &
Readout, Filter, \\
ID\tablefootmark{b} & ID & & (ks) & (counts s$^{-1}$) & Position \\
\hline
0135720801 & PNS001  &   2001-12-25   &   21.5  &  12.68  $\pm$  2.4e-02   &  SW, thin, centred \\
0135721101 & PNS001  &   2002-10-13   &   \phantom{2}7.5  &  12.69  $\pm$  4.2e-02   &  SW, thin, centred  \\
0135721301 & PNS001  &   2002-12-14   &   \phantom{2}7.7  &  12.61  $\pm$  4.1e-02   &  SW, thin, centred  \\
0135721401 & PNU002  &   2003-04-20   &   \phantom{2}8.6  &  12.31  $\pm$  3.8e-02   &  SW, medium, centred  \\
0135722401 & PNS001  &   2004-10-14   &   21.5  &   9.09  $\pm$  2.1e-02   &  SW, thick, centred  \\
0135722601 & PNS001  &   2005-11-05   &   21.0  &  12.27  $\pm$  2.4e-02   &  SW, medium, centred  \\
0135722701 & PNS001  &   2006-04-20   &   21.0  &  12.89  $\pm$  2.5e-02   &  SW, thin \\
0412980101 & PNS001  &   2006-11-05   &   22.4  &  12.19  $\pm$  2.3e-02   &  SW, medium, centred  \\
0412980201 & PNS001  &   2007-04-25   &   24.7  &  12.85  $\pm$  2.3e-02   &  SW, thin \\
0412980301 & PNS001  &   2007-10-26   &   25.7  &  12.22  $\pm$  2.2e-02   &  SW, medium, centred  \\
0412980501 & PNS001  &   2008-04-19   &   20.6  &  12.75  $\pm$  2.5e-02   &  SW, thin \\
0412980701 & PNS001  &   2008-11-14   &   19.9  &  12.42  $\pm$  2.5e-02   &  SW, medium \\
0412980801 & PNS001  &   2009-04-13   &   14.1  &  12.75  $\pm$  3.0e-02   &  SW, thin \\
0412980901 & PNS001  &   2009-10-21   &   20.0  &  12.34  $\pm$  2.5e-02   &  SW, medium \\
0412981001 & PNS001  &   2010-04-21   &   20.6  &  12.86  $\pm$  2.5e-02   &  SW, thin \\
0412981401 & PNS001  &   2011-04-20   &   23.1  &  12.52  $\pm$  2.3e-02   &  SW, thin \\
0412981701 & PNS001  &   2012-12-06   &   10.5  &  12.52  $\pm$  3.5e-02   &  SW, thin, centred \\
0412981701 & PNS012  &   2012-12-06   &   11.8  &  12.11  $\pm$  3.2e-02   &  SW, medium, centred \\
0412981701 & PNS013  &   2012-12-07   &   14.7  &   9.49  $\pm$  2.6e-02   &  SW, thick, centred \\
0412982101 & PNS001  &   2013-11-07   &   22.3  &  12.60  $\pm$  2.4e-02   &  SW, thin, centred \\
0412982201 & PNS001  &   2014-10-20   &   23.5  &  12.10  $\pm$  2.3e-02   &  SW, medium, centred \\
0412982301 & PNS001  &   2014-10-20   &   30.4  &  12.34  $\pm$  2.0e-02   &  SW, medium \\
0412982501 & PNS001  &   2015-10-28   &   23.4  &  12.28  $\pm$  2.3e-02   &  SW, medium \\
0412982401 & PNS001  &   2015-10-30   &   26.1  &  12.17  $\pm$  2.2e-02   &  SW, medium, centred \\
\noalign{\smallskip}\hline
\end{tabular}
\tablefoot{\tablefoottext{a}{Single-pixel events in the 0.3 $-$ 3.0
    keV band.} \\
\tablefoottext{b}{All pn observations were included in the comparison of
    effective areas discussed in Section~\ref{fitresults}. }
}

\end{center}
\label{tab-obs-epicpn}
\end{table*}


\subsubsection{Processing}\label{pnprocessing}

The data were processed with \xmmn SAS version 14.0.0 and we extracted
spectra using single-pixel events (PATTERN=0 and FLAG=0) to obtain
highest spectral resolution. Response files were generated using {\tt rmfgen} and {\tt arfgen},
assuming a point source for PSF 
corrections. Due to the extent of E0102, the standard PSF correction for the lost flux outside the extraction region 
introduces systematic errors, leading to different fluxes from spectra using different extraction radii.
To utilize the observations with the target placed at the nominal boresight position, 
we extracted spectra from the SW-centered observations with 30\arcsec\ and 75\arcsec\ radius. For the large 
extraction radius, PSF losses are negligible and, from a comparison of the two spectra, an average correction 
factor 1.0315 was derived to account for the PSF losses in the smaller extraction region.

\begin{figure}
   \resizebox{\hsize}{!}{\includegraphics[clip=,angle=-90]{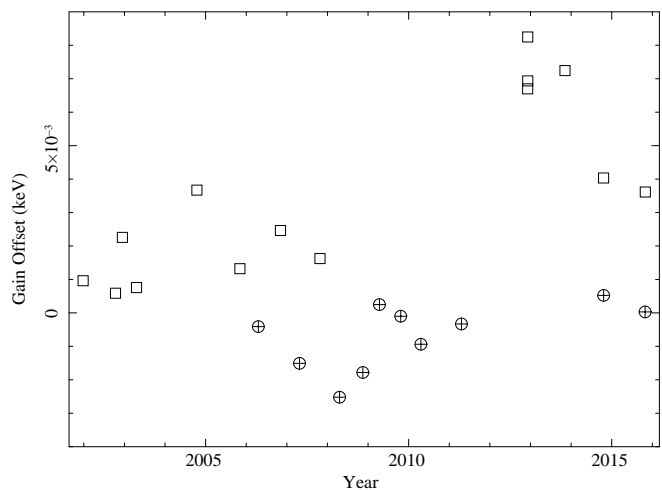}}
   \caption{Energy shift applied to the EPIC-pn spectra of
     E0102 as function of time.   Observations with the SNR
     placed at the centre of the readout window are marked with a square, 
     those at the nominal boresight position with a circle and cross.}
   \label{fig-pnshift}
\end{figure}

In order to derive reliable line fluxes from the EPIC-pn spectra, the lines must be at their nominal energies 
as accurately as possible. Otherwise, the high statistical quality
leads to bad fits and wrong line normalisations. 
Energy shifts of generally less than 5~eV in the EPIC-pn spectra of E0102 
lead to increases in $\chi^2$ from typical values of
600-700 to 700-800 and changes in line normalisations by $\sim2$\%. 
Only for the highest required gain shifts 
of 7--8 eV (Fig.~\ref{fig-pnshift}) errors in the line normalisations reach $\sim5$\%. 
Therefore, we created for each observation a set of event files with the energies of the events (the PI value)
shifted by up to $\pm$9 eV in steps of 1 eV. In order to do so the initial event file was produced with 
an accuracy of 1 eV for the PI values (PI values are stored as integer numbers with an accuracy of 5 eV by default) 
using the switch {\tt testenergywidth=yes} in {\tt epchain}. Spectra were then created from the 19 event files with 
the standard 5 eV binning. The 19 spectra from each of the SW mode
observations were fitted using the model described in Sect.~\ref{model}
with five free parameters (the overall 
normalisation factor and 4 line normalisations representing the  
\ion{O}{VII}~He$\alpha$~{\em f}, \ion{O}{VIII}~Ly$\alpha$,
\ion{Ne}{IX}~He$\alpha$~{\em r} and \ion{Ne}{X}~Ly$\alpha$).
The best-fit spectrum was then used to obtain the energy adjustment and the line normalisations for each observation.
In parallel we determined the energy adjustment using the {\tt gain fit} command in {\tt XSPEC} with the standard spectrum,
only allowing the shift as free parameter and fixing the slope to 1.0.
A comparison of the required energy adjustments obtained from the two methods shows no significant differences. 
Therefore, we proceeded to use the {\tt gain fit} in {\tt XSPEC} because it is simpler to use.

In Fig.~\ref{fig-pnshift} we show the derived energy shift (using {\tt
  gain fit}) for the SW mode observation as function of time. 
No clear trend is visible with an average shift of $+1.8 \pm 1.0$ eV 
(1$\sigma$ confidence), which is well within the instrument channel
width of 5 eV. However, a systematic difference between the two sets
of observations (boresight, centered) is revealed. For the 
observations with the target placed at boresight (centered) the average 
shift was determined to $-0.7 \pm 0.5$ eV ($+3.6 \pm 1.2$ eV).
The boresight position is the best calibrated which is supported by the 
small average energy shift. The centre location corresponds to different 
RAWX and RAWY coordinates on the CCD. It is closer to the CCD read out 
(lower RAWY) and therefore charge transfer losses are reduced. On the other hand
the gain depends on the read-out column (RAWX). Therefore, it is not clear
if uncertainties in the gain or charge transfer calibration or both are 
responsible for the difference in energy scale of about 4 eV at the two positions.
Similar position-dependent effects were found from observations of the isolated 
neutron star RX\,J1856.5-3754 \citep{sartore2012}.

\begin{figure}
   \resizebox{\hsize}{!}{\includegraphics[clip=,angle=-90]{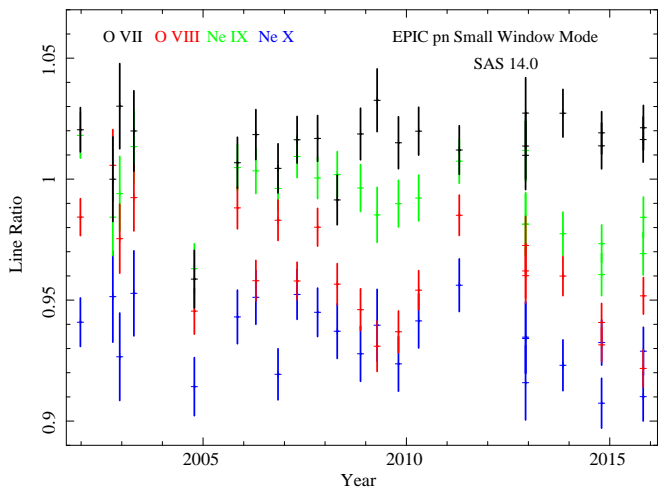}}
   \caption{Relative line normalisations (compared to the IACHEC
     model) derived from the EPIC-pn SW mode spectra of E0102 as function of time.}
   \label{fig-pnnorms}
\end{figure}

The line normalisations of the four line complexes after the gain fit
(multiplied by the overall normalisation and corrected to the large 
extraction radius) relative to the model normalisations are shown in 
Fig.~\ref{fig-pnnorms}. 
For each line complex the derived line normalisations are consistent
with being constant in time. The largest deviations are seen from the 
fifth observation, one for which the thick filter was used.  On the
other hand, during revolution 2380 (2012-12-06) three observations
were performed with
the three different filters yielding consistent results.  The average 
values (fitting a constant to the normalisations) are 1.014$\pm$0.004 
(\ion{O}{vii}), 0.959$\pm$0.003 (\ion{O}{viii}), 0.991$\pm$0.003
(\ion{Ne}{ix}) and 0.933$\pm$0.004 (\ion{Ne}{x}).   The \ion{O}{viii} 
and \ion{Ne}{x} ratios are significantly lower by about 5\% than the 
ratios from their corresponding lower ionization lines. A possible
error in the calibration over such relatively narrow energy bands is 
difficult to understand and needs further investigation.


\subsection{\xmmn\/  EPIC-MOS}\label{epicmos}
\subsubsection{Instruments}

\xmmn~\citep{jansen2001} has three X-ray telescopes
each with a European Photon Imaging Camera (EPIC) at the focal
plane. Two of the cameras have seven MOS CCDs (henceforth MOS1 and
MOS2) \citep{turner2001}  and the third has twelve pn CCDs (see
Sect. \ref{epicpn}). Apart from the characteristics of the
detectors, the telescopes are differentiated by the fact that the
MOS1 and MOS2 telescopes contain the reflection grating arrays which
direct approximately half the X-ray flux into the apertures of the
reflection grating spectrometers (RGS1 and RGS2).


\begin{table*}
\centering

\caption[ ]{MOS Observations of E0102}

\label{tab:mosobs}
\begin{tabular}{lllccl}
\hline
\hline
OBSID & Instrument & DATE & Exposure & Counts  & Mode  \\
 &  &  & (ks) & (0.5-2.0 keV) &  \\
\hline
0123110201\tablefootmark{a} & MOS1 & 2000-04-16 &  17.4 &   \phantom{1}60354 & LW, 0.9s frametime, thin filter \\
0123110201\tablefootmark{a} & MOS2 & 2000-04-16 &  17.4 &   \phantom{1}60633 & LW, 0.9s frametime, thin filter \\
0123110301\tablefootmark{a} & MOS1 & 2000-04-17 &  12.1 &   \phantom{1}40092 & LW, 0.9s frametime, medium filter \\
0123110301\tablefootmark{a} & MOS2 & 2000-04-17 &  12.1 &   \phantom{1}40893 & LW, 0.9s frametime, medium filter \\
0135720601 & MOS1 & 2001-04-14 &  18.6 &   \phantom{1}65561 & LW, 0.9s frametime, thin filter \\
0135720601 & MOS2 & 2001-04-14 &  18.6 &   \phantom{1}62927 & LW, 0.9s frametime, thin filter \\
0135720801 & MOS1 & 2001-12-25 &  28.0 &  102340 & LW, 0.9s frametime, thin filter \\
0135720801 & MOS2 & 2001-12-25 &  28.0 &   \phantom{1}99855 & LW, 0.9s frametime, thin filter \\
0135721301 & MOS1 & 2002-12-14 &  27.2 &   \phantom{1}93882 & LW, 0.9s frametime, thin filter \\
0135721301 & MOS2 & 2002-12-14 &  27.2 &   \phantom{1}93392 & LW, 0.9s frametime, thin filter \\
0135721501 & MOS1 & 2003-10-27 &  21.0 &   \phantom{1}76062 & LW, 0.9s frametime, thin filter \\
0135721501 & MOS2 & 2003-10-27 &  21.0 &   \phantom{1}70285 & LW, 0.9s frametime, thin filter \\
0135721901 & MOS1 & 2004-04-28 &  31.0 &  106535 & LW, 0.9s frametime, thin filter \\
0135721901 & MOS2 & 2004-04-28 &  31.0 &  104510 & LW, 0.9s frametime, thin filter \\
0135722401 & MOS1 & 2004-10-14 &  29.4 &   \phantom{1}84006 & LW, 0.9s frametime, thick filter \\
0135722401 & MOS2 & 2004-10-14 &  29.4 &   \phantom{1}80836 & LW, 0.9s frametime, thick filter \\
0135722501 & MOS1 & 2005-04-17 &  29.4 &   \phantom{1}98331 & LW, 0.9s frametime, thin filter \\
0135722501 & MOS2 & 2005-04-17 &  29.4 &   \phantom{1}97993 & LW, 0.9s frametime, thin filter \\
0135722601 & MOS1 & 2005-11-05 &  29.1 &   \phantom{1}98813 & LW, 0.9s frametime, thin filter \\
0135722601 & MOS2 & 2005-11-05 &  29.1 &   \phantom{1}96880 & LW, 0.9s frametime, thin filter \\
0412980101 & MOS1 & 2006-11-05 &  31.0 &  102286 & LW, 0.9s frametime, thin filter \\
0412980101 & MOS2 & 2006-11-05 &  31.0 &   \phantom{1}99904 & LW, 0.9s frametime, thin filter \\
0412980301 & MOS1 & 2007-10-26 &  35.0 &  119372 & LW, 0.9s frametime, thin filter \\
0412980301 & MOS2 & 2007-10-26 &  35.0 &  111364 & LW, 0.9s frametime, thin filter \\
0412981401 & MOS1 & 2011-04-20 &  30.3 &   \phantom{1}84751 & LW, 0.9s frametime, thin filter \\
0412981401 & MOS2 & 2011-04-20 &  30.3 &  100949 & LW, 0.9s frametime, thin filter \\
0412981701 & MOS1 & 2012-12-06 &  12.8 &   \phantom{1}39780 & LW, 0.9s frametime, thin filter \\
0412981701 & MOS2 & 2012-12-06 &  12.8 &   \phantom{1}43069 & LW, 0.9s frametime, thin filter \\
0412981701 & MOS1 & 2012-12-06 &  16.6 &   \phantom{1}47453 & LW, 0.9s frametime, medium filter \\
0412981701 & MOS2 & 2012-12-06 &  16.6 &   \phantom{1}53615 & LW, 0.9s frametime, medium filter \\
0412982101 & MOS1 & 2013-11-07 &  31.3 &   \phantom{1}95929 & LW, 0.9s frametime, thin filter \\
0412982101 & MOS2 & 2013-11-07 &  31.3 &   \phantom{1}98882 & LW, 0.9s frametime, thin filter \\
0412982201 & MOS1 & 2014-10-20 &  32.8 &  104184 & LW, 0.9s frametime, thin filter \\
0412982201 & MOS2 & 2014-10-20 &  32.8 &  103544 & LW, 0.9s frametime, thin filter \\

\hline

\end{tabular}
\tablefoot{\tablefoottext{a}{Observation included in the comparison of
    effective areas discussed in Section~\ref{fitresults}. } }

\end{table*}


\subsubsection{Data}

E0102 was first observed by \xmmn\/ quite early in the
mission in April 2000 (orbit number 0065). The MOS observations
are listed in (Table~\ref{tab:mosobs}).
 This first look at the
target in orbit 0065 was split into two observations, each
approximately 18 ks in duration, with a different choice of
optical filter. Observation 0123110201 had the THIN filter and 0123110301 had
the MEDIUM filter. Both filters have a 1600~$\AA$ polyimide film with
evaporated layers of 400~$\AA$ and 800~$\AA$ respectively of Aluminum. The
EPIC-MOS readout was configured to the Large Window (LW) imaging mode
(in the central CCD only the inner $300 \times 300$ pixels of the
total available $600 \times 600$ pixels are read out). LW mode is the
most common imaging mode used in EPIC-MOS observations of this target as
the faster readout (0.9~s compared with 2.6~s in full frame (FF) mode)
minimises pile-up whilst retaining enough active area to contain the
whole remnant for pointings up to around 2 arcminutes from the centre
of the target. This is useful for exploring the response of the instrument
for off-axis angles near to the boresight.

\subsubsection{Processing}

The EPIC-MOS data were first processed into calibrated event lists with SAS
version 12.0.0 and the current calibration files (CCFs) as of May 2013
and later with SAS version 13.5.0 and the CCFs as of December
2013. The signficant differences between the SAS and CCF versions are
dealt with in Sect.~\ref{epicmostrend}.

Source spectra were extracted from a circular region of radius
80\arcsec\ centered on the remnant. Background spectra were taken
from source-free regions on the same CCD. The event selection filter
in the nomenclature of the SAS was (PATTERN==0)\&\&(\#XMMEA\_EM). This
selects only mono-pixel events and removes events whose reconstructed
energy is suspect due, for example, to proximity to known bright
pixels or CCD boundaries which can be noisy.

Mono-pixel events are chosen over the complete X-ray pattern library
because it minimises the effects of pile-up with little loss of
sensitivity over the energy range of interest. The effects of pile-up
on the mono-pixel spectrum can be shown to be small. The
mono-pixel pile-up fraction, the fraction of events lost to higher
patterns or formed from two (or more) X-rays detected in the same
pixel within a frame (the former is more likely by a factor of about
8:1), can be estimated from the observed fraction of diagonal bi-pixel
events which arise almost exclusively from the pile-up of two
mono-pixel events. By default the SAS splits these events
(nominally pattern classes 26 to 29) back into two separate
mono-pixels although this action can be switched off. Less than
1.0\% of events within the source spectra are diagonal bi-pixels which
is approximately the same fraction of mono-pixel events lost to
horizontal or vertical bi-pixels (event pattern classes 1 to 4).

We employ a simple screening algorithm to detect flares in the
background due to soft protons. Light curves of bin size 100s were
created from events with energies greater than 10 keV within the whole
aperture. Good time intervals were formed where the observed rate was
less than $\rm 0.4 \ cts \ s^{-1}$. The cut-off limit was chosen by
manual inspection of the light curves. Typically after this procedure
the observed background is less than 1\% of the total count rate below
2.0~keV.

All spectra were extracted with a 5.0~eV bin size. Response (RMF)
files were generated with the SAS task {\em rmfgen} in the energy
range of interest with an energy bin size of 1.0~eV. This is
comparable to the accuracy with which line centroids can be determined
for the stronger lines in this source for typical exposures in the
EPIC-MOS. Although the source is an extended, but compact, object, the
effective area (ARF) file was calculated with the SAS task {\em
  arfgen} assuming a point-source function model with the switch
PSFMODEL=ELLBETA. 

\begin{table*}
\centering

\caption{EPIC-MOS fitting results}             
\label{table:mosfits}      
\begin{tabular}{lcccc}        
\hline
\hline
  & \multicolumn{2}{c}{MOS1} & \multicolumn{2}{c}{MOS2} \\
\hline
 Parameter\tablefootmark{a} & Thin & Medium & Thin & Medium \\
\hline
\multicolumn{5}{c}{Without Gain Fit} \\
\hline
 Global & 1.063 (0.009) & 1.057 (0.011) & 1.091 (0.009) & 1.059 (0.011)\\
\ion{O}{VII} & 1.380 (0.026) & 1.428 (0.032) & 1.385 (0.026) & 1.460 (0.033) \\
\ion{O}{VIII} & 4.576 (0.068) & 4.459 (0.081) & 4.487 (0.068) & 4.623 (0.085) \\
\ion{Ne}{IX} & 1.375 (0.022) & 1.369 (0.026) & 1.348 (0.022) & 1.404 (0.027) \\
\ion{Ne}{X}  & 1.393 (0.025) & 1.371 (0.029) & 1.373 (0.025) & 1.419 (0.031) \\
 C-stat/dof & 450.5/334 & 413.9/334 & 432.3/334 & 419.6/334 \\
\hline
\multicolumn{5}{c}{With Gain Fit} \\
\hline
 Global & 1.060 (0.009) & 1.044 (0.011) & 1.081 (0.009) & 1.059 (0.011) \\
\ion{O}{VII} & 1.383 (0.026) & 1.410 (0.032) & 1.383 (0.027) & 1.449 (0.034) \\
\ion{O}{VIII} & 4.609 (0.069) & 4.508 (0.082) & 4.536 (0.069) & 4.650 (0.086) \\
\ion{Ne}{IX} & 1.386 (0.022) & 1.378 (0.026) & 1.361 (0.022) & 1.409 (0.027) \\
\ion{Ne}{X}  & 1.400 (0.026) & 1.383 (0.029) & 1.385 (0.025) & 1.426 (0.031) \\
 Offset & -4.358 & -6.293 & -6.297 & -3.641 \\
 Slope & 1.0061 & 1.0059 & 1.0081 & 1.0035 \\
 C-stat/dof & 422.7/332 & 376.9/332 & 387.8/332 & 408.7/332 \\
\hline                                   
\end{tabular}
\tablefoot{\tablefoottext{a}{Line normalisations for O~VII, O~VIII, Ne~IX and Ne~X are multiplied by $10^{-3}$}}
\end{table*}


We justify this over attempting to accurately account for the extended
nature of the object in the generation of the ARF because to do so is
mathematically much more complex and the end result can be predicted
to produce a result which would be much closer to the point-source
approximation than the basic uncertainties in the calibration.
To formally account for the extended nature of the object
would require deconvolving the image with the telescope point-spread
function to get the true input spatial distribution relative to the
mirror and then estimating for each point in the image both the
encircled energy fraction (EEF) relative to the applied spectral
extraction region and also the vignetting function. The final ARF
would then be a counts weighted average of the ARF derived at each
point.

As the remnant is approximately a ring like structure $25''$ in radius
then the bulk of the input photons have an angular distance relative
to the circular $80''$ extraction region which varies between $55''$
to $105''$, but has a mean of about $82''$. The EEF for a point source
is approximately 91\%, 94\% and 96\% at $55"$, $80''$ and $105''$
respectively (in the energy range of interest). Hence, the adoption of
a single EEF for an $80''$ radius is estimated to be approximately
within 1\% of the value that would be derived if one adopted the
technically more accurate method outlined previously. Similarly, the
calibrated vignetting variation across the remnant is less than 1\%
hence assuming the representative value at the centre is a justifiable
approximation. Overall, the accuracy of the calculated ARF is clearly
dominated more by the absolute uncertainty in the calibration of the
vignetting and EEF than the point-source assumption employed here.

Fitting of the model to the source spectra followed the recipe
described earlier. Table~\ref{table:mosfits} shows the fitting results
from each of the EPIC-MOS observations from Orbit 0065. 
The Thin and Medium filter
results are consistent within the errors and the global results for
MOS1 and MOS2 (see Sect.~\ref{fitresults}) are the weighted averages of results
from each filter. Shifts in the calibration of the event energy scale
were investigated using the {\tt gain fit} command in {\tt XSPEC} with an
improvement in the fit statistic arising from shifts of around $\sim
5$~eV at 1.0 keV. This is typical of the calibration accuracy of the
event energy scale in the EPIC-MOS detectors. The values of the parameter
normalisations are relatively insensitive to gain shifts of this
magnitude. 
This was confirmed by applying a reverse gain shift, using
the values indicated by {\tt XSPEC}, to the calibrated energies of each
event, then re-extracting and re-fitting the spectra.

EPIC-MOS differs from other instruments in this paper because we
explicitly use the source model described here to constrain the
redistribution model of our RMF. Within a few years of launch it was
noticed that the redistribution properties of the central CCD (in both
MOS1 and MOS2) in a spatial region centered around the telescope
boresight (within $\sim 1$~arcminute) were evolving with time. The
change was consistent with an evolution of the strength and shape of
the low-energy charge-loss component of the redistribution profile.  As
we do not have a physical model of this effect accurate enough to
describe the changing RMF, it is calibrated using a method of varying
the parameters of a phenomenological model of the RMF to provide a
{\em best fit} solution to a joint simultaneous fit to spectra from
our onboard calibration source and several astrophysical sources,
including E0102, using fixed spectral models  \citep{Sembay2011}. 
In the case of the astrophysical sources the
input spectral models are derived primarily from the RGS and
EPIC-pn. The fitting procedure allows variation in the global
normalisation of each spectral model otherwise all parameters within
the model are fixed. The consequence of this is that the RMF
parameters are driven someway towards a result which gives an energy
independent cross-calibration between the EPIC-MOS and RGS. This in part
explains why the relative line-to-line normalisations are consistent
with the RGS although there is a global offset between the instruments
(see Sect. \ref{fitresults}).

\subsection{\chan\/  ACIS}
\subsubsection{Instruments}

The ACIS is an X-ray imaging-spectrometer consisting of the ACIS-I and ACIS-S CCD
arrays. The imaging capability is unprecedented with a half-power diameter (HPD) of
$\sim$1\arcsec\ at the on-axis position. We use data from one of the 
back-illuminated CCDs (ACIS-S3) in 
the ACIS-S array in this analysis since the majority of imaging data
have been collected using this CCD and its response at
low energies is significantly higher than the front-illuminated CCDs 
in the ACIS-I array. There are also more observations of E0102 on S3 than on
the I array which allows a better characterization of the time
dependence of the response.
The ACIS-S3 chip is sensitive in the 0.2--10~keV band.
The chip has
1024$\times$1024 pixels covering a 8\farcm4$\times$8\farcm4 area.  The
spectral resolution is $\approx150$~eV in the 0.3--2.0~keV bandpass.

\subsubsection{Data}
 
The high angular resolution of \chan\/ compared to the
other observatories is apparent in Fig.~\ref{fig:image4} as
evidenced by the fine structure apparent in this SNR.
The majority of the observations of E0102 early in the
\chan\/ mission were executed in full-frame mode with 3.2~s exposures.
Unfortunately the bright parts of the ring are significantly piled-up when
ACIS is operated in its full-frame mode. In 2003, an observation was
conducted in subarray mode that showed the line fluxes were depressed
compared to the observations in full-frame mode.  In 2005, the \chan\/
calibration team switched to using subarray modes with readout times
of 1.1~s and 0.8~s as the default modes to observe E0102 resulting in
a reduction in the pileup level. There have been 14 subarray
observations of E0102 on the S3 CCD within one arcminute of the
on-axis position, twelve in node 0 and two in node 1 (see 
Table~\ref{tab:acisobs}).
There are other observations of E0102 on S3 at larger off-axis angles
which we exclude from the current comparison to the other instruments
close to on-axis.
 We have selected the two earliest OBSIDs for comparison to the other
instruments and discuss the analysis of all 14 observations in \S\ref{fitresults}.

\subsubsection{Processing}

The data were processed with the {\em Chandra X-ray Center}(CXC) analysis
SW {\tt CIAO} v4.7 and the CXC calibration database CALDB~v4.6.8.
We followed the standard {\tt CIAO} data analysis threads to select
good events, reject times of high background, and extract source and 
background spectra in PI channels.  Background spectra were extracted 
from regions
off of the remnant that were specific to each observation since the
region of the sky covered varied from observation to observation due
to the roll angle of the observation. Response matrices were produced
using the standard {\tt CIAO} tool {\tt mkacisrmf} with PI channels
and auxiliary response files were produced using {\tt mkwarf} to
account for the extended nature of E0102. These tools were called as 
a \chan\/ Guest Observer would using the {\tt CIAO} script {\tt
specextract.pl}.

There are several time-dependent effects which the analysis SW
attempts to account for \citep{plucinsky2003,marshall2004,depasq2004}.  
The most important of these is the
efficiency correction for the contaminant on the ACIS optical-blocking
filter which significantly reduces the efficiency at energies around
the O lines. We chose the earliest two OBSIDs to compare to the other
instruments since the contamination layer was thinnest at that time.
The analysis SW also corrects for the CTI of the BI CCD
(S3), including the time-dependence of the gain.  Even with this
time-dependent gain correction, some of the observations exhibited
residuals around the bright lines that appeared to be due to gain
issues. We then fit allowing the gain to vary and noticed that some of
the observations had significant improvements in the fits when the
gain was allowed to adjust.  The adjustments were small, about 5~eV
which corresponds to one ADU for S3.  We derived a non-linear gain
correction using the energies of the  \ion{O}{vii}~He$\alpha$ triplet, 
the \ion{O}{viii}~Ly$\alpha$ line, the \ion{Ne}{ix}~He$\alpha$ triplet
\& \ion{Ne}{X}~Ly~$\alpha$ line, requiring the gain adjustment to go to
zero at 1.5~keV.  These gain adjustments were applied to the events
lists and spectra were re-extracted from these events lists.  The 
modified spectra were used for subsequent fits. This ensures that the
line flux is attributed to the correct energy and the appropriate
value of the effective area is used to determine the line normalization.

\subsection{Suzaku XIS}
\subsubsection{Instruments}

The XIS is an X-ray imaging-spectrometer equipped with four X-ray CCDs
sensitive in the 0.2--12~keV band. One CCD is a back-illuminated
(XIS1) device and the others are front-illuminated (XIS0, 2, and 3)
devices.  The four CCDs are located at the focal plane of four co-aligned
X-ray telescopes with a half-power diameter (HPD) of $\sim$2\farcm0. Each
XIS sensor has 1024$\times$1024 pixels and covers a
17\farcm8$\times$17\farcm8 field of view.  The XIS instruments, constructed
by MIT Lincoln Laboratories, are very similar in design to the ACIS CCDs
aboard {\em Chandra}.  They are fully described by \citet{koyama2007}.
Due to expected degradation in the power supply system, {\suzaku\/}
lost attitude control in June 2015, and the science mission was declared
completed in August 2015.\footnote{See
http://global.jaxa.jp/press/2015/08/20150826\_suzaku.html.}

The XIS2 device suffered a putative micro-meteorite hit in November 2006
that rendered two-thirds of its imaging area unusable, and it has been
turned off since that point.  XIS0 also suffered a micro-meteorite hit in
June 2009 that affected one-eighth of the device.  Since this region is
near the edge of the chip, the device was still used for normal
observations  until the cessation of science
operations in August 2015.  The other two CCDs continued to
operate normally. 

Unlike the ACIS devices, the XIS CCDs possess a charge injection capability
whereby a controlled amount of charge can be introduced via a serial
register at the top of the array.  This injected charge acts to fill CCD
traps that cause charge transfer inefficiency (CTI), mitigating the effects
of on-orbit radiation damage \citep{ozawa2009}.  In practice, the XIS
devices were operated with spaced-row charge injection (SCI) on
starting in
August 2006.  A row of fixed charge is injected every 54 rows; the injected
row is masked out on-board, slightly reducing the useful detector area.
The level of SCI in the FI chips has been set to about 6 keV for the
duration of the mission.  The level in the BI chip was initially set to 2
keV to reduce noise at soft energies, however in late 2010 and early 2011
this level was raised to 6 keV.

\begin{table*}
\centering

\caption[ ]{XIS Observations of E0102}

\label{tab:xisobs}
\begin{tabular}{lccccl}
\hline
\hline
OBSID & Instrument & DATE & Exposure & Counts & Mode\tablefootmark{b}
\\
 &  &  & (ks)\tablefootmark{a} & (0.5-2.0 keV) &  \\
\hline
100014010\tablefootmark{c} & XIS0 & 2005-08-31 & 22.1 & \phantom{2}33078 & full window,SCI off   \\
100014010\tablefootmark{c} & XIS1 & 2005-08-31 & 22.1 & \phantom{2}71394 & full window,SCI off   \\
100014010\tablefootmark{c} & XIS2 & 2005-08-31 & 22.1 & \phantom{2}33475 & full window,SCI off   \\
100014010\tablefootmark{c} & XIS3 & 2005-08-31 & 22.1 & \phantom{2}31569 & full window,SCI off   \\
100044010 & XIS0 & 2005-12-17 & 52.6 & \phantom{2}70904 & full window,SCI off   \\
100044010 & XIS1 & 2005-12-17 & 94.4 & 224811 & full window,SCI off   \\
100044010 & XIS2 & 2005-12-17 & 52.6 & \phantom{2}65054 & full window,SCI off   \\
100044010 & XIS3 & 2005-12-17 & 52.6 & \phantom{2}58182 & full window,SCI off   \\
101005030 & XIS0 & 2006-06-27 & 21.0 & \phantom{2}24879 & full window,SCI off   \\
101005030 & XIS1 & 2006-06-27 & 18.5 & \phantom{2}36029 & full window,SCI off   \\
101005030 & XIS2 & 2006-06-27 & 21.0 & \phantom{2}21734 & full window,SCI off   \\
101005030 & XIS3 & 2006-06-27 & 18.5 & \phantom{2}17858 & full window,SCI off   \\
102002010 & XIS0 & 2007-06-13 & 24.0 & \phantom{2}24632 & full window,SCI on   \\
102002010 & XIS1 & 2007-06-13 & 24.0 & \phantom{2}40526 & full window,SCI on   \\
102002010 & XIS3 & 2007-06-13 & 24.0 & \phantom{2}21898 & full window,SCI on   \\
103001020 & XIS0 & 2008-06-05 & 17.5 & \phantom{2}15843 & full window,SCI on   \\
103001020 & XIS1 & 2008-06-05 & 17.5 & \phantom{2}28543 & full window,SCI on   \\
103001020 & XIS3 & 2008-06-05 & 17.5 & \phantom{2}15646 & full window,SCI on   \\
104006010 & XIS0 & 2009-06-26 & 17.4 & \phantom{2}14915 &  full window,SCI on   \\
104006010 & XIS1 & 2009-06-26 & 17.4 & \phantom{2}27072 &  full window,SCI on   \\
104006010 & XIS3 & 2009-06-26 & 17.4 & \phantom{2}15278 &  full window,SCI on   \\
105004020 & XIS0 & 2010-06-19 & 15.5 & \phantom{2}12629 & full window,SCI on   \\
105004020 & XIS1 & 2010-06-19 & 15.5 & \phantom{2}24207 & full window,SCI on   \\
105004020 & XIS3 & 2010-06-19 & 15.5 & \phantom{2}12993 & full window,SCI on   \\
106002020 & XIS0 & 2011-06-29 & 27.4 & \phantom{2}21361 &  full window,SCI on   \\
106002020 & XIS1 & 2011-06-29 & 27.4 & \phantom{2}44156 &  full window,SCI on   \\
106002020 & XIS3 & 2011-06-29 & 27.4 & \phantom{2}23490 &  full window,SCI on   \\
107002020 & XIS0 & 2012-06-25 & 29.6 & \phantom{2}23722 & full window,SCI on   \\
107002020 & XIS1 & 2012-06-25 & 29.6 & \phantom{2}45564 & full window,SCI on   \\
107002020 & XIS3 & 2012-06-25 & 29.6 & \phantom{2}26527 & full window,SCI on   \\
108002020 & XIS0 & 2013-06-27 & 33.1 & \phantom{2}28615 & full window,SCI on   \\
108002020 & XIS1 & 2013-06-27 & 33.1 & \phantom{2}56869 &  full window,SCI on   \\
108002020 & XIS3 & 2013-06-27 & 33.1 & \phantom{2}31354 &  full window,SCI on   \\
109001010 & XIS0 & 2014-04-21 & 29.6 & \phantom{2}25982 & full window,SCI on   \\
109001010 & XIS1 & 2014-04-21 & 29.6 & \phantom{2}48706 & full window,SCI on   \\
109001010 & XIS3 & 2014-04-21 & 29.6 & \phantom{2}26135 & full window,SCI on   \\
\hline

\end{tabular}

\tablefoot{\tablefoottext{a}{Exposure for XIS is for the filtered event data.} \\
\tablefoottext{b}{'SCI' stands for spaced-row charge injection.} \\
\tablefoottext{c}{Observation included in the comparison of
    effective areas discussed in Section~\ref{fitresults}. } }

\end{table*}


\subsubsection{Data}

The XIS observations in this work include representative datasets over the
course of the mission. E0102 was a standard calibration source for \suzaku,
with 74 separate observations during the life of the mission,
including the very first observation when the detector doors were opened.
We have chosen eleven observations each taken about one year apart and
typically 20--30 ksec in duration. 
One of these observations was taken shortly after launch on 17 Dec
2005, and is the longest single observation of E0102 with the XIS (94 ksec
of clean data from the BI CCD and 50 ksec from each of the FI CCDs).
However, these data were taken at a time when the molecular contamination
on the optical blocking filters was rapidly accumulating.  Since the
calibration at this epoch is uncertain, we have chosen an earlier, somewhat
shorter observation (31 Aug 2005) to compare to the other instruments.
The observations are summarized in Table
\ref{tab:xisobs}.
Three of these observations (in 2005 and 2006) were taken with SCI off, the
remainder with SCI on. Observations starting in 2011 were taken with the
XIS1 SCI level set to 6 keV. Only three observations have been included for
XIS2, which ceased operation in late 2006. Normal, full-window observing
mode was used for all analyzed datasets.

\subsubsection{Processing}

The data were reprocessed to at least v2.7 of the XIS pipeline.  In
particular, the CTI, charge trail, and gain parameters were applied from
v20111018 or later of the makepi CALDB file, which reduced the gain
uncertainty to less than 10 eV for all of the observations.  Further gain
correction was performed during the spectral analysis, in a similar way to
\chan\/ ACIS-S3 and as described in Sect.~\ref{fitmethod}.

During the data processing, we found a large variation in the \suzaku\/
pointing accuracy, with an average astrometric offset of 20 arcsec but
ranging up to 1.5 arcmin.  In the worst case this is quite a bit larger
than the published astrometric accuracy of 20 arcsec, although
smaller than the PSF of the \suzaku\/ XRT mirrors ($\sim$ 2 arcmin HPD).
Given this pointing error, and the presence of a contaminating point source
(RXJ0103.6-7201) projected 2 arcmin from E0102, we corrected the pointing
by applying a simple offset in RA and Dec to the attitude data.  This
offset was calculated from a by-eye comparison of the \suzaku\/ centroids of
E0102 (in the 0.4--2 keV band) and RXJ0103.6-7201 (in the 2--7 keV band) to
the source locations in a stacked \chan\/ ACIS-S3 image.  The offset for each XIS
was determined separately and then averaged to produce the attitude offset
for a single observation.  From the dispersion of these measurements, we
estimate that the RMS uncertainty in the corrected astrometry is 5 arcsec,
or about 5 unbinned CCD pixels.  We note that this correction is different
from the \suzaku\/ XRT thermal wobble, which is corrected in the
pipeline\footnote{ftp://legacy.gsfc.nasa.gov/suzaku/doc/xrt/suzakumemo-2007-04.pdf};
and the attitude control problem which plagued the satellite between Dec
2009 and June 2010, which has not been corrected and effectively produces a
smearing of the
PSF\footnote{ftp://legacy.gsfc.nasa.gov/suzaku/doc/general/suzakumemo-2010-04.pdf}.

Spectra were extracted from a 3 arcmin radius aperture, which would contain
95\% of the flux from a point
source
We excluded a 1 arcmin radius region around RXJ0103.6-7201.  Background
spectra were extracted from a surrounding annulus encompassing 5.6--7.4
arcmin.  The redistribution matrix files (RMFs) were produced with the
\suzaku\/ FTOOL xisrmfgen (v20110702), using v20111020 of the CALDB RMF
parameters. The ancillary response files (ARFs) were produced with the
Monte Carlo ray-tracing FTOOL xissimarfgen (v20101105). The ARF includes
absorption due to OBF contamination (Koyama et al. 2007), using v20130813
of the CALDB contamination parameters.  To ensure the ARF properly
accounted for the partially-resolved extent of the source, a \chan\/ ACIS-S3
broad-band image of the inner 30 arcsec of E0102 was used as an input
source for the ray-tracing.

This X-ray binary RXJ0103.6-7201, projected 2 arcmin from E0102, shows up
clearly in \chan\/ ACIS-S3 observations, and it is well-modeled by a power law
with spectral index 0.9 plus a thermal mekal component with kT = 0.15 keV,
with a strong correlation between the component normalizations (Haberl \&
Pietsch 2005).  By masking it out in the spectral extraction with a 1
arcmin radius circle, we reduce its contribution by 50\%.  In the region
below 3 keV, we expect E0102 thermal emission to dominate the residual
contaminating flux by several orders of magnitude.

\subsection{Swift XRT}
\subsubsection{Instruments}

The {\em Swift} X-ray Telescope (XRT) comprises a Wolter-I telescope,
originally built for JET-X, which focuses X-rays onto an e2v CCD22
detector, similar to the type flown on the \xmmn\/ EPIC-MOS
instruments \citep{burrows2005}.
The CCD, which was responsive to $\sim 0.25-10.0 \keV$ X-rays at
launch, has
dimensions of $600\times 600$ pixels, giving a 
$23.6\arcmin \times 23.6\arcmin$ field of view. The mirror has a HPD
of $\sim 18\arcsec$ and
can provide source localization accurate to better than 2\arcsec\
\citep{evans2009}.

\begin{table*}
\caption{{\em Swift}-XRT observation log. The data reported here were
  taken under target ID number 50050. The reported rates are from
  grade 0 events and are not corrected for potential loss of exposure
  due to the location of the source with respect to the detector
  bad-columns.}
\label{tab:swiftxrt_obs}
\centering
\begin{tabular}{llccr}
\hline
Start Date & Stop Date & Exposure & 0.3-1.5 keV rate & Offset\\
 &  & (ks) & (${\rm count\,s^{-1}}$) & (eV)\phantom{0} \\
\hline
PC mode : &  & & & \\
2005-02-18\tablefootmark{a} & 2005-05-22 & 24.2 & 0.87 &  +2\phantom{00} \\
2006-03-11 & 2006-05-05 &  \phantom{2}8.5 & 0.81 &  0\phantom{00} \\
2007-06-08 & 2007-06-13 & 20.8 & 0.73 &  +4\phantom{00} \\
2007-09-25 & 2007-10-02 & 28.4 & 0.73 & -6\phantom{00} \\
2008-10-01 & 2008-10-04 & 20.4 & 0.80 & +8\phantom{00} \\
2009-10-18 & 2009-11-27 & 20.8 & 0.74 & -6\phantom{00} \\
2010-03-16 & 2010-09-11 & 40.0 & 0.68 & -1\phantom{00} \\
2011-03-19 & 2011-09-14 & 35.3 & 0.64 & -3\phantom{00} \\
2012-03-09 & 2012-09-13 & 35.1 & 0.70 & -8\phantom{00} \\
2013-03-15 & 2013-10-19 & 43.1 & 0.61 & -11\phantom{00} \\
2014-03-13 & 2014-09-27 & 49.9 & 0.73 & -11\phantom{00} \\
\hline
WT mode : & & & & \\
 2005-02-23\tablefootmark{a} & 2005-03-01 & 25.2 & 1.09 & +1\phantom{00} \\
 2006-03-11 & 2006-04-27 &  \phantom{2}7.4 & 0.91 & -2\phantom{00} \\
 2007-06-07 & 2007-06-20 & 20.7 & 0.94 & +4\phantom{00} \\
 2007-09-30 & 2007-10-01 & 15.4 & 0.95 & -1\phantom{00} \\
 2008-08-24 & 2008-10-09 & 17.3 & 0.87 & +13\phantom{00} \\
 2009-10-15 & 2009-10-20 & 22.1 & 0.98 & +4\phantom{00} \\
 2010-03-21 & 2010-09-25 & 40.1 & 0.95 & -7\phantom{00} \\
 2011-03-18 & 2011-10-15 & 41.5 & 0.90 & -1\phantom{00} \\
 2012-03-10 & 2012-08-21 & 38.2 & 0.90 & +4\phantom{00} \\
 2013-03-24 & 2013-10-24 & 38.5 & 0.81 & +2\phantom{00} \\
 2014-03-15 & 2014-10-14 & 46.7 & 0.91 & +2\phantom{00} \\
\hline
\end{tabular}

\tablefoot{\tablefoottext{a}{Observation included in the comparison of
    effective areas discussed in Section~\ref{fitresults}. } }

\end{table*}


Since its launch in 2004 November, {\em Swift}'s primary science goal has
been to rapidly respond to gamma-ray bursts (GRBs) and other targets
of opportunity (TOOs). To achieve
this, the XRT was designed to operate autonomously, so that it could
measure GRB light curves and spectra over several orders of magnitude in
flux. In order to mitigate the effects of pile-up, the XRT can
automatically switch between different CCD readout modes depending
on the source brightness. The two most frequently-used modes are:
Windowed Timing (WT) mode, which provides 1D spatial information and 
spectroscopy in
the central 7.8 arcminutes of the CCD with a time resolution of 1.8 ms, 
and Photon Counting (PC) mode, which allows full 2D
imaging-spectroscopy with a time resolution of 2.5 s 
\citep[see][for further details]{hill2004}.

The CCD charge transfer inefficiency (CTI) was seen to increase
approximately threefold a year after launch and has steadily worsened
since then. The location and depth of the deepest charge traps
responsible for the CTI in the central 7.8 arcminutes of the CCD have been
monitored since 2007 September and methods have been put in place to
minimize their effect on the spectral resolution \citep{pagani2011},
however, even with such trap corrections, the intrinsic resolution of
the CCD has slowly deteriorated with time.

A description of the XRT CCD initial in-flight calibration can be
found in \citet{godet2009}.  However, since this paper, the XRT
spectral calibration has been completely reworked, with both PC and WT
RMFs generated from a newly rewritten CCD Monte Carlo simulation code
(Beardmore et al., in prep.). 
The recalibration included a modification to the 
low energy quantum efficiency (above the oxygen edge at $0.545 \keV$), 
in order to improve the modelling of the E0102 line normalisations
compared with the IACHEC  model for the first epoch XRT
WT observation (i.e. 2005; see Sect.~\ref{sect:xrt_refdata}).

\subsubsection{Data}\label{sect:xrt_refdata}

E0102 is used as a routine calibration source by the
{\em Swift}-XRT, with $\sim 20$ ks observations taken every 6 to 12 months in
both PC and WT mode. The data are used to check the low
energy gain calibration of the CCD, as well as to monitor the
degradation in energy resolution below $1 \keV$.
The observations are performed under target ID 50050.

As {\em Swift} has a flexible observing schedule, often interrupted by
GRBs or TOOs, observations consist of one or more snapshots on the
target, where each snapshot has a typical exposure of $1-2$ ks, 
assigned to a unique observation identification number (ObsID).  When
observing E0102, it is necessary to accumulate data from different
ObsIDs for any given epoch to build up sufficient statistics in the
spectra.  We chose to divide the data by year\footnote{There are two
  epochs in 2007 corresponding to observations taken before and after
  a change made to the CCD substrate voltage.}  which gives ample
temporal resolution for monitoring the CCD spectral response
evolution.  The observation summary is shown in
Table~\ref{tab:swiftxrt_obs}.  The data were taken over 68 ObsIDs in
PC and 61 in WT, with exposures totalling 276.6 ks and 266.4 ks in
each mode, respectively.

We select observations taken in 2005 for comparison with the other
instruments, as these data were taken when the CCD charge transfer
efficiency and spectral resolution were at their best, and the
observations took place before a micrometeoroid struck the CCD,
introducing bad-columns which make the absolute flux calibration more
uncertain. We discuss the results from
the following  epochs (2006--2013) in Sect.~\ref{swiftxrt_temporal}.

\subsubsection{Processing}

The data were processed with the latest {\em Swift}-XRT pipeline
software (version 4.3) and   
CALDB release 2014-Jun-10 was used, which includes the
latest epoch dependent RMFs that  track the
ever broadening response kernel of the CCD.
We selected grade 0 events for the spectral
comparison, as this minimises the effects of pile-up on the PC mode
data. Due to its faster readout, the WT data are free from pile-up.
A circular region of radius 30 pixels ($70.7\arcsec$) was used for the
spectral extraction for both modes, though due to the 1D nature of the
WT readout this effectively becomes a box of size $60\times 600$
pixels (in detector coordinates) in this mode.  Background spectra
were selected from suitably sized annular regions. 
The source and background spectra from each  ObsID in
a particular epoch were then summed.  For WT mode, the background is
$\sim 10 \times$ larger than that in PC mode and dominates the WT
source spectrum above $\sim 3 \keV$.
(Note, the WT
background spectrum BACKSCAL keyword has to be modified to ensure the
correct 1D proportional sizes of the extraction regions are used, otherwise
insufficient background is subtracted when the spectra are read into
{\tt XSPEC}.)

For PC mode, exposure corrected ancillary response files (ARFs) were
created by taking the \chan\/ ACIS-S3 image (which has a superior spatial
resolution to the XRT) and convolving it with the XRT point spread
function. Then, by comparing the exposure corrected convolved counts
in the extraction region to the total number in the image, an ARF
correction factor could be estimated. The ARFs were then corrected
for vignetting by supplying the exposure weighted average source
offaxis angle position to the {\tt xrtmkarf} task using the 
{\tt offaxis} option.

For WT mode, per snapshot ARFs were created assuming point source
corrections can be applied --- the convolved \chan\/ ACIS-S3 image
analysis showed that the $70.7\arcsec$ extraction region contains 95
percent of the SNR encircled energy fraction, which is consistent at
the $1-2$ percent level to that obtained if it is treated as a
point source. However, the latter corrections can become inaccurate
(at the 10 percent level) when the remnant is situated on the CCD
bad-columns (e.g. for the data taken after 2005).  The ancillary
response files were then averaged, weighted by the snapshot exposure
time.

The spectral gain calibration was checked for energy scale
offsets in two ways. First, a gain offset fit was performed in {\tt XSPEC}
when the standard model was applied to the data. In the second
method, the data were reprocessed applying offsets in 1 eV steps
(i.e. a tenth the size of the nominal PI channel width) and the resulting 
spectra were then fit to find the one which minimized the
C-statistic. Both methods gave consistent results and the measured offsets
are reported in Table~\ref{tab:swiftxrt_obs}, which shows the energy scale is
good to better than $11\eV$ for most epochs. The
gain corrected spectra were used in the analysis which follows.

\section{Analysis \& Results}\label{analysis}

\subsection{Time Variability}\label{timevar}

E0102 has an estimated age of 1000-2000~yr based on the expansion
studies of \citet{hughes2000} and \citet{finkelstein2006}. It is 
possible that there might be
discernible changes in the integrated X-ray spectrum of a
SNR this young over a time span of  $\sim15$~yr. In order to place
an upper limit on any changes in the integrated X-ray spectrum, we
examined the total count rate from E0102 with the EPIC-pn instrument
in the 0.3--2.0~keV band.  The EPIC-pn instrument has proven to be the
most stable instrument included in our analysis.  The total count rate
as measured by the EPIC-pn has varied by less than 1.3\% over the 14 years
of measurements. We conclude that whatever changes might be occurring
in E0102, the effect on the integrated X-ray spectrum is small.  

We also examined the \chan\/ images over a 7.5~yr timeframe for
differences.  The images were exposure-corrected to account for the
time-variable absorption of the contamination layer on the ACIS filter
and difference images were created. We limited the time span to 7.5~yr
to reduce the impact of the uncertainty in the correction for
time-variable contamination layer.
We calculated the percentage difference between the two observations 
in narrow bands
around the bright emission line complexes of \ion{O}{vii}~He$\alpha$, 
\ion{O}{viii}~Ly$\alpha$, \ion{Ne}{ix}~He$\alpha$ \& 
\ion{Ne}{X}~Ly~$\alpha$. The largest differences are on the
order of 2\% in a $0.5\arcsec\times0.5\arcsec$ pixel. 
Some parts of the remnant have apparently brightened
while other parts have dimmed.  The total flux change is consistent
with the value measured with the EPIC-pn; however we note that the ACIS-S3
value has a much larger uncertainty given the relatively large
correction that must be applied for the contamination layer. An
analysis of flux changes on arcsecond spatial scales would require a
detailed registration of the \chan\/ images and is beyond the scope of
this paper.  It is possible that some of the changes we observe in the
difference images at the few percent level are due to a less accurate
registration of the images at the two epochs.
For the current analysis, we can conclude that the flux
in a few arcsecond region might be changing by as much as 2\% over a
7.5~yr timespan, but the effect on the integrated spectrum is less
than 1.3\%.

\begin{table*}
\centering

\caption[ ]{Fit Statistics for Data Sets Included in the Comparison}

\label{tab:fit1}
\begin{tabular}{lcccc}
\hline
\hline
Instrument & DOF  & C Statistic &  $\chi^2$ & reduced $\chi^2$  \\
\hline
ACIS-S3        & 227 & 444.9 & 463.3 & 2.04 \\
MOS1        & 332 & 415.1 & 421.6 & 1.27 \\
MOS2        & 332 & 422.9 & 431.6 & 1.30 \\
pn          & 337 & 761.2 & 762.7 & 2.26  \\
XIS0        & 461 & 713.8 & 683.1 & 1.48 \\
XIS1        & 461 & 864.6 & 898.2 & 1.99 \\
XIS2        & 461 & 742.7 & 688.6 & 1.49 \\
XIS3        & 461 & 904.6 & 835.1 & 1.81 \\
XRT WT mode & 106 & 178.2 & 178.0 & 1.68 \\
XRT PC mode & 106 & 140.9 & 143.4 & 1.35 \\
\hline

\end{tabular}

\end{table*}


\subsection{Spectral Fits to the Reference Data}\label{fits}

The details of our spectral fitting methodology are described in 
Sect.~\ref{fitmethod}. The key points are that we fit in the
0.3--2.0~keV energy range, we do not subtract
background, we do not bin our spectra, and we use the so-called ``W
statistic'' in {\tt XSPEC} which is a modified version of the C statistic  as
the fit statistic. The one exception to this is that the $\chi^2$
statistic was used for the fits to the EPIC-pn spectra as described in 
Sect.~\ref{fitmethod}.
The spectra for each of the CCD instruments were first compared to the
IACHEC model without allowing any of the parameters to vary.
It was noticed
that a global offset to account for different size extraction regions
would help to reconcile the overall normalization of the spectra.
The spectra were then fit with 5 free parameters: 
a constant factor multiplying the entire spectrum which acts as a
global normalization, and the \ion{O}{vii}~He$\alpha$~{\em f}, 
\ion{O}{viii}~Ly$\alpha$, \ion{Ne}{ix}~He$\alpha$~{\em r} \& 
\ion{Ne}{X}~Ly$\alpha$ line normalizations.  After the best fit had
been found by minimizing the C statistic, we freeze the parameters
at their best-fit values and compute the $\chi^2$ using the model
value as the weight instead of the data.  We report these   $\chi^2$
values for readers who might be more accustomed to using the
$\chi^2$ statistic.

\begin{figure}[t]
 \begin{center}
  \resizebox{\hsize}{!}{\includegraphics[angle=270]{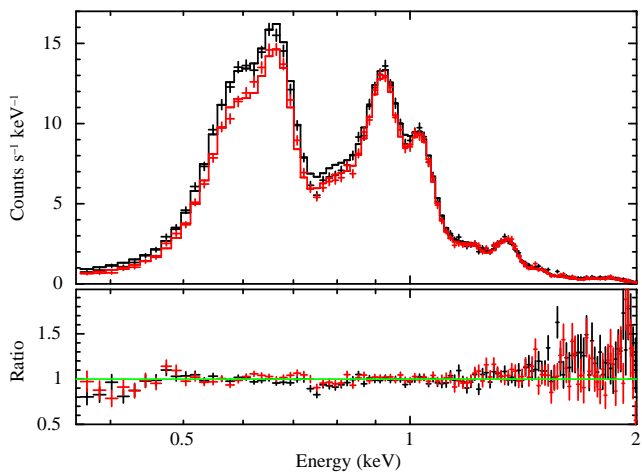}}
 \end{center}
 \caption[ACIS]
 { \label{fig:acis_spec} ACIS-S3 spectra from OBSIDs 3545(black) \&
   6765(red) with the best-fitted model and residuals.
}
 %
\end{figure}

\begin{figure}[hbtp]
  \resizebox{\hsize}{!}{\includegraphics[angle=270.0]{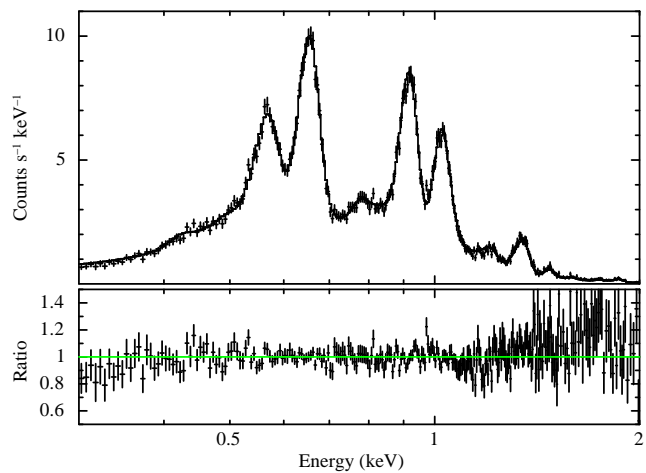}}
\caption{ \label{fig:mos1_spec}
Representative fit to EPIC-MOS data. The spectrum shown is the MOS1 thin filter observation from Orbit 0065. The fit is shown without any gain correction applied.}
\end{figure}

\begin{figure}[hbtp]
 \begin{center}
  \resizebox{\hsize}{!}{\includegraphics[angle=270]{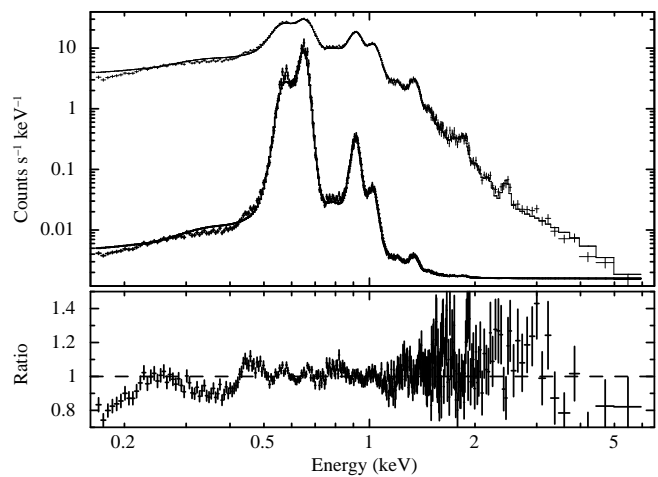}}
 \end{center}
 \caption[pn]
 { \label{fig:pn_spec} EPIC-pn spectrum from OBSID~0412980301. The second (lower)
curve shows the same data but with a linear axis which has been
shifted downwards for clarity. Note the high count rate and the
pattern in the residuals which might indicate an issue with the
spectral redistribution function.
}
 %
\end{figure}

\begin{figure}[hbtp]
 \begin{center}
  \resizebox{\hsize}{!}{\includegraphics[angle=270]{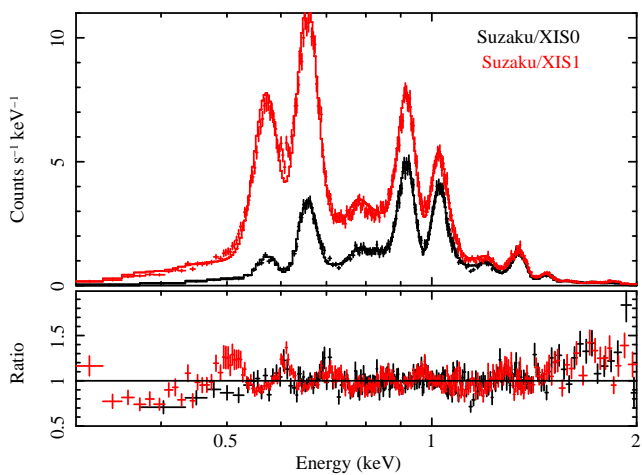}}
 \end{center}
 \caption[xis]
 {\label{fig:xis_spec} {XIS0 (black) and XIS1 (red) spectra from OBSID
 100014010, shown with the best-fit model and residuals.
} }
\end{figure}

\begin{figure}
\resizebox{\hsize}{!}{\includegraphics[width=\textwidth,angle=270]{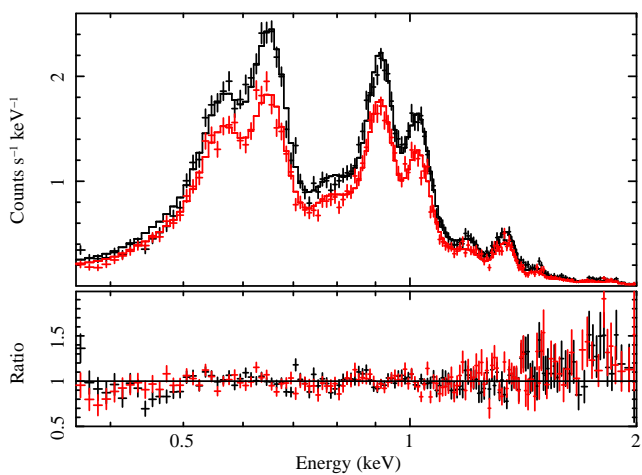}}
\caption{{\em Swift}-XRT grade 0 spectra from 2005 with the best fit
model and data/model ratio. WT is in black (upper spectrum) and PC in
red (lower spectrum).}

\label{fig:swiftxrt2005}
\end{figure}

Table~\ref{tab:fit1} lists the fit statistics for the various CCD
instruments for the representative data sets for each instrument
discussed previously.  None of the
fits are formally acceptable and the quality of the fit around the
bright lines varies.
The spectral fits with the line normalizations free are
displayed in Figures~\ref{fig:acis_spec}, \ref{fig:mos1_spec}, 
\ref{fig:pn_spec}, \ref{fig:xis_spec}, and \ref{fig:swiftxrt2005}
for the ACIS-S3, EPIC-MOS, EPIC-pn, XIS, and XRT data respectively.
The fit to the ACIS-S3 data has a reduced $\chi^2$ of 2.04 and appears to
fit the data reasonably well with the largest residuals below 0.5~keV
and above 1.5~keV.  There also appears to be significantly more flux
in the model around 0.8~keV than in the ACIS-S3 data. This could indicate
a deficiency in the ACIS contamination model in this energy range or a
problem with the weaker lines in the IACHEC model.
Nevertheless, the bright O and Ne lines appear to be
well-fitted and the line normalizations well-determined.  

The MOS1 data have the lowest reduced $\chi^2$ of 1.27
and the data appear to be well-fitted with some small systematics in
the residuals. The MOS2 data are almost equally well-fitted with a reduced 
$\chi^2$ of 1.30.  The EPIC-MOS has the highest spectral resolution of any
of the CCD instruments in this bandpass as shown by the details
visible in Fig.~\ref{fig:mos1_spec}.  The bright lines appear to be
well-fitted in the EPIC-MOS spectra.
 The EPIC-pn data are fitted with a reduced $\chi^2$ of
2.26 with large residuals on the low-energy side of the 
\ion{O}{VII}~triplet. The peak of the \ion{O}{VII}~triplet does not
appear to be well-fitted by the IACHEC model.  We suspect this is an
issue with the spectral redistribution function for the EPIC-pn at these
energies and the effect is under investigation.  The EPIC-pn spectra do
have the largest number of counts, making it somewhat easier to
identify issues with the calibration.
Nevertheless, the \ion{O}{viii} Ly$\alpha$ line, 
the \ion{Ne}{ix}~He$\alpha$~{\em r} line, and the \ion{Ne}{X}~Ly$\alpha$ line 
appear to be well-fitted by the IACHEC model.

The XIS spectral fits are reasonably good ranging from a reduced
$\chi^2$ of 1.48 for XIS0 to 1.99 for XIS1.  The residuals in
Fig.~\ref{fig:xis_spec} around the
O lines appear larger than the residuals around the Ne lines for
both the XIS0 and XIS1. Nevertheless, the \ion{O}{viii} Ly$\alpha$ line, 
the \ion{Ne}{ix}~He$\alpha$~{\em r} line, and the \ion{Ne}{X}~Ly$\alpha$ line 
appear to be well-fitted by the IACHEC model. The XIS spectral fit is 
complicated by the rapidly increasing molecular contamination. Hence
the selection of this early data set when the contamination layer was
still relatively small.

Figure~\ref{fig:swiftxrt2005} shows the 2005 {\em Swift}-XRT reference
data and applied IACHEC model, obtained with the overall constant
factor and line normalisations free to vary in the fit. As expected,
the measured WT line normalisations (shown in
Fig.~\ref{fig:comp_norms}) agree well with the model, as the WT
spectrum from this epoch was used to improve the XRT QE calibration
above the O-K edge at $0.545\keV$.  Prior to the calibration change,
the normalisations showed a strong energy dependence \citep[e.g., see
  Fig. 3 in][]{plucinsky2012}, indicating that the instrumental O-K edge was
too shallow.  The depth of this edge was subsequently increased, by
thickening the silicon dioxide layer in the model of the CCD electrode
structure, in order to bring the measured normalisations into better
agreement with the IACHEC model.  Also, the CCD spectral resolution
had been slightly underestimated in the RMFs used in the earlier E0102
analysis \citep[see the residuals in Fig. 8 in][]{plucinsky2012} and this
was improved during the XRT response recalibration.

The PC mode line normalisations are lower that the WT ones by $5-10$
percent.  We suspect the origin of this is pile-up, as an image of the
PC mode data formed of diagonal events (i.e. events with grades
$26-29$) clearly show the SNR and these events are indicative of
pile-up. We have also performed a detailed simulation of the data
using the XRT event simulator described by Beardmore et al. (in
prep.).  The simulations, which used the \chan\/ ACIS-S3 image to define
the spatial distribution and the IACHEC model to provide the spectral
distribution, confirmed that the PC mode line normalisations are
suppressed by between 5 and 10 percent, whereas WT mode is unaffected.


\subsection{Comparison of the Fit Results }\label{fitresults}

 The fitted line normalizations for the \ion{O}{vii}~He$\alpha$~{\em
   r} line, the
\ion{O}{viii}~Ly$\alpha$ line, the \ion{Ne}{ix}~He$\alpha$~{\em r} line, \& 
\ion{Ne}{X}~Ly$\alpha$ line are listed in Table~\ref{tab:fit2}.
We report the normalizations for the \ion{O}{vii}~He$\alpha$~{\em
   r} line but the normalization of the \ion{O}{vii}~He$\alpha$~{\em
   f} line was the free parameter in the fit. As described in 
Sect.~\ref{fitmethod}, the \ion{O}{vii}~He$\alpha$~{\em
   f} and {\em r} line normalizations are linked together so there is
 a constant factor relating the two values.
The first row lists the normalization in the IACHEC model to
facilitate comparison.  The results for the spectral fits to the
reference data for each instrument are 
presented in groups of three rows.  Within a group of three rows for a
given instrument,
the first row gives the best-fitted value, the second row gives the 
$1.0\sigma$ lower and upper confidence limits, and the third row 
gives the ``scaled'' value where the
best-fitted value has been multiplied by the constant factor.
The second and third groups of rows include the results for the RGS1 and RGS2.
Note that the best-fitted values and the scaled values are the same
  for the RGS1 since the scale factor is 1.0. This is due to the fact
  that the RGS data were a primary input in the development of the
  IACHEC model for E0102. The best-fitted values
  and the scaled values are different for the RGS2 since the scale
  factor is 0.96. The different scale factors for RGS1 and RGS2
  indicate that there is a systematic 4\% offset
between RGS1 and RGS2 effective areas.

\begin{table*}
\centering

\caption[ ]{Fitted Values for Constant Factor and Line Complex Normalizations}

\label{tab:fit2}
\begin{tabular}{lccccc}
\hline
\hline
Instrument & Constant & \ion{O}{vii}~He$\alpha$~r & \ion{O}{viii}~Ly$\alpha$ & \ion{Ne}{ix}~He$\alpha$~r & \ion{Ne}{x}~Ly$\alpha$  \\
           &          & Norm  & Norm & Norm & Norm \\
           &          & ($10^{-3} {\rm ph~cm^{-2} s^{-1}}$) 
& ($10^{-3} {\rm ph~cm^{-2} s^{-1}}$) 
& ($10^{-3} {\rm ph~cm^{-2} s^{-1}}$)
& ($10^{-3} {\rm ph~cm^{-2} s^{-1}}$) \\
\hline
IACHEC     &  &    &    &         & \\
model     &  & 2.745   &  4.393  &   1.381      & 1.378\\
value     &  &    &    &         & \\
\hline

RGS1     & 1.0 & 2.700   &  4.404  &   1.400      & 1.415\\
1$\sigma$ CL &   & [2.652,2.746] & [4.341,4.466] & [1.380,1.420] & [1.378,1.452] \\
Scaled  &       & 2.700  &  4.404   &  1.400      & 1.415 \\
\hline

RGS2     & 0.96 &  no data  & 4.445       &   1.371      & 1.409\\
1$\sigma$ CL &   & [----] & [4.330,4.561] & [1.331,1.410] & [1.350,1.468]\\
Scaled  &       & ----  &   4.267            &  1.316      & 1.353 \\
\hline

MOS1     & 1.054 & 2.914   &      4.567         & 1.383        & 1.392 \\
1$\sigma$ CL &   & [2.872,2.955] & [4.514,4.620] & [1.366,1.400] & [1.373,1.411] \\
Scaled  &       & 3.071   &      4.814         & 1.458       & 1.467 \\
\hline

MOS2     & 1.072 & 2.943   &      4.581         & 1.380        & 1.401 \\
1$\sigma$ CL &   & [2.899,2.987] & [4.527,4.635] & [1.363,1.397] & [1.382,1.420] \\
Scaled  &       & 3.155   &      4.911         & 1.479       & 1.502 \\
\hline

pn     & 0.969 & 2.786  &       4.213        &  1.368       & 1.286 \\
1$\sigma$ CL &   & [2.778,2.794] & [4.200,4.226] & [1.364,1.372] & [1.281,1.291] \\
Scaled  &       & 2.700   & 4.082   &   1.326   & 1.246 \\
\hline

ACIS-S3    & 1.072 & 2.230   &      3.803         & 1.311        & 1.375 \\
1$\sigma$ CL &   & [2.176,2.285] & [3.728,3.879] & [1.283,1.340] & [1.342,1.410] \\
Scaled  &       &  2.391  &    4.077   &   1.405   &  1.474  \\  
\hline

HETG    & 1.037 & 2.736   &  4.656    & 1.281    &  1.325 \\
1$\sigma$ CL &   & [2.696,2.778] & [4.604,4.707] & [1.272,1.290] & [1.313,1.337] \\
Scaled  &       &  2.837  & 4.827      &  1.328    &  1.374  \\  
\hline

XIS0         & 1.034  & 2.468         &  4.128        & 1.335         &  1.375  \\
1$\sigma$ CL &        & [2.385,2.556] & [4.044,4.213] & [1.311,1.358] & [1.349,1.402] \\
Scaled       &        & 2.552         &  4.268        &  1.380        & 1.422  \\  
\hline

XIS1         & 1.141  & 2.475         &  4.030        & 1.316         &  1.349  \\
1$\sigma$ CL &        & [2.439,2.512] & [3.981,4.080] & [1.298,1.335] & [1.328,1.371] \\
Scaled       &        & 2.824         &   4.598       &  1.502        & 1.539    \\  
\hline

XIS2         & 1.029  &  2.596        &  4.001        & 1.323         &  1.379  \\
1$\sigma$ CL &        & [2.512,2.679] & [3.920,4.082] & [1.300,1.347] & [1.352,1.405] \\
Scaled       &        & 2.671         &  4.117        & 1.361         & 1.419   \\  
\hline

XIS3         & 1.017  & 2.134         &  3.659        & 1.324         &  1.361  \\
1$\sigma$ CL &        & [2.057,2.213] & [3.581,3.738] & [1.300,1.348] & [1.334,1.388] \\
Scaled       &        & 2.170         &  3.721        &  1.347        &   1.384  \\  
\hline

XRT-PC    & 0.935  &  2.700   & 3.972   & 1.284   &  1.303 \\
1$\sigma$ CL & & [2.606,2.792] & [3.859,4.087] & [1.247,1.323] & [1.260,1.346] \\
Scaled   &    & 2.523     & 3.714    & 1.201      & 1.218   \\
\hline

XRT-WT    & 0.971   & 2.705    & 4.501   & 1.346   & 1.386   \\
1$\sigma$ CL & & [2.606,2.792] & [4.390,4.615] & [1.311,1.382] & [1.346,1.427] \\
Scaled   &    & 2.626   & 4.371    &  1.307     & 1.346   \\
\hline

\end{tabular}

\end{table*}

\begin{figure*}
 \begin{center}
 \includegraphics[width=11.0cm,angle=90]{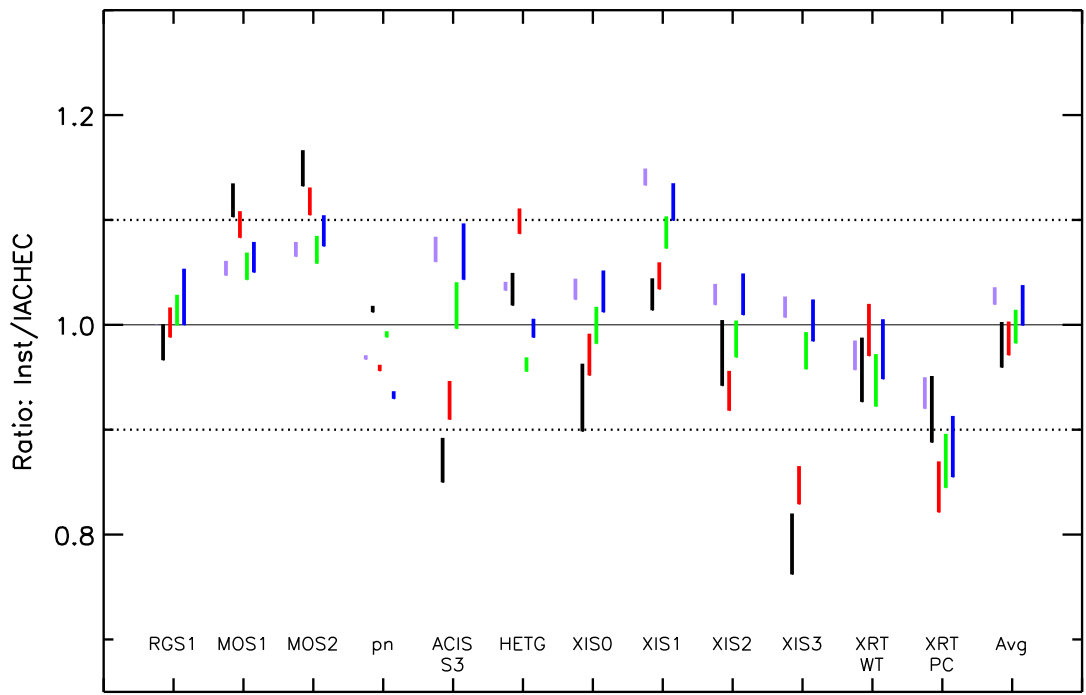}
 \end{center}
 \caption{\label{fig:comp_norms} {Comparison of the scaled
 normalizations for each instrument to the IACHEC model values and the average.
 There are four or five  points for each
instrument which are from left to right, global normalization (purple),
\ion{O}{vii}~He$\alpha$~{\em r} (black), \ion{O}{VIII}~Ly$\alpha$ (red), \ion{Ne}{ix}~He$\alpha$~{\em r} (green), \& \ion{Ne}{X}~Ly$\alpha$ (blue). The
length of the line
indicates the $1.0\sigma$ CL for the scaled normalization.
 } }
\end{figure*}

Figure~\ref{fig:comp_norms} presents the data in Table~\ref{tab:fit2}
in a graphical manner. The scaled normalizations for each instrument
are compared to the IACHEC values and the average of all
instruments is also plotted.  There are several interesting trends
visible in this plot.  The MOS1 and MOS2 data appear to be 5-15\%
higher than the IACHEC model, with the O line normalizations being
10-15\% higher.  The EPIC-pn values have the smallest uncertainties since
they are derived from a joint fit of all of the data, whereas the
other instruments used a small number of observations. The EPIC-pn data
agree with the IACHEC model values to better than 10\% with the 
\ion{Ne}{ix}~He$\alpha$~{\em r} being the most discrepant at 7\%.  The
ACIS-S3 data are
most discrepant at the O lines but agree better at the Ne lines. One
explanation for the O line normalizations being lower than the IACHEC
values could be that there is still some residual pileup in the ACIS-S3
data suppressing the line fluxes at low energies and enhancing the
line fluxes at higher energies. Another possible explanation is that
the contamination model is under-predicting the absorption at the
energies of the O lines at the time of these observations. As shown
later in Sect.~\ref{acistrend}, the fluxes of the O lines are in
better agreement with the IACHEC model for later observations.
The XIS
results agree to within 10\% of the IACHEC model values with the
exception of the \ion{O}{vii}~He$\alpha$~{\em r} line which can be discrepant by as
much as 20\% for XIS3.  The XRT WT mode agree to within 5\% of the IACHEC model
whereas the XRT PC mode data can be more than 10\% discrepant. 
As mentioned above,  the
XRT PC mode data most likely suffer from pileup which reduces the
observed count rate.

Most of the scaled normalizations 
agree with the IACHEC model values with $\pm10\%$.  Specifically, 38 of the 48
normalizations are within  $\pm10\%$ of the IAHCEC model values.
The scaled normalizations agree to
within 10\%  for the  \ion{Ne}{ix}~He$\alpha$~{\em r} and 
\ion{Ne}{X}~Ly$\alpha$ lines with the exception of the XRT PC mode
data (which are affected by pileup) and the XIS1 for the
\ion{Ne}{X}~Ly$\alpha$ line.
The agreement is significantly worse for the O lines, $\pm15\%$ for the
\ion{O}{viii}~Ly$\alpha$ line and $\pm25\%$ for the 
\ion{O}{vii}~He$\alpha$~{\em r}
line.  The fact that the lowest energy line produces the worst
agreement is indicative of the difficulty in calibrating the
spectral redistribution function and the
time-dependent response of CCDs instruments at an energy as low at
$\sim570$~eV.  The good agreement at higher energies for these different
instruments is indicative of the quality of the calibration provided
by the various instrument teams.  Therefore, we conclude that the
absolute effective areas of the
combined systems of mirrors plus detectors agrees to within $\pm12\%$
at 0.9 and 1.0~keV for \chan~ACIS-S3, \xmmn~EPIC-MOS, \xmmn~EPIC-pn, \suzaku~XIS0,
XIS2, \& XIS3, and \swift~XRT for these representative E0102 spectra.




\subsection{Time Dependence of the Response}\label{timedep}

Each of the CCD instruments included in this study has a significant
time dependence in its response except for the EPIC-pn. As stated in 
Sect.~\ref{timevar}, the EPIC-pn has recorded a constant count rate
from E0102 in the 0.3--2.0~keV band to better than 1.3\% over the
course of the mission.  All of the other instruments have had a
variable response for one of several reasons.  Some of the instruments
have developed a contamination layer that produce an additional,
time-variable absorption. Some of the instruments have had significant
changes in their CTI due to radiation damage that produces a
time-variable spectral response.  The EPIC-MOS CCDs have experienced a
change in response that appears to be related to the X-ray photon does
near the aimpoint of the telescope.
In the following sections we discuss
the time-variable response of each instrument individually.

\subsubsection{\xmmn\/ EPIC-MOS}\label{epicmostrend}

\begin{figure*}
\includegraphics[width=6.4cm,angle=90.0]{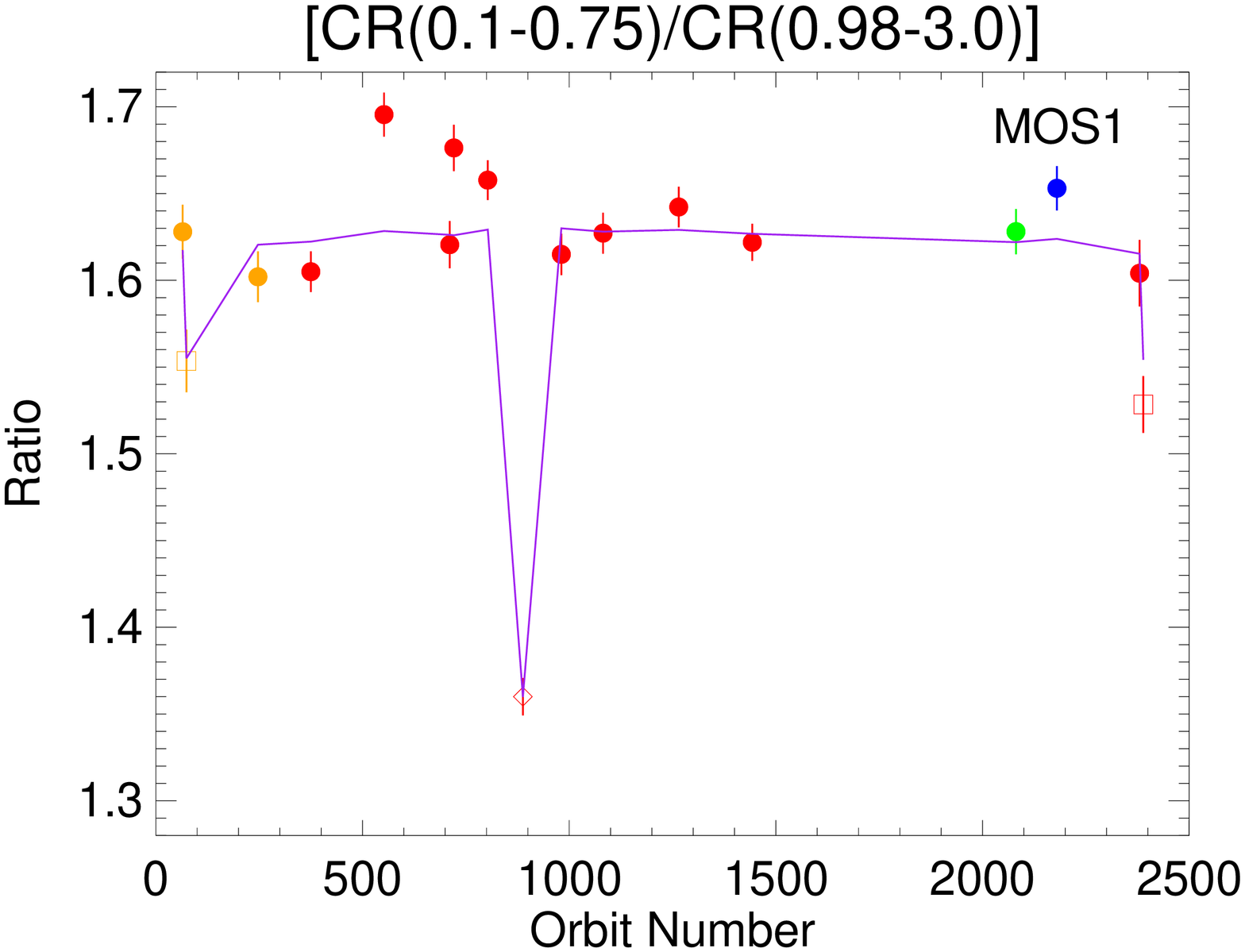}
\includegraphics[width=6.4cm,angle=90.0]{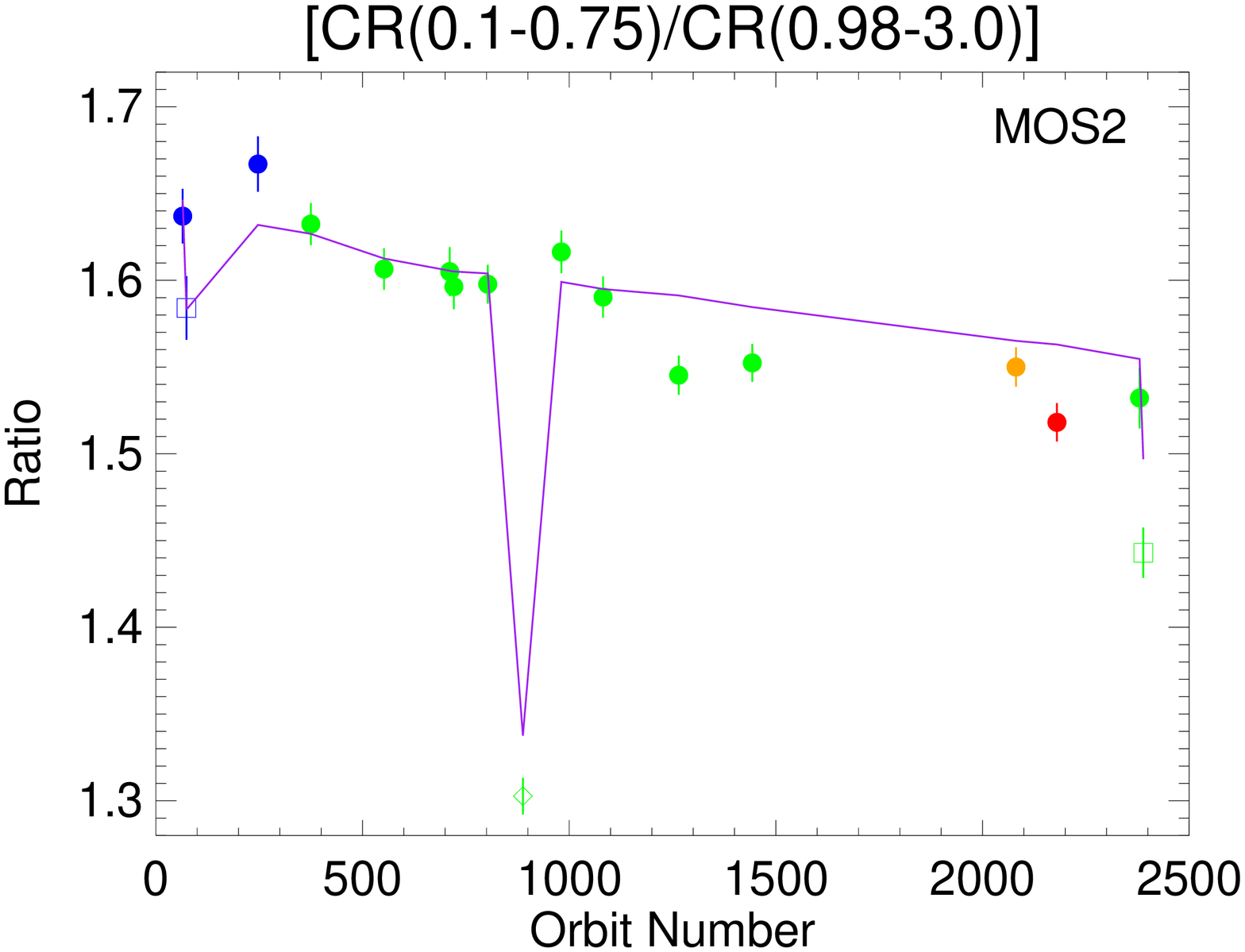}
\caption{Observed (0.1-0.75)/(0.98-3.0) keV count rate ratio as a
  function of the Orbit number. Left panel is MOS1, right panel is
  MOS2. Data points are color-coded to reflect the position of the
  source on the detector relative to the boresight as viewed on the
  image plane in detector coordinates: blue (above boresight), orange
  (to the right), green (below) and red (to the left.). All
  observations are thin filter with the exception of two medium filter
  (Orbit 0065 and 2380, squares) and one thick filter (Orbit 0888,
  diamond). The solid
  line on each plot is the predicted ratio derived by folding the
  standard spectral model through the EPIC-MOS response modified by a
  contamination model as described in the text.}
\label{figure:mos-ratios}
\end{figure*}

A subset of the available EPIC-MOS observations of E0102 (Table~\ref{tab:mosobs}) was
used to investigate potential trends in the effective area calibration. To minimize 
variance due to calibration uncertainties all observations with the same observing
mode (LW mode) were selected and, with the exception of the first observation, were
all positioned around 1 arcminute or so from the telescope
boresight. The bulk of the observations are therefore unaffected by
the redistribution patch (see Sect.~\ref{epicmos}). The last two
observations were positioned such that the edge of the remnant just
crosses the spatial area of the patch as defined within the
calibration. In addition to the analysis procedure outlined in
Sect.~\ref{epicmos} we were careful to individually examine the
spectrum from every column on the CCDs passing through the
remnant. Large traps which cause sections of a given column to shift
in energy by 10's of eV are generally detected by the calibration
software and a spatial correction applied to realign the reconstructed
event energy to the correct scale. The frequency of these traps is
increasing throughout the mission, however, and it is apparent that
the current event energy calibration as used in this paper does not
properly calibrate a few columns which pass through the remnant
in the later observations. These columns were individually removed
from the analysis and the global normalisation in the fit has been
corrected for the loss of flux due to their exclusion by the same
method used for bad pixels and columns excluded by the filtering
process as discussed in Sect.~\ref{epicmos}.

The data were initially processed with SAS12.0.0 and on examining the data 
sample it was noticed that there was a significant trend in the spectral hardness 
derived from an examination of the background subtracted source count rates. In
Fig.~\ref{figure:mos-ratios} we show the spectral hardness ratio
defined as the source count rate in the 0.1 to 0.75~keV band divided
by the rate in the 0.98 to 3.0~keV band. The actual energy boundaries
used to derive this ratio has been adjusted for each observation by
the gain fit parameters to ensure that the same portion of the spectrum is
used in each in case.  The data points have been color coded to
reflect the source position relative to the boresight as the ratio is
influenced by the mirror vignetting function. The ratio is also
obviously dependent on the filter used in the observation. Most of the
observations use the thin filter with the exception of two in medium
filter (Orbit 0065 and 2380) and one observation in thick filter
(Orbit 0888). Examining observations with the same filter there is a
clear trend in the MOS2 data which can only plausibly be explained by
the existence of a thin but growing contaminant.

\begin{figure}
\begin{center}
\includegraphics[width=8.0cm,angle=0.0]{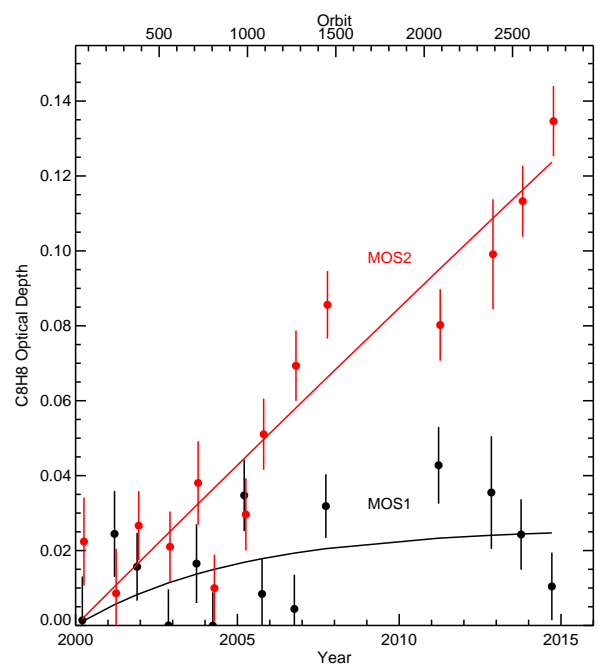}
\end{center}
\caption{EPIC-MOS contamination model as optical depth of C8H8 as a
  function of time. Data points are derived by finding the depth
  of C8H8 which minimises the best-fit parameter when fitting the
  standard E0102 model to the observed EPIC-MOS spectra. The solid curves
  represent best-fit exponential models. MOS1 is equally consistent 
with a fixed value of the contaminant (see text).}
\label{figure:mos-contamination}
\end{figure}

Also show on Fig.~\ref{figure:mos-ratios} is the predicted count
ratio derived by folding the model E0102 spectrum through the
instrument response modified by a contaminant model as described
below. The model at this time consists of pure Carbon and no other
compounds.
The RGS detectors also suffer from a gradual loss of effective area which is presumed to be
due to a carbon or carbon plus hydrogen contaminant (no edges are detected within the RGS
energy passband). It is assumed the contaminant on the RGS detectors arises
from outgassing of the telescope tube structure. 

As EPIC shares this structure with the RGS it is plausible that the
EPIC-MOS contaminant arises from the same source, however, the complicated
nature of the EPIC-MOS RMF and uncertainties in its calibration make an
unambiguous identification of low energy edges, even at Oxygen, very
difficult. The pure Carbon model currently adopted for the EPIC-MOS is the
simplest contaminant model which reasonably fits the data.

The contaminant model was derived directly from the E0102 data by
finding the contaminant depth which gave the best reproduction of the
observed hardness ratio (derived by folding the E0102 standard model
through a response modified by the contaminant absorption) for each
observation and fitting an exponential time-dependent model to the
depth parameters. The models for MOS1 and MOS2 are shown in
Fig.~\ref{figure:mos-contamination}. Our methodology was to
initially use the thin filter data only (which comprises the bulk of
the observations) then compare the contamination model with
observations taken with the medium and thick filters.

\begin{figure*}
\begin{center}
\includegraphics[width=11.0cm,angle=90.0]{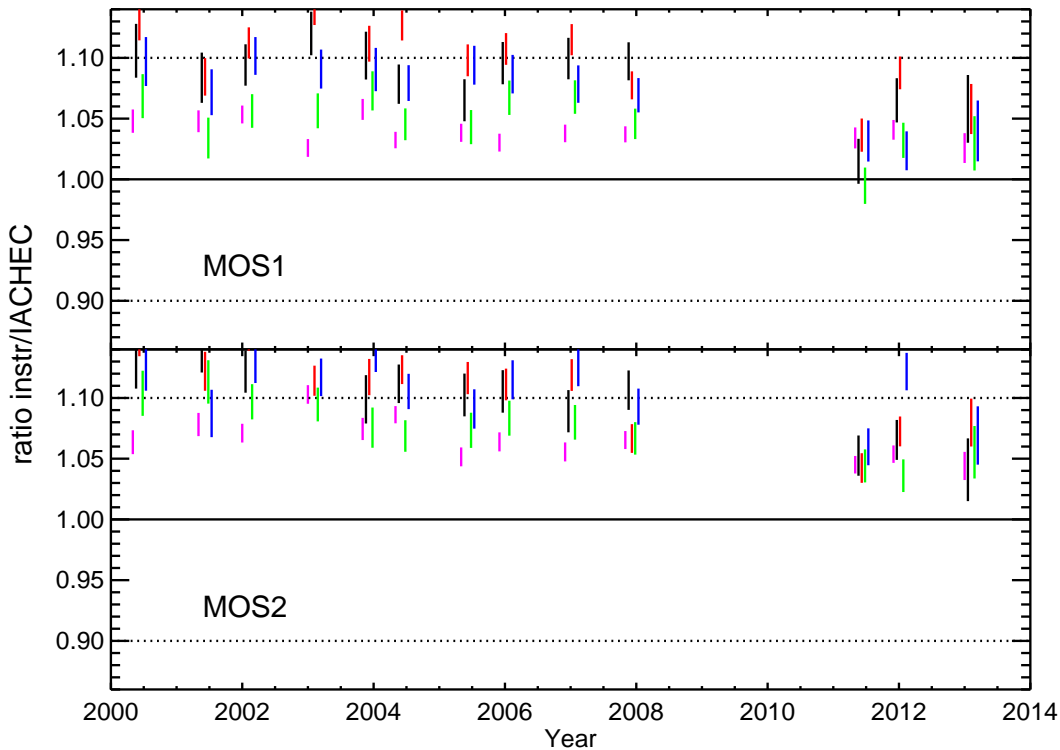}
\end{center}
\caption{ Ratio of the normalizations of the 
\ion{O}{vii}~He$\alpha$~{\em r} line (black), \ion{O}{viii}~Ly$\alpha$ line
(red), \ion{Ne}{IX}~He$\alpha$~{\em r} line (green),  \ion{Ne}{X}~Ly$\alpha$
line (blue), and the global normalization (magenta) 
compared to the IACHEC model value as a function of time for MOS1 and
MOS2 for a sample of observations from Table~\ref{tab:mosobs}. Some 
observations taken close in time to others are
  not shown for clarity. 
}
\label{figure:mos-parameters}
\end{figure*}

The evidence for a contaminant on MOS2 is significant but very
marginal for MOS1. There is also evidence for a discontinuity in the
evolution of the contamination from around orbit 1200
onwards. Further monitoring of E0102 and analysis of additional
suitable calibration targets may lead to a refinement of the
contamination model at a later date.

It should be noted that the predicted contaminant
depth on MOS2 (around $0.04 \ \mu$m) for the latest observation in the sample 
was $\sim 20$\% of that on the RGS at that epoch. The contaminant is most likely on the CCD detector plane
being the coldest surface (currently $-120^{\circ}$) on the
instrument. The filter wheel assembly (at $\sim -20^{\circ}$) provides
a warm barrier but has gaps through which molecules from the telescope
tube could reach the CCDs. There is no evidence as yet for a spatial
dependence of the contaminant at least within the central few
arcminutes. The contaminant is not expected to be on the filters and
in fact, as shown in Fig.~\ref{figure:mos-ratios} the model provides
an equally good representation of the medium and thick filter
ratios. In addition, dedicated calibration observations taken with the
thin, medium and thick filter {\it at the same epoch} show no evidence
for any change in the {\it relative} transmission of the filters since
launch. The contamination model is therefore currently applied as a
time-dependent adjustment to the overall EPIC-MOS detector
efficiency. Finally, there is no evidence for contamination on the
EPIC-pn, however, unlike the EPIC-MOS, the EPIC-pn has a cold finger trap above
the detector plane. As the EPIC-MOS contaminant layer is currently
relatively thin there are no immediate plans to bake-off the
contaminant by heating the focal plane.

The model has been shown to improve the cross-calibration between EPIC-MOS
and EPIC-pn at low energies for later epoch continuum sources and was implemented 
in the EPIC calibration from SAS version 13.5.0 onwards. Formally the RMF solution 
depends on the assumed effective area for any given standard calibration target and a new set
of self-consistent RMFs were released in conjunction with the
contamination model.

We have analyzed the observations listed in Table~\ref{tab:mosobs} with
SAS version
13.5.0 and derived the five parameters from the standard IACHEC model
in each case. These are shown graphically in
Fig.~\ref{figure:mos-parameters} although only regular spaced
observations are shown for clarity. Significant global trends in the
data are now largely absent. There are some common features of the
fitted parameters. Most noticeable of course is that both EPIC-MOS cameras
return a higher predicted flux than the IACHEC standard model by about
10\% (0.3 to 2.0~keV) although the global normalisation parameter is
relatively higher in MOS2 than in MOS1 by around $\sim 3$\%. In both
cameras the \ion{Ne}{IX}~He$\alpha$~{\em r} normalisation typically has the 
lowest relative value compared with the IACHEC model.

\subsubsection{\chan\/ ACIS}\label{acistrend}

\begin{figure*}
\begin{center}
\includegraphics[width=16.0cm,angle=0]{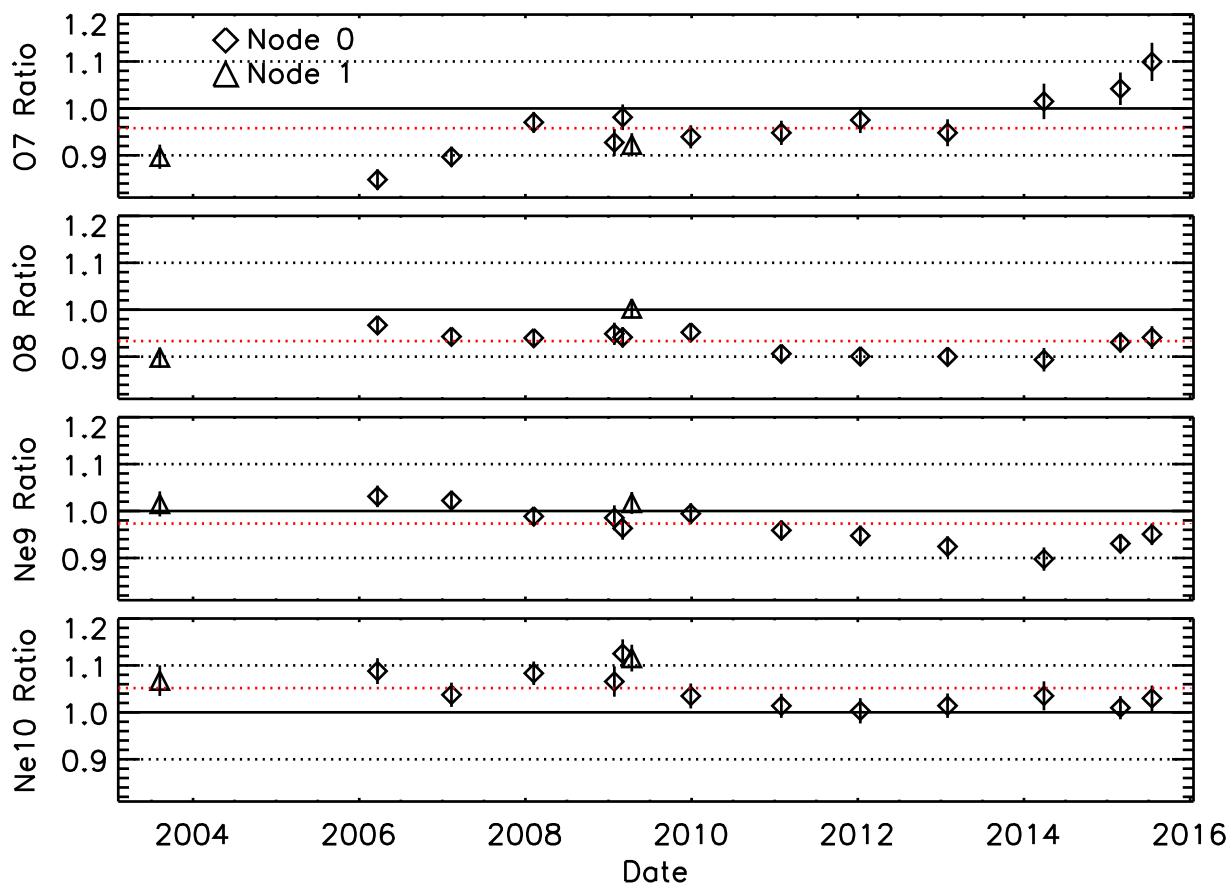}
\end{center}
 \caption{\label{fig:acis_line_norms} {Ratio of the
     normalizations of the
\ion{O}{vii}~He$\alpha$~{\em r}, \ion{O}{viii}~Ly$\alpha$ line,
\ion{Ne}{IX}~He$\alpha$~{\em r}, and 
\ion{Ne}{X}~Ly$\alpha$ line compared to the
IACHEC model value as a function of time for 
ACIS-S3 subarray observations near the aimpoint with the
contamination model N0009 in CALDB 4.6.2. The different symbols
indicate data collected in different nodes on the CCD. The solid,
black line is the IACHEC value (normalized to 1.0), the dashed,
red line is the average for each line normalization, and the black,
dashed lines are $\pm10\%$ of the IACHEC value.}}
\end{figure*}

  The time dependence of the low energy response of ACIS-S3 is determined
primarily by the contamination layer on the filter in front of the
CCDs.  There are time dependent changes in the S3 CCD response to X-rays,
but this is a much smaller effect than the contamination.  The model
for the absorption due to the ACIS contamination layer has a time dependence, a spatial
dependence, and a spectral dependence.  The model has been revised
four times over the course of the mission, most recently in July
2014 with the release of CALDB 4.6.2. The contamination model is
developed based on data from the external calibration source and
celestial calibration targets other than E0102.  The E0102
observations are then used to validate the contamination model as an
independent check.  The normalizations for the bright line complexes
of the \ion{O}{vii}~He$\alpha$~{\em r},
the \ion{O}{viii}~Ly$\alpha$ line, the \ion{Ne}{ix}~He$\alpha$~{\em r}, and
the \ion{Ne}{X}~Ly$\alpha$ are determined using the IACHEC model as
described above.  If the contamination model is accurate, the line
fluxes should be constant in time within the uncertainties. In the
past, clear downward trends in the E0102 line fluxes with time have
prompted revisions of the contamination model.

 We have determined the lines fluxes for all of the E0102 observations
 listed in Table~\ref{tab:acisobs} using the latest version of the
ACIS contamination model (N0009) released in CALDB 4.6.2 on 9
July 2014 (we used CALDB 4.6.8 for this analysis but the contamination
model has not changed since the
CALDB 4.6.2 release).  These observations are on the 
S3 CCD in subarray mode (to mitigate pileup), near the aimpoint which
is in the middle of the CCD. The line fluxes versus time are displayed in
Fig.~\ref{fig:acis_line_norms} for the \ion{O}{vii}~He$\alpha$~{\em r},
the \ion{O}{viii}~Ly$\alpha$ line, 
the \ion{Ne}{ix}~He$\alpha$~{\em r}, and
the \ion{Ne}{X}~Ly$\alpha$ line. 
The line fluxes for the \ion{O}{vii}~He$\alpha$~{\em r} are lower than the
average from 2003-2007, are rather consistent with the average from
2007 until 2014, and then appear to be increasing after 2015. 
The average \ion{O}{vii}~He$\alpha$~{\em r} normalization is lower than the IACHEC 
value by $\sim4\%$. Recall
that the first two observations were used for the comparison in
Fig.~\ref{fig:comp_norms} to the line normalizations for the other 
instruments.  If later observations were used, the 
\ion{O}{vii}~He$\alpha$~{\em r} would be higher by $\sim10\%$ and in better agreement 
with the IACHEC
model.  The  \ion{O}{viii}~Ly$\alpha$ normalization does not show a
clear trend in time. The values are mostly consistent with the average
although there appears be scatter on the order of $\pm5\%$ which is
larger than statistical uncertainties. 
The average \ion{O}{viii}~Ly$\alpha$  normalization is lower than the 
IACHEC value by $\sim7\%$.
The \ion{Ne}{ix}~He$\alpha$~{\em r} normalization shows a gradual decrease in
time after 2010 but the effect is only significant at the $2\sigma$
level. 
The average \ion{Ne}{ix}~He$\alpha$~{\em r} normalization is lower than the IACHEC 
value by only $\sim2\%$.
Finally, the \ion{Ne}{X}~Ly~$\alpha$ line appears mostly
constant in time within the uncertainties.  
The average \ion{Ne}{X}~Ly~$\alpha$ normalization is higher than the IACHEC 
value by $\sim5\%$.
It should be noted that
there are a handful of E0102 observations towards the bottom and top
of S3 that show significantly lower line normalizations than the
measurements presented here in the middle of S3.  The contamination
layer is thicker towards the bottom and top of the S3 CCD and it
appears that the current contamination model is under-predicting this
gradient. This effect is under investigation and may be addressed in a
future release of the contamination file.

  The E0102 results on S3 indicate that the current contamination
model is returning fluxes constant to within $\pm5\%$
for the \ion{O}{viii}~Ly$\alpha$ line, the
\ion{Ne}{ix}~He$\alpha$~{\em r} line,
and the \ion{Ne}{X}~Ly$\alpha$ line and to within  $\pm10\%$
for the \ion{O}{vii}~He$\alpha$~{\em r} line.  These results hold
for observations near the middle of S3. The fluxes are constant 
to this level  from the beginning of the 
\chan\/ mission until 2016, with the
exception perhaps of the \ion{O}{vii}~He$\alpha$~{\em r} which
exhibits a possible trend with time.  The effective area
at the energy of the \ion{O}{viii}~Ly~$\alpha$ (654~eV)line has
decreased by about 80\% since the beginning of the mission.
Therefore, the contamination model must make a large correction as a
function of time for these line fluxes. E0102 will continue to be used
to monitor the accuracy of the ACIS contamination model in the future.

\begin{figure}[hbtp]
 \begin{center}
  \resizebox{\hsize}{!}{\includegraphics[angle=0]{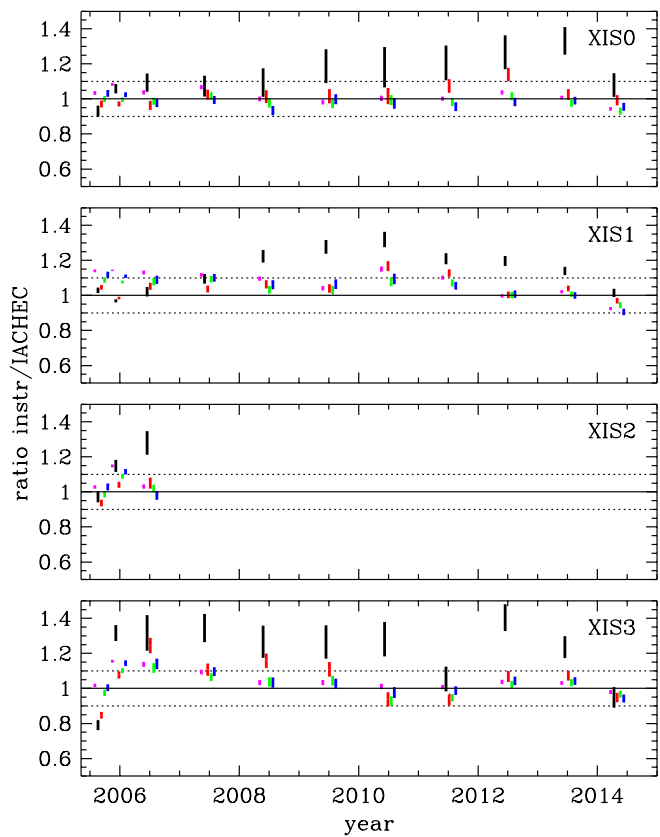}}
 \end{center}
 \caption{\label{fig:xis_time} {Ratio of the normalization of the
\ion{O}{vii}~He$\alpha$~{\em r} (black), \ion{O}{viii}~Ly$\alpha$ line
(red), \ion{Ne}{ix}~He$\alpha$~{\em r} (green),  \ion{Ne}{X}~Ly$\alpha$ line
(blue) and the global normalization (purple) compared to the
IACHEC model value as a function of time for the XIS0, XIS1, XIS2, \&
XIS3 detectors.}}

\end{figure}

\subsubsection{Suzaku XIS}

The line normalizations for the \suzaku\/ XIS depend strongly on the
amount of molecular contamination, which began building up on the
optical blocking filters shortly after launch.  E0102 is one of three
primary calibration sources for measuring the on-axis contamination
build-up and chemical composition, and since the IACHEC model
presented in this work is used in that analysis, there is an inherent
conflict in drawing conclusive comparisons between XIS and the other
instruments.  Nevertheless, the quality of the contamination model can
be explored here.

As can be seen in Fig.~\ref{fig:xis_time}, except for very early in 
the mission,  the \ion{O}{vii}~He$\alpha$~{\em r} line normalization 
for each XIS is 20--40\% higher than the
IACHEC model.  The other line normalizations and overall normalization
are generally within 10\% of the IACHEC values, and show no strong
trend except early in the mission, when the normalization increases
with line energy (from \ion{O}{vii} to \ion{Ne}{X}).  This broad-band 
(0.5--1~keV)
energy dependence at early times possibly indicates that the column
density or chemical composition of the contamination is not correct,
as the increasing line normalization with energy mimics the effects of
an underestimated oxygen absorption edge.  At later times, only \ion{O}{VII}
is discrepant, which could result from an incorrectly modeled
low-energy redistribution tail becoming apparent as the X-ray
sensitivity decreases.

\begin{figure}[hbtp]
\begin{center}
\resizebox{\hsize}{!}{\includegraphics{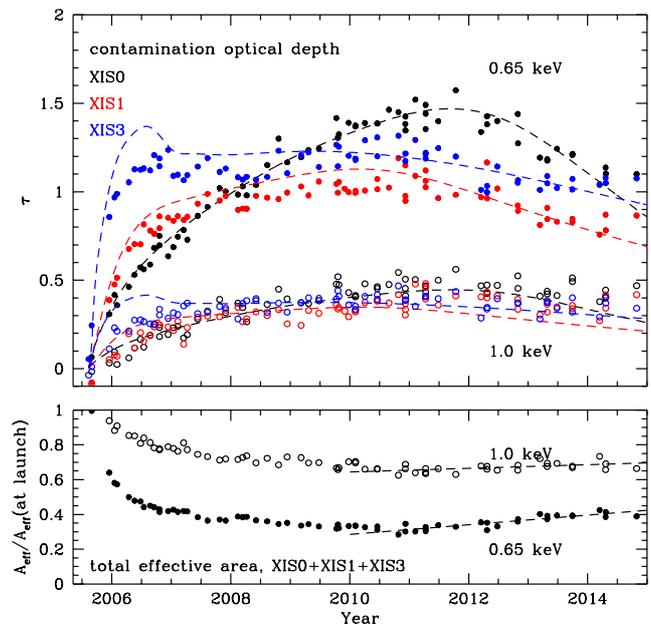}}
\end{center}
\caption{ \label{fig:contam_xis}  Contamination history of the Suzaku
XIS.  (top) The optical depth at two energies as inferred from E0102
observations (points) compared to v20140825 of the Suzaku
contamination CALDB trend (dashed lines).  E0102 is one of the
calibration sources used to measure the on-axis contamination,
although differences are seen between the inferred values and the
CALDB, especially early in the mission.  (bottom) Relative combined
effective area of the three working CCDs at two energies as measured
from E0102.  This assumes no contamination at the opening of the XIS
doors in July 2005.  The dashed lines are from a linear fit to the
data after 2010, and indicate a decrease in the contaminant at later
times.}
\end{figure}

\begin{figure*}
\resizebox{\hsize}{!}  
{\includegraphics[width=11.0cm,angle=0]{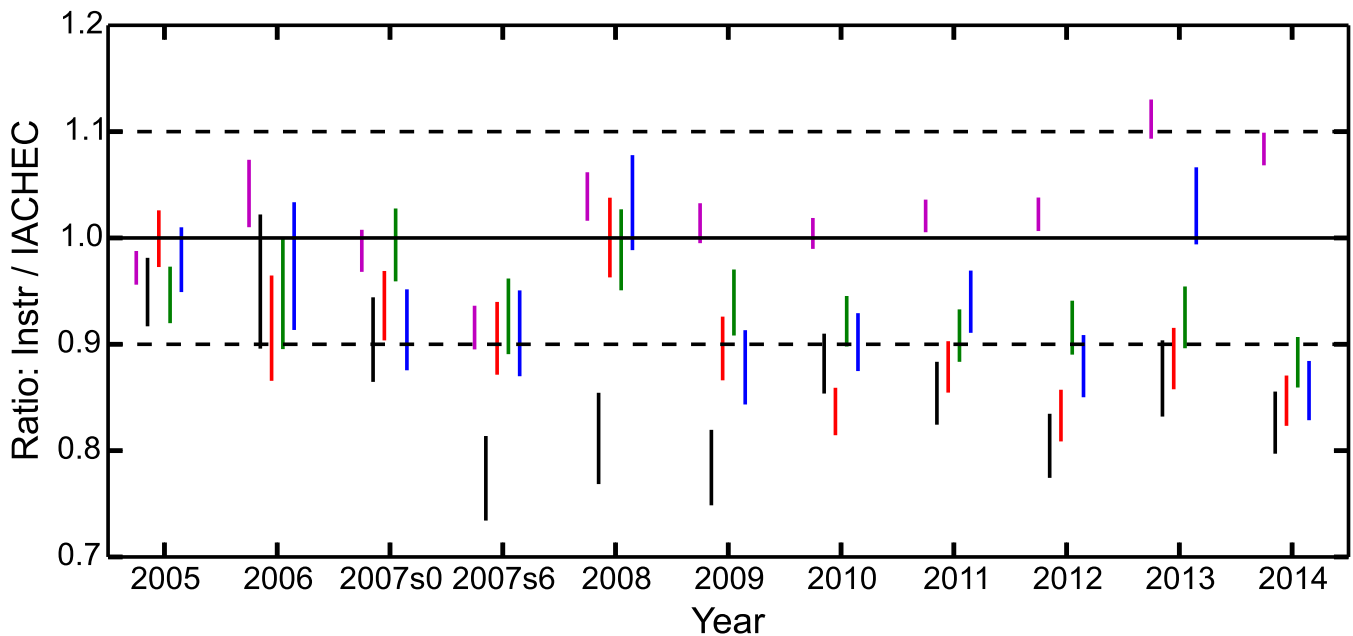}}
\caption{Ratio of the normalization of the
\ion{O}{vii}~He$\alpha$~{\em r} (black), \ion{O}{viii}~Ly$\alpha$ line
(red), \ion{Ne}{ix}~He$\alpha$~{\em r} (green),  \ion{Ne}{X}~Ly$\alpha$ line
(blue) and the global normalization (purple) compared to the
IACHEC model value as a function of time for the XRT in WT mode.}
\label{fig:swiftxrtwtepochs}
\end{figure*}

The quality of the calibration can be seen by comparing 
Fig.~\ref{fig:xis_time}
with Fig.~\ref{fig:contam_xis}, which shows the optical depth and
fractional effective area at two energies, 0.65~keV and 1.0~keV.
These estimates are made by fitting simple Gaussians to the bright
\ion{O}{VIII}~Ly$\alpha$ and \ion{Ne}{X}~Ly$\alpha$ lines, along with a 
simple model for the
continuum and other lines, and comparing those line normalizations to
the IACHEC values.  The dashed lines in the top panel show the current
contamination CALDB for an on-axis source, produced from multiple
calibration sources including E0102.  Several things are apparent from
this figure.  First, the contamination built up very quickly, reaching
nearly a maximum on XIS3 within 6 months.  This is about the level
that ACIS reached after several years.  Modeling the build-up early is
hampered by a lack of calibration observations during this time.
Second, the measured line normalizations for XIS3 (and to a lesser
extent XIS1) do not match the CALDB well between 2005 and 2008, with a
bump-like feature in the CALDB trend that is not reflected in the data
points.  These calibration issues will be addressed in future work.

\subsubsection{Swift XRT}

\label{swiftxrt_temporal}

\begin{figure}
\resizebox{\hsize}{!}
{\includegraphics[width=\textwidth,angle=270]{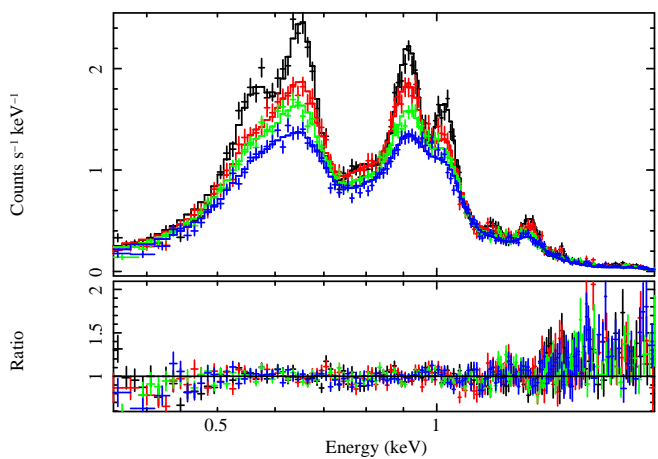}}
\caption{{\em Swift}-XRT WT spectra  from  from (top to bottom)  2005 (black),
2009 (red), 2011 (green), 2013 (blue) and their data/model ratios.}
\label{fig:swiftxrtwt3epochs_spec}
\end{figure}

Figure~\ref{fig:swiftxrtwtepochs} show the temporal behavior of the
line normalisations for the {\em Swift}-XRT WT data, with
representative spectra (from 2009, 2011, 2013) shown in
Fig.~\ref{fig:swiftxrtwt3epochs_spec}. The latter, when compared
with Fig.~\ref{fig:swiftxrt2005}, illustrates how the spectral
resolution has degraded with time, even with CCD charge trap
corrections applied.
The line normalisations are initially well behaved, but from the
second half of 2007 show an occasional loss in line flux, particularly
in the lowest energy line (i.e. the \ion{O}{vii}~He$\alpha$ near $0.57\keV$)
in 2007(s6), 2008 and 2009.
 
This discrepant behavior appears to occur when the main emitting ring
of the remnant (radius $10-18\arcsec$) falls on the CCD bad-column gaps.
\cite{gaetz2000} and \cite{flanagan2004}
show that the line components have different spatial
origins, with \ion{O}{vii}~He$\alpha$~{\em r} arising predominantly from the south-east
quadrant of the ring, while \ion{Ne}{x}~Ly$\alpha$~{\em r} is more symmetrically formed
in the ring.  Individual snapshots are quite frequently placed with
parts of the remnant on the CCD bad columns. Depending on the
orientation of the remnant with respect to the bad-columns we can see
a suppression of all four line fluxes compared to the IACHEC model, or
sometimes a more preferentially loss of the \ion{O}{vii}~He$\alpha$ component.
At this point in time, it is not possible to account for these effects
as no method exists to correct for such spatial-spectral variation
when the ARFs are generated.

Other factors could contribute to the normalisation variations, such
as the existence of deep charge traps which remove events below the
onboard threshold, or a response kernel width which does not quite
match the data. While other missions report time variable low energy
contamination, we do not think that such a problem exists for the {\em
  Swift}-XRT, because observations of low column density sources show
no significant increases in the inferred column density with time.

\section{Discussion}\label{disc}

\subsection{Comparison with Other Cross-Calibration Studies}
\subsubsection{\xmmn\/  Internal Cross-calibration Studies}


  There have been several studies of the cross-calibration amongst the
  instruments on {\xmmn}. \citet{mateos2009} used {\tt 2XMM} sources
  that were bright enough to have sufficient counts but not bright
  enough to have significant pileup to compare the derived fluxes in
  several energy bands: 0.2--0.5, 0.5--1.0, 1.0--2.0, 2.0--4.5, and
  4.5--12.0~kev. The data were processed with SAS~v7.1 and the
  calibration files available at that time. Of interest for our study,
  \citet{mateos2009} found that on-axis sources returned a MOS1/MOS2
  flux that was 5-7~\% higher than the EPIC-pn flux in the 0.5--1.0~\keV\/
  band. \citet{stuhlinger2010} conducted a similar analysis after
  processing the data with SAS~v10.0 and compared the EPIC-MOS and EPIC-pn
  fluxes in several narrow bands using {\tt 2XMM} sources that were not
  piled-up. In their 0.54--0.85~keV band, they found that the MOS1 and
  EPIC-pn returned nearly identical fluxes, while the MOS2 returned fluxes
  that were on average 5\% higher than EPIC-pn fluxes. An update of this
  analysis with SAS~v15.0 shows MOS1/MOS2 returning 4/7\% higher
  fluxes in the 0.54--0.85~keV band than EPIC-pn; however, the distribution
  of values spans the range from 0--12\%.  

  \citet{read2014}
  conducted a similar analysis with on-axis  {\tt 2XMM} sources after
  processing the data with SAS~v12.0. They used the ``stacked
  residuals'' method to quantify the derived flux differences between
  EPIC-MOS and EPIC-pn on a finer energy scale. They find that the flux
  difference between MOS1/MOS2 and  EPIC-pn peaks at around 5\% at an energy of
  $\sim0.6$~keV in the 0.5--1.0 keV  band (with the MOSs returning
  higher fluxes), while at other energies in the 0.5--1.0 keV  band
  the difference is typically a few percent. For the E0102 line fluxes
  described in this paper, EPIC-MOS and EPIC-pn differ by 10-15\% in the
  0.5--1.0~keV bandpass. It should be noted that the RGS effective
  area below 0.5~keV was adjusted to agree better with the EPIC-pn in
  SAS~v10.0. It is not clear why the results are
  apparently significantly different for {\tt 2XMM} sources on-axis
  compared to E0102 on-axis.  E0102 is obviously an extended source
  but given that the diameter is only 45\arcsec, it is difficult to
  understand how the effective area could be that different at such 
  small off-axis angles compared to on-axis. The flux in the E0102 spectrum is
  mostly concentrated in the four bright line complexes that we have
  fit in this paper, while the {\tt 2XMM} sources have spectra that
  more closely resemble a continuum in the 0.5--1.0~keV band. There
  are several effects contributing at the $\sim1\%$ level, 
  that might contribute to this discrepancy.  For example, there is
  still pileup in the EPIC-MOS and EPIC-pn, the use of a point source PSF is an 
  approximation, the vignetting function is also approximated, the
  gain fit effects the line normalizations, and the point source
  analysis still does not include 100\% of the encircled energy fraction.
  This apparent discrepancy remains under
  investigation.

\subsubsection{External Cross-calibration Studies}

There have been several analyses initiated by the IACHEC to compare
the cross-calibration of the effective areas of the current generation
of X-ray instruments. \cite{tsujimoto2011} used the Galactic SNR G21.5-0.9
to compare the effective area in the 2.0--8.0~keV range for \chan,
\xmmn, \swift, and \suzaku\/ using the calibration products available at
the writing of that paper. G21.5-0.9 has a highly absorbed spectrum
with an \NH\/ of $\sim3.0\times10^{22}~{\rm cm^{-2}}$ such that there is
little flux below ~2.0~keV. \cite{tsujimoto2011} found that the
2.0--8.0~keV fluxes generally agreed to within $\pm10\%$, with the
maximum discrepancy being between ACIS-S3 and EPIC-pn with ACIS-S3 producing a
flux $\sim16\%$ higher than EPIC-pn. \citet{ishida2011} used the BL Lac
object PKS~2155-304 to compare the derived spectral parameters
(power-law index and column density) for XIS, EPIC-MOS, EPIC-pn, and the  
{\em Low-Energy Transmission Grating} (LETG) on \chan\/ used with both
ACIS and the {\em High Resolution Camera} (HRC).  Since E0102 has not
been observed with the LETG on \chan,  we can not compare the results of
\citet{ishida2011} to our results.  Although the
primary objective of this study was to compare the consistency of the
fitted parameters that determine the shape of the spectrum,
\citet{ishida2011} did compute fluxes in narrow bands.  Since
PKS~2155-304 is variable, both in intensity and spectral shape, the
observations had to be executed simultaneously or as simultaneous as
possible given the scheduling constraints of the observatories.  
They computed
the fluxes in several bands from 0.5 to 10.0~keV, the most relevant
for comparison to our results is the 0.5--1.0~keV band.  There were
three simultaneous observations with \xmmn\/ and \suzaku\/ in 2005, 2006,
and 2008.  The EPIC-pn, EPIC-MOS, and XIS fluxes agree to within 10\% in the
0.5--1.0~keV band, although the order of which instrument produces the
highest flux is not the same for the three epochs.  The EPIC-pn produces
higher fluxes than the MOS1 in all three epochs, in disagreement with
the studies based on {\tt 2XMM} sources.  The XIS1 produces the lowest
average flux, while the XIS0 is higher than the XIS1.  The order of
which instrument  produces the higher flux is opposite to what we observe 
with E0102. They observe the EPIC-pn to produce higher fluxes than the EPIC-MOS and 
XIS0 to produce higher fluxes than XIS1.  The
differences are on the order of 5\% so this may be an indication of the 
systematic uncertainties that persist in our analyses.  Or it might be 
indicating
something fundamentally different in the calibration of extended
sources with line-dominated spectra compared to point sources with
continua spectra.

  There have a been a series of papers motivated by the IACHEC using
  cluster of galaxies to compare the calibration of the current
  generation of instruments.  \citet{nevalainen2010} used a sample
of relaxed clusters to compare the temperatures and fluxes derived by
\xmmn, \chan, and {\em BeppoSAX}. \citet{nevalainen2010} computed
fluxes in a hard band (2.0--7.0~keV) and a soft band
(0.5-2.0~keV). They found that the ACIS fluxes were on average 11\%
higher than the EPIC-pn fluxes in the hard band, similar to the findings of
\cite{tsujimoto2011} in a similar band. In the soft band, they find
that the ACIS and EPIC-pn fluxes agree on average to within 2\%, however
the scatter is large with outliers at -10\% and +14\%. It is not clear
what the explanation is for the relatively large scatter in this
cluster sample. \citet{kettula2013} expanded this study to include the
XISs on \suzaku.  They noted discrepancies between the XIS0, XIS1, and
XIS3 in the derived spectral shape that could be addressed with a
modification to the \suzaku\/ contamination model. Later releases of the
contamination model brought the three XISs into better agreement.
\citet{schellenberger2015}
analyzed 63 clusters from the HIFLUGCS sample \citep{reiprich2002}
performing a similar analysis as \citet{nevalainen2010} using the most
recent software and calibration at the time of writing that paper, 
CIAO~4.5 and CALDB 4.5.5.1 for \chan\/
and SAS~v12.0.1 and CCF dated 14.12.2012 for \xmmn. 
\citet{schellenberger2015}'s primary objective was to characterize any
systematic difference in the derived temperatures of clusters for
cosmology, but they did conduct a stacked residuals analysis similar to
\citet{read2014} to characterize differences in the effective area
calibration.  They find in the 0.5--1.0~keV band that ACIS produces
0--10\% higher fluxes than EPIC-pn and MOS1/MOS1 produce 0--5\% higher
fluxes in rough agreement with \cite{read2014}.

The studies described above used sources with different properties (point 
vs. extended, thermal vs. non-thermal spectra, etc.) and used different
versions of the calibration files and analysis software.  Therefore, it is
difficult to know if the results would be more consistent if the same
calibration files and analysis software were used on the same types of sources.
Nevertheless, the majority of the studies indicate that ACIS produces the
highest fluxes in the 0.5--1.0~keV band, the EPIC-MOS produces the second
highest fluxes, and the EPIC-pn produces the lowest fluxes.

\subsection{The E0102 Cross-calibration Results}

  The cross-calibration results using E0102 presented in
  Table~\ref{tab:fit2} and Fig.~\ref{fig:comp_norms}
differ from the previous
  studies in several ways.  It is clearest to discuss the differences
  line by line or equivalently, energy by energy. At the energy of the
  \ion{O}{vii}~He$\alpha$~{\em r} ($\sim 570~eV$), MOS1/MOS2 produce
  significantly higher fluxes than EPIC-pn, and ACIS-S3 produces the lowest
  fluxes.  All XISs produce fluxes consistent with the IACHEC model
  value  except for  XIS3 which is $\sim20\%$ lower
  than the IACHEC value.  The XRT WT mode data are within 5\% of the
  IACHEC value (for the remainder of this discussion we will limit the
  discussion to the XRT WT mode data since they agree better with the
  IACHEC model than the PC mode data).  The relative ordering of ACIS-S3,
  EPIC-pn, and MOS1/MOS2 is different than what the stacked residuals
  method with clusters and {\tt 2XMM} sources determined.  It should
  be noted that the E0102 analysis is sampling the response of the
  instruments in a narrower range of energies (essentially a line) for
  the CCD instruments while the stacked residuals approach must
  necessarily sum over a range of pulse-heights or energies.  The
  agreement at the energy of \ion{O}{viii}~Ly$\alpha$ line ($\sim
  654~eV$) is better in terms of the magnitude of the difference.  The
  XIS instruments agree better with the IACHEC model value except for
  XIS3 which is  $\sim15\%$ lower than the IACHEC value.
  The ACIS-S3 value is in agreement with the EPIC-pn value and both are only
  slightly lower than the IACHEC value and the XRT WT mode data are in
  excellent agreement with the IACHEC value. The EPIC-MOS values are $\sim15\%$
  higher than the EPIC-pn value.  But the relative order
  of ACIS-S3, EPIC-pn, and EPIC-MOS are different again compared to the stacked residuals
  approach. At the energy of the \ion{Ne}{ix}~He$\alpha$~{\em r} ($\sim
  922~eV$), the agreement is within 10\% for all instruments.   At the
  energy of the \ion{Ne}{X} ~Ly~$\alpha$ line ($\sim 1022~eV$), the
  agreement is within 15\% for all instruments. It is somewhat
  surprising that the MOS1/MOS2 and EPIC-pn disagree by 10-15\% at this
  energy.

  The fact that ACIS-S3 and XIS disagree with the IACHEC model
  the most at the lowest energy
  line for which we conducted a comparison could indicate a deficiency
  in the contamination models for each instrument.  ACIS-S3 and the XIS
  have the largest correction for a contamination layer of any of the
  instruments. As noted earlier, if later ACIS-S3 observations were 
  used for this comparison, the \ion{O}{vii}~He$\alpha$~{\em r} normalization  would be larger than it
  was for the first two ACIS-S3 observations used and in better agreement
  with the IACHEC model.  This might also indicate that there is some
  residual pileup which is depressing the fluxes of the lowest energy
  lines in the ACIS-S3 data.
  The latest XIS observations indicated that the contamination layer
  on \suzaku\/ was getting thinner with time.  Both of these facts indicate how
  challenging it is to develop an accurate contamination model for all
  times of a mission.  

  E0102 is fundamentally different from the point sources and clusters
  used in the stacked residuals analyses. First, E0102 has a
  line-dominated spectrum where the majority of the flux in the
  0.5--1.0 keV band is produced by 4 lines/line complexes.  The
  response of these instruments is changing rapidly over the
  0.5--1.0~keV band.  Therefore, it is possible that the E0102 data
  might reveal issues with the calibration as a function of energy
  more clearly than a
  continuum source.  Second, E0102 is an extended source with a diameter
  of 45\arcsec.  The point sources from the {\tt 2XMM} analyses sample
  a different part of the detectors and the point-spread function of
  the telescope, although one hopes that the telescope response
  changes little over 45\arcsec.  There may still be some residual
  pileup affecting the {\tt 2XMM} sources.  The cluster analysis
  samples a much larger region on the detector and larger off-axis
  angles than the E0102 analysis. For these reasons, each of these
  studies is probing different aspects of the instruments'
  calibrations.

  The time-dependent line normalizations we presented highlight the
  challenges in providing an accurate calibration throughout a mission
  as the CCD response changes due to radiation damage and the
  accumulation of a contamination layer. The EPIC-pn instrument has the
  most stable response and in this respect has the simplest job in
  providing a time-dependent calibration.  Both ACIS-S3 and the XIS have
  a time-variable contamination layer that has a complex time dependence.
  For the XIS, the contamination layer grew at different rates
  on the different detectors and appeared to be decreasing at the end
  of the mission.  For ACIS, the accumulation rate has varied over the
  mission and the chemical composition has also changed with
  time. Given how much the ACIS-S3 and XIS response have changed in the
  0.5--1.0~keV over the mission, it is encouraging that the line
  normalizations are as stable as they are with time.  The EPIC-MOS
  response has changed because of the ``patch'' effect in the CCDs and
  the accumulation of a contamination layer.  Fortunately for the EPIC-MOS,
  the contamination layer is much thinner than on ACIS or the XIS and
  it has been simpler to model.  The XRT WT mode data have exhibited
  significant evolution with time, with the lowest energy line
  normalizations decreasing the most with time. This behavior might be
  explained by the placement of E0102 on bad columns that are
  increasing with time.
  The E0102 analysis
  presented here is the first of these cross-calibration studies to
  characterize the time-dependent response of these instruments. It is
  our hope that the Guest Observer community for these missions can
  use these results to assess the reliability of their results.

\section{Conclusions}\label{conc}

We have used the line-dominated spectrum of the SNR E0102 to test the
response models of the ACIS-S3, EPIC-MOS, EPIC-pn, XIS, and XRT CCDs below
1.5~keV. We have fitted the spectra with the same model in which the
continuum and absorption components and the weak lines are held fixed
while allowing only the normalizations of four bright lines/line
complexes to vary.  We have compared the fitted line normalizations 
of the \ion{O}{vii}~He$\alpha$~{\em r} line, the \ion{O}{viii}~Ly$\alpha$ line, 
the \ion{Ne}{ix}~He$\alpha$~{\em r} line, and \ion{Ne}{X}~Ly~$\alpha$ line
to examine the consistency of the effective area models for the
various instruments in the energy ranges around 570~eV, 654~eV,
915~eV, and 1022~eV.
We find that the instruments are in general 
agreement with 38 of the 48 scaled normalizations within $\pm10\%$
of the IACHEC model values.
However, the agreement is better for the higher energy lines than
the low energy lines.  We find that the scaled line
normalizations agree with the IACHEC model normalization to 
within  $\pm9\%$ \& $\pm12\%$ for the \ion{Ne}{ix}~He$\alpha$~{\em r} \&
\ion{Ne}{X}~Ly$\alpha$ line complexes when all instruments are considered 
(if we adopt the {\em Swift} XRT WT mode results and exclude the PC
mode results).  The agreement is significantly worse for the low
energy lines, as the scaled line
normalizations agree with the IACHEC model normalization to 
within $\pm20\%$ \& $\pm15\%$ for the
\ion{O}{vii}~He$\alpha$~{\em r} and \ion{O}{viii}~Ly$\alpha$~{\em r}
line complexes.  This difference with energy emphasizes the challenges
presented by the low energy calibration of these CCD instruments.
When only \chan\/ and 
\xmmn\/ are considered, we find that the fitted line
normalizations agree with the IACHEC normalization to within $\pm15\%$, 
$\pm12\%$, $\pm8\%$, \& $\pm9\%$ for
\ion{O}{vii}~He$\alpha$~{\em r}, \ion{O}{viii}~Ly$\alpha$, 
\ion{Ne}{ix}~He$\alpha$~{\em r}, \& \ion{Ne}{X}~Ly$\alpha$.
Therefore, the absolute effective areas of \chan\/ ACIS-S3, 
\xmmn\/ EPIC-pn,
  and \xmmn\/-MOS agree to better than $\pm10\%$ at 0.9 and 1.0~keV for
  the time intervals and data modes considered in this analysis.

The time dependence of each of the CCD instruments was presented. 
All of the CCD instruments have a significant variation in response
with time except for the \xmmn\/ EPIC-pn which measures a flux in the
0.3--2.0~keV band consistent to 1.3\% over the course of the mission.
The derived normalizations with time for ACIS-S3, EPIC-MOS, XIS, and XRT for
the  \ion{O}{vii}~He$\alpha$~{\em r} line, the \ion{O}{viii}~Ly$\alpha$ line, 
the \ion{Ne}{ix}~He$\alpha$~{\em r} line, and \ion{Ne}{X}~Ly$\alpha$ line can
be used to assess the reliability of the effective area calibration at
a given point in time for the respective mission/instrument.

\begin{acknowledgements}
This work was supported by NASA contract NAS8-03060.
APB acknowledges support from the UK Space Agency.
We thank Daniel Dewey for the analysis of the HETG data that was
critical in the development of the IACHEC standard model for E0102.
We thank Joseph DePasquale and Jennifer
Posson-Brown for their years of assistance in analyzing the \chan\/
data of E0102 that developed the methods and software for the ACIS
analysis. We thank Randall Smith for helpful discussions about the use
of the APEC models. We thank Herman Marshall and Alexey Vikhlinin
for helpful discussions
on the analysis and the modifications to the ACIS contamination model.
We thank Marcus Kirsch who took a leadership role in setting
up the IACHEC  and Matteo Guainazzi for leading the IACHEC over
the last several years. 
\end{acknowledgements}

\bibliographystyle{aa}
\bibliography{ms}

\end{document}